\documentclass[a4paper,11pt]{article}
\pdfoutput=1
\usepackage{jheppub}
\usepackage{placeins}
\usepackage{adjustbox}
\usepackage[table]{xcolor}
\usepackage{amsfonts,algorithmic,algorithm,subcaption,multirow}
\graphicspath{{graphics/},{graphics/qq/},{graphics/gg/},{graphics/qq2qq/},{graphics/qg2qg/},{graphics/gg2gg/},{graphics/qqqq/},{graphics/diagrams/}}

\newcommand{\td}{\mathrm{d}}
\renewcommand{\>}{\rangle}
\newcommand{\<}{\langle}
\newcommand{\Nc}{N_{\mathrm{c}}}

\newcommand{\LC}{LC$'$}
\newcommand{\CVolver}{\texttt{CVolver}}
\newcommand{\unitop}{\mathbb{I}}

\title{Exact colour evolution for jet observables}

\author[a]{Jeffrey R. Forshaw,}
\author[b,c]{Simon Pl\"atzer,}
\author[a]{and Fernando Torre Gonz\'alez}
\affiliation[a]{Department of Physics and Astronomy, University of
  Manchester, Manchester M13 9PL, United Kingdom}
\affiliation[b]{Institute of Physics,
  NAWI Graz, University of Graz, Universit\"atsplatz 5, A-8010 Graz,
  Austria}
\affiliation[c]{Particle Physics, Faculty of Physics, University of
  Vienna, Boltzmanngasse 5, A-1090 Wien, Austria}

\emailAdd{jeffrey.forshaw@manchester.ac.uk}
\emailAdd{simon.plaetzer@uni-graz.at}
\emailAdd{fernando.torre@manchester.ac.uk}

\abstract{We perform a systematic and comprehensive analysis of sub-leading colour corrections in perturbative QCD processes involving multiple soft gluon emissions. This necessitates going beyond the standard parton shower paradigm in order to incorporate interference effects and is accomplished using the \CVolver\ program, which simulates parton showers at the amplitude level. We can compute cross-sections with full-colour precision and also broken down explicitly in terms of their $N_c$ dependence. In this paper, we focus on the jet cross-section with a veto of additional jets in some fixed region of phase-space, since this is sensitive to wide-angle, soft gluon radiation. We consider $Z \to q \bar{q}$, $H \to gg$, $q\bar{q} \to q\bar{q}$, $qg \to qg$, $gg \to gg$ and $ZZ \to q\bar{q}q\bar{q}$. We find that non-trivial sub-leading colour effects are generally important at the 5 -- 30\% level and much more than this for certain interference contributions. Remarkably, for this observable, we find that for all of the $t$-channel gluon exchange processes that we consider, the strictly leading colour approximation, which includes the replacement $C_F \to C_A/2$, is an excellent approximation to the full colour result (up to an overall colour factor).}

\begin{document}

\maketitle
\flushbottom

\section{Introduction}
\label{sec:Introduction}

Standard parton shower algorithms, as implemented in several well-established Monte Carlo simulation codes \cite{Bewick:2023tfi,Sherpa:2024mfk,Bierlich:2022pfr}, have been immensely successful in simulating high-energy particle collisions and are extremely useful for analysing data. Considerable ingenuity has allowed the precision of these implementations to steadily improve to the point that one can now contemplate the notion of a shower with next-to-next-to-leading logarithmic accuracy \cite{vanBeekveld:2024wws}. However, there is a fundamental limitation in accuracy that cannot be resolved within the standard paradigm, which is built around classical parton branching. To properly include interference effects in the shower requires evolving a density matrix. Density matrix evolution is also required in order to include corrections suppressed by powers of $1/N_c$ \cite{Nagy:2012bt,Nagy:2015hwa,Nagy:2019pjp}. Necessarily, these corrections are absent in the standard approach.

We have developed a framework and simulation code, \CVolver, which is able to include interference effects for the evolution of general hard-scattering processes involving any number of coloured partons \cite{Platzer:2013fha,DeAngelis:2020rvq}. In this paper, we will present extensive results from a \CVolver\ plugin providing soft gluon evolution for a wide range of partonic processes evolved at the amplitude level. We are able to track powers of $N_c$ and thereby compute at any specified accuracy in the $1/N_c$ expansion. This is a highly non-trivial exercise due to the rapid increase in the complexity of the colour structure as more gluons are emitted. Using the colour-flow basis, we can count powers of $\Nc$ in such a way that we can dynamically sample the most important colour flows and thereby perform calculations with modest computational effort. Throughout this paper, we present results obtained using \CVolver\ in a mode dedicated to computing jet veto cross sections. \CVolver\ can also be operated in a more general mode, as a partonic event generator, and we will report results from operating in this way in a partner paper \cite{CVolver:eventgenerator}.

The remainder of this paper is organised as follows. In Section~\ref{sec:evolution} we briefly summarise the colour evolution algorithm that has been implemented in \CVolver\ including a brief discussion of the colour flow basis, the all-orders summation of the virtual corrections and the roles of dipoles, rings and strings. We also discuss some of the specific details of the algorithm, specifically, how we are able to track the powers in $N_c$ and how we handle collinear singular regions. Section~\ref{sec:FinalState} presents results on primary dijet production from a colour singlet initial state, as might occur in $Z \to q\bar{q}$ or $H \to gg$. We focus our attention on the jet veto cross-section for fixed kinematics of the jets, i.e. we veto the production of additional gluons in some fixed angular region if their energy exceeds some threshold. Our goal is to provide a systematic study of sub-leading colour effects in a clear and controlled manner. In Section~\ref{sec:2to2} we move on to study dijet production in hadron-hadron collisions, again for fixed parton kinematics. Finally, we look at the colour singlet production of $q\bar{q}q\bar{q}$, as might occur following the hadronic decay of two $Z$ bosons.

\section{Soft gluon evolution}
\label{sec:evolution}

\begin{figure}[htbp!]
\centering
\begin{subfigure}[t]{0.4\textwidth}
\includegraphics[width=0.9\textwidth]{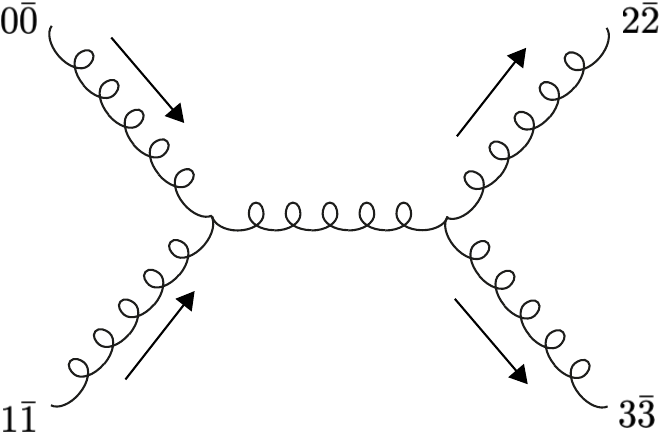}
\caption{The hard process diagram corresponding to $s$-channel gluon exchange.}
\label{fig:ggFeynman}
\end{subfigure} \hfill
\begin{subfigure}[t]{0.4\textwidth}
\includegraphics[width=0.9\textwidth]{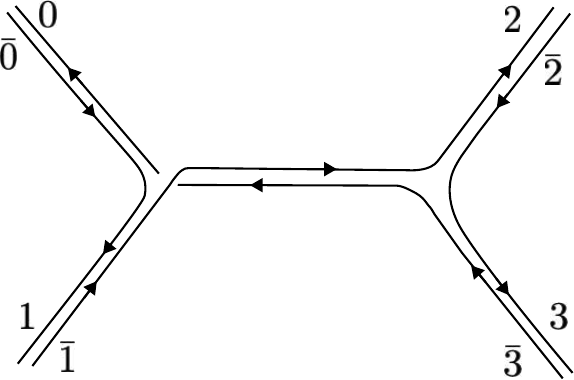}
\caption{The diagram of a possible configuration in the colour flow basis.}
\label{fig:ggFlow}
\end{subfigure} \\ \vspace*{0.5cm}
\begin{subfigure}[t]{0.4\textwidth}
\centering
\includegraphics[width=0.28\textwidth]{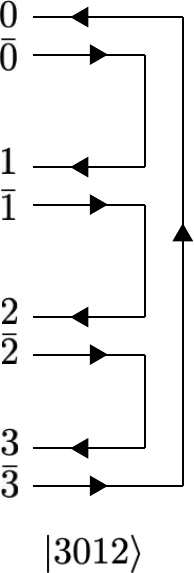}
\caption{The corresponding colour flow diagram.}
\label{fig:ggPerm1}
\end{subfigure} \hfill
\begin{subfigure}[t]{0.4\textwidth}
\centering
\includegraphics[width=0.51\textwidth]{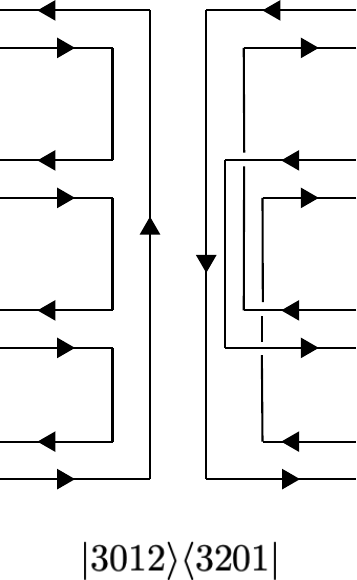}
\caption{An interference contribution to the hard scattering density matrix. This contribution is relevant in the case of $st$ and $su$ channel interference.}
\label{fig:ggPerm2}
\end{subfigure} \\ \vspace*{0.5cm}
\begin{subfigure}[t]{1\textwidth}
\centering
\includegraphics[width=1\textwidth]{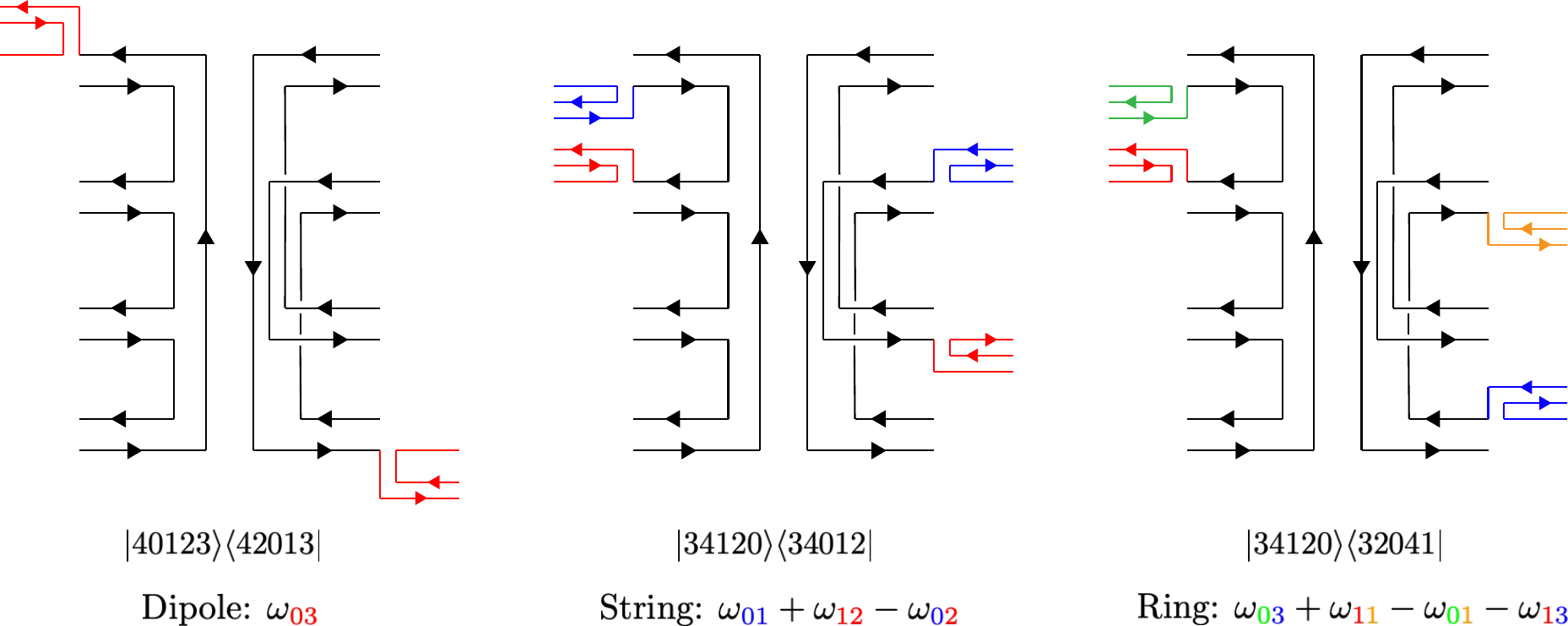}
\caption{The emission of a gluon via the $\mathbf{D}$ operator from a dipole, a string and a ring. The kinematic factor and updated density matrix after the emission are also indicated. For the string and ring, only one gluon is emitted from a colour line in the amplitude and conjugate. The different possibilities are coloured differently. }
\label{fig:ggDipoleStringRing}
\end{subfigure} 
\caption{An example to illustrate the colour flow basis and how we classify real emissions in terms of dipoles, strings and rings.}
\label{fig:ggFigure}
\end{figure}

The leading soft gluon logarithms, due to the emission of wide-angle, low-energy gluons, can be summed to all orders in QCD perturbation theory using a simple recursive algorithm that is well-suited to numerical implementation. What follows is a brief summary of the salient features of the algorithm since we have presented it in detail elsewhere \cite{AngelesMartinez:2018cfz,DeAngelis:2020rvq}. The algorithm can be derived from the more general approach given in \cite{Platzer:2022jny}. The density matrix evolves from an initial hard scattering matrix, $\mathbf{H} = |\mathcal{M}\>\< \mathcal{M}|$, where $\mathcal{M}$ labels a partonic scattering amplitude, according to
\begin{align}
\mathbf{A}_n(E) = \mathbf{V}_{E,E_n} \mathbf{D}_n^\mu \mathbf{A}_{n-1}(E_n) \mathbf{D}_{n\mu}^\dag\mathbf{V}_{E,E_n}^\dag \Theta(E \leq E_n),
\end{align}
where $\Theta$ is the Heaviside function. In our current study, we can interpret the evolution variable $E$ directly as the energy of an emitted gluon, though in general this scale should merely be referred to as a factorization scale \cite{Platzer:2022jny}. $\mathbf{D}$ is the operator that adds one gluon emission into the final state with energy $E_n$ and $\mathbf{V}_{E,E_n}$ is the virtual evolution operator that inserts any number of gluon exchanges between the external legs. The soft gluon evolution is recursive starting from $\mathbf{A}_0(E) = \mathbf{V}_{E_,Q} \mathbf{H} \mathbf{V}_{E,Q}^\dag$. Boldface is used to represent operators in colour space (the spin evolution is trivial in the soft approximation) and for numerical purposes the operators are expressed in the colour flow basis, whose basis vectors can be represented by elements of the permutation group. We refer to one such element as a colour flow and the emission operator acts to increase the number of colour lines in a flow by one whilst the virtual evolution operator acts to re-arrange the colour lines in a flow. An example illustrating how the colour flow basis is used in \CVolver\ is illustrated in Fig.~\ref{fig:ggFigure}. The figure also illustrates how gluon emissions are classified in terms of dipoles, rings and strings \cite{Forshaw:2021mtj}. Dipole emission is when the gluon is emitted off the same colour line in the amplitude and conjugate amplitude. String emission occurs when the gluon is emitted off different dipoles that share one colour line and ring emission occurs when the gluon is emitted off disconnected dipoles. Dipole emissions are associated with a single kinematic factor, 
\begin{align}
\omega_{ij} = \frac{p_i \cdot p_j}{ p_i\cdot k \, p_j \cdot k},	
\end{align}
where $k$ is the emitted gluon momentum. Strings are associated with three distinct $\omega_{ij}$ terms that all lead to the same final colour flow, and rings come with four distinct terms. By classifying emissions as occurring from dipoles, rings or strings we are able to directly account for collinear cancellations that occur within individual rings and strings, which significantly improves the numerical stability of our evolution. The leading colour approximation involves only dipole emissions. 

To construct an observable, $\Sigma$, we use
\begin{align}
	\Sigma = \int \sum_n \text{Tr} \mathbf{A}_n \, \td \Pi_n \, u_n(\{ k_i \})
\end{align}
where $u_n(\{ k_i \})$ is the measurement function that depends on the final state momenta ($\{ k_i \}$) and $\td \Pi_n$ is the $n$-parton phase space. In the colour flow basis, which is not an orthogonal basis, we compute the trace with the aid of a scalar product matrix $\mathbf{S}$, such that $\text{Tr} \mathbf{A}_n = \text{Tr} \underline{\mathbf{A}}\underline{\mathbf{S}}$ and the underline denotes matrices in the colour flow basis. The $ab$ matrix element of the scalar product matrix is equal to $\Nc^{n-\#(a,b)}$, where $\#(a,b)$ is the minimum number of pairwise swaps required to map colour flow $a$ onto colour flow $b$ and $n$ is number of colour flows in the amplitude.

Real emission is relatively simple to implement. For gluon emission off a quark we must include the possibility of singlet gluon emission, which is suppressed by a factor $1/\Nc$ at the amplitude level. Singlet gluons have the property that the gluon is described by a single colour line. As such they are inert, which means they do not subsequently radiate nor do they participate in the virtual evolution. On the other hand, the virtual evolution is a challenge to implement, the relevant object is the colour flow matrix element $[\tau | \mathbf{V}_{E_1,E_2}|\sigma \rangle $, which involves the exponentiation of the single gluon exchange operator (anomalous dimension), i.e. $\mathbf{V} = \exp(\mathbf{\Gamma})$. To handle this we use a result presented first in \cite{Platzer:2013fha} in which this complicated matrix exponential is expressed in terms of $c$-number exponentials, as we now explain. The single gluon exchange operator has matrix elements
\begin{align}
[\tau |\mathbf{\Gamma} |\sigma \rangle = \delta_{\tau \sigma} \left(\Nc \Gamma_\sigma+\frac{1}{\Nc} \rho\right) + \Sigma_{\sigma \tau},	
\end{align}
where $\sigma$ and $\tau$ are two colour flows. These matrix elements depend upon the types of particles between which the gluon is exchanged, though we suppress this dependence for brevity. For explicit expressions for the $\mathbf{D}$ and $\mathbf{V}$ operators, and for more details on the basis algebra, see \cite{AngelesMartinez:2018cfz}.
The $\Gamma$ and $\rho$ contributions are colour diagonal and very easy to implement. In fact, $\rho$ is only non-zero for exchange between pairs of quarks and/or anti-quarks (it is responsible for correcting $C_A/2 \to C_F$ in the virtual evolution). The challenge is to include the $\Sigma$ operator, which is able to swap colour lines. For example, if a gluon is exchanged between gluons 0 and 2 in Fig.~\ref{fig:ggPerm1} then the colour flow could evolve from $|3012\rangle$ to $|3102\rangle$, which corresponds to a single swap in the permutation. There are other possible swaps depending on how the virtual gluon attaches to the colour lines of gluons 0 and 2. In contrast, the colour diagonal exchange of a gluon occurs between two gluons that are colour connected (these are the $\Gamma$ and $\rho$ terms). The result in \cite{Platzer:2013fha} states that

\begin{align}
\label{eq:simon}
[\tau|e^{\mathbf{\Gamma}}|\sigma\rangle &= \delta_{\tau \sigma}R(\{ \sigma \} ,\Gamma) + \sum_{l=1}^{d}\frac{(-1)^l}{\Nc^l} \sum_{k=0}^l \frac{(-\rho)^k}{k!} \nonumber \\ & \times \sum_{\{ \sigma_0, \sigma_1, \cdots, \sigma_{l-k}\}} \delta_{\tau \sigma_0} \delta_{\sigma_{l-k} \sigma} \left( \prod_{\alpha=0}^{l-k-1} \Sigma_{\sigma_\alpha \sigma_{\alpha+1}}\right) R(\{ \sigma_0, \sigma_1, \cdots, \sigma_{l-k}\},\Gamma) ,
\end{align}
where $d$ is taken to infinity. In practise $d=2$ is sufficient to achieve convergence in the $1/\Nc$ expansion for all of the results we present in this paper. The $R$ factors are no longer matrix exponentials (see~\cite{Platzer:2013fha} for details), for example, for $d=1$ we need
\begin{align}
R(\{ \sigma \},\Gamma) &= e^{\Nc \Gamma_\sigma} \\
\text{and~~~} R(\{ \sigma_0,\sigma_1 \},\Gamma) &= \frac{e^{\Nc \Gamma_{\sigma_0}} - e^{\Nc \Gamma_{\sigma_1}}}{\Gamma_{\sigma_0} - \Gamma_{\sigma_1}} \ .
\end{align}
The colour suppressed but diagonal terms involving $\rho$ that correspond to singlet exchange between quarks and anti-quarks can easily be summed to all orders into the $R$ functions simply by shifting the argument of the exponentials from $\Gamma \to \Gamma + \rho$ and taking only the $k=0$ term in the sum in Eq.~\eqref{eq:simon}. We usually choose to do this and, when no other sub-leading colour effects are included in the virtual evolution, we refer to it as the \LC\ approximation. Eq.~\eqref{eq:simon} allows us to count the powers of $1/\Nc$ as the evolution proceeds: each factor of $\Sigma$ is associated with a single swap in the permutation and a corresponding factor $1/\Nc$.
\vspace*{1ex}

\noindent \CVolver\ tracks the $1/\Nc$ factors as follows:
\begin{enumerate}
	\item Select an initial pair of flows from the hard scatter matrix. This may be associated with an explicit factor $1/\Nc^p$.
	\item Emit a gluon in the amplitude and its conjugate. If the gluon is a singlet, it is associated with a factor $1/\Nc$, which means any emission is associated with a factor $1/\Nc^q$ where $q = 0, 1$ or 2. After the emission, count the minimum number of swaps ($r$) by which the colour flows in the amplitude and conjugate amplitude differ. If this would be the final emission, there would then be a colour suppression of $1/\Nc^r$ from the scalar product matrix, $\underline{\mathbf{S}}$, and the total suppression factor would be $1/\Nc^{q+r}$. Crucially, if this is not the final emission it is not possible for further emissions to reduce the degree of colour suppression. This means that the cumulative suppression after the first real emissions is $p+(q+r)$.
	\item Operate with the virtual evolution operator in the amplitude and its conjugate. This is done according to the $d$-expansion, i.e. for $d=0$ this will involve zero swaps in the colour flows, for $d=1$ it will involve up to one swap (i.e. one factor of $\Sigma$) in each of the amplitude and conjugate (i.e. potentially two swaps in total) etc. Each swap is associated with a factor $1/\Nc$ giving a total suppression of $1/\Nc^s$. As in the case of the real emissions, there is a further factor $1/\Nc^t$ arising from the scalar product matrix (i.e. the swap pushes the colour flows apart so they differ by at least $t$ swaps). Again, subsequent evolution can never reduce the degree of colour suppression and so the cumulative suppression as this phase is $p+(q+r)+(s+t)$. 
	\item The entire process repeats for subsequent emissions and we veto the event if the cumulative suppression exceeds the required accuracy. This veto on events if the colour wanders too far from the diagonal is vital in ensuring we can handle multiple gluon emissions.      
\end{enumerate}

The soft gluon plugin of \CVolver\ must handle the divergences in $\omega_{ij}(k)$ when $k$ is parallel to $i$ or $j$. For the collinear safe observables which we consider in this study, these divergences cancel between the real and virtual evolution but need regulating in the numerical implementation. To do this we make the replacement
\begin{align}
\frac{n_i \cdot n_j}{n_i \cdot n\ n_j \cdot n} \to \frac{n_i \cdot n_j}{n_i \cdot n\ n_j \cdot n} \Theta\left( \text{min}\left( \frac{n\cdot n_i}{n \cdot n_j},\frac{n\cdot n_i}{n \cdot n_j} - \frac{\lambda}{n_i \cdot n_j}\right) \right),
\end{align}
where $\lambda$ is the collinear cutoff and the $n$ vectors are the dimensionless vectors that specify the directions of partons $i$, $j$ and the emitted gluon, $k$. This form of the regulator guarantees that\footnote{It also guarantees that rings integrate to the correct conformal cross ratios, which are free from divergences and therefore independent of $\lambda$.}
\begin{align}
	\int \frac{\td \Omega}{4\pi} \frac{n_i \cdot n_j}{n_i \cdot n\ n_j \cdot n} \Theta\left( \text{min}\left( \frac{n\cdot n_i}{n \cdot n_j},\frac{n\cdot n_i}{n \cdot n_j} - \frac{\lambda}{n_i \cdot n_j}\right) \right) = 2 \ln\frac{n_i \cdot n_j}{\lambda},
\end{align} 
where the integral is over the full solid angle of the emitted gluon. This integral is central to computing the virtual evolution. For observables which are not collinear safe, an evolution in the collinear cutoff will in general also need to be considered \cite{Platzer:2022jny}. We leave this study to future work.

\section{Colour singlet production of dijets}
\label{sec:FinalState}

\begin{figure}[htbp!]
\centering
\includegraphics[width=0.5\textwidth]{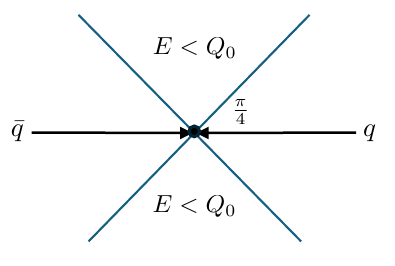}
\caption{Two partons (here a quark and anti-quark) are produced back-to-back and radiation is vetoed in between the corresponding jets.}
\label{fig:GBJ}
\end{figure}

We start with the production of a pair of jets from a colour singlet initial state, such as might be produced in $Z \to q \bar{q}$ or $H \to gg$. We fix the hard process kinematics such that the two jets are produced back-to-back, each with energy $Q$. Additional radiation is vetoed if it falls in the interjet region, which is defined by cones centred on each jet with an opening angle of $\pi/2$, and has energy $E>Q_0$, as illustrated in Fig.~\ref{fig:GBJ}. The relevant large logarithm is $\ln(1/\rho)$ where $\rho = Q_0/Q$. For this process the hard process density matrix is simply $H = \Nc$, corresponding to a single colour flow.

\begin{figure}[htbp!]
\centering
\includegraphics[width=0.9\textwidth]{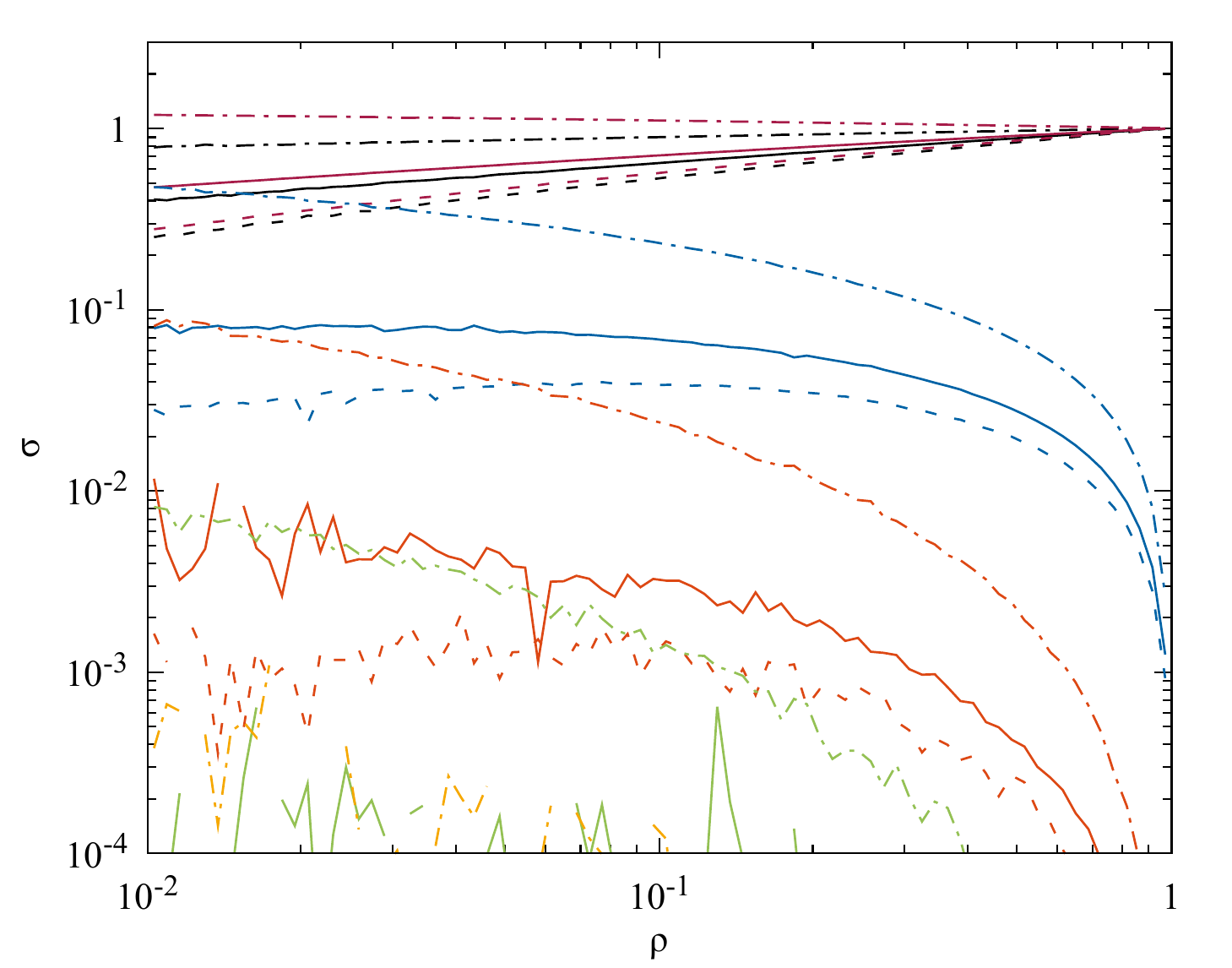}
\caption{The veto cross-section as a function of veto scale for $Z \to q \bar{q}$. Solid: $\Nc = 3$, Dash-dotted: $\Nc = \sqrt{2}$, Dashed: $\Nc = 4$.
Black: Full colour, Red: \LC, Blue: -NN\LC, Orange: N$^4$\LC, Green: -N$^6$\LC, Yellow: N$^8$\LC.}
\label{fig:qq-N8LC}
\end{figure}

In Fig.~\ref{fig:qq-N8LC} we show how \CVolver\ is able to compute the terms in the $1/\Nc$ expansion. The figure shows the cross-section (normalized to unity at $\rho = 1$) for different values of $\Nc$ (these are the different line types). The different colours show the contributions from each colour order. We could present results strictly in the expansion of $1/\Nc$ but since it is very easy to sum the colour diagonal virtual corrections to all orders (our \LC\ approximation) we choose to do so. Specifically, the black curves are computed in the \LC\ approximation, the blue curves are the $1/\Nc^2$ suppressed contributions to the \LC\ approximation, the orange are the $1/\Nc^4$ suppressed contributions etc. The solid green curve indicates that the $1/\Nc^6$ contribution is suppressed by 3 orders of magnitude, which is as one would naively anticipate. The figure also illustrates the effect of changing the numerical value of $\Nc$ and, as expected, we see an increase in the size of the $\Nc$ suppressed terms as $\Nc$ is reduced. In what follows we shall have much more to say about the nature of the sub-leading colour contributions. For now we simply note that the subleading colour contributions to the \LC\ approximation are substantial ($\approx 20\%$ for $\Nc = 3$ at $\rho = 10^{-2}$).

\begin{figure}[htbp!]
\centering
\begin{subfigure}[t]{0.9\textwidth}
\centering
\includegraphics[width=1.0\textwidth]{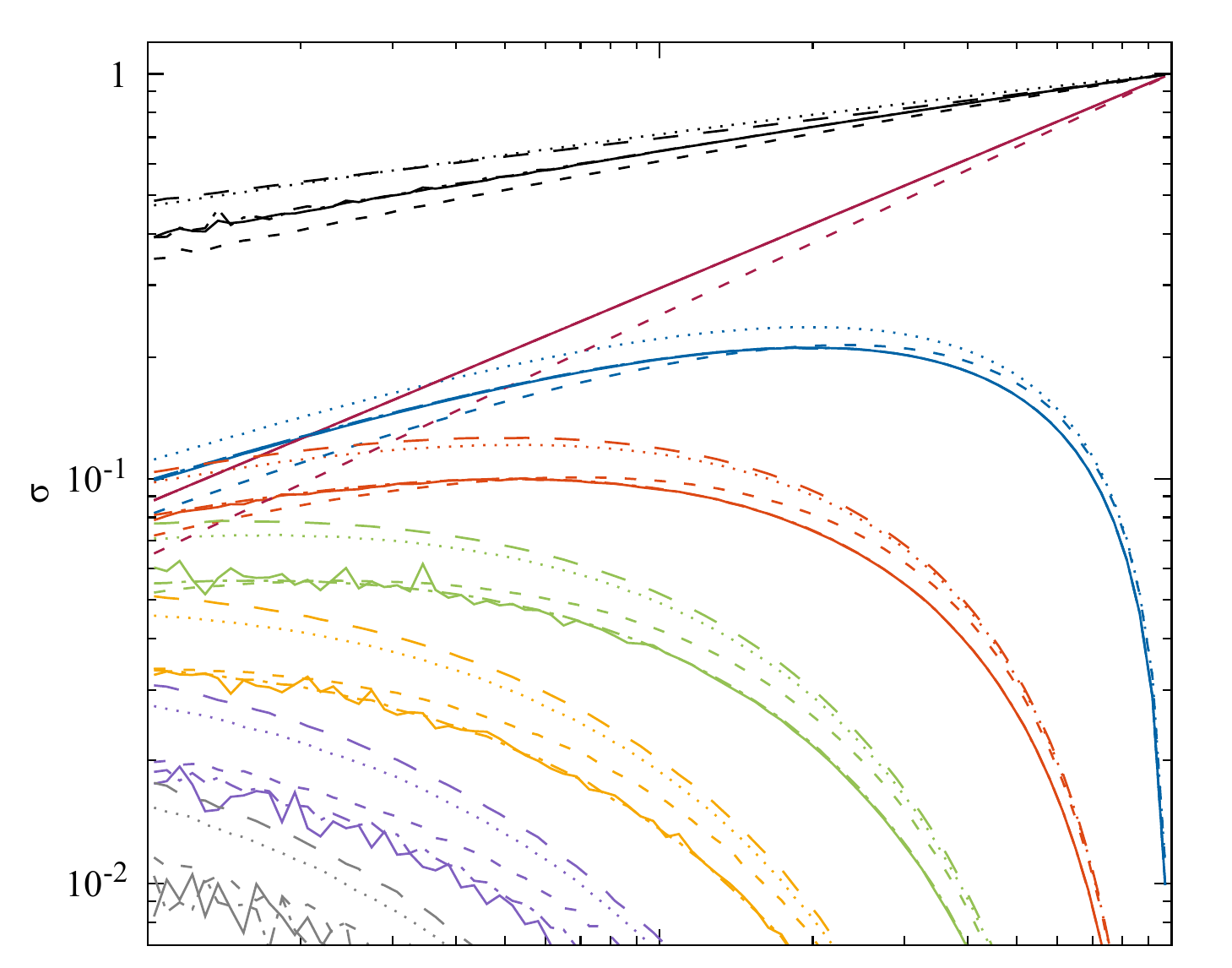}
\end{subfigure} \hfill
\begin{subfigure}[t]{0.9\textwidth}
\centering
\includegraphics[width=1.0\textwidth]{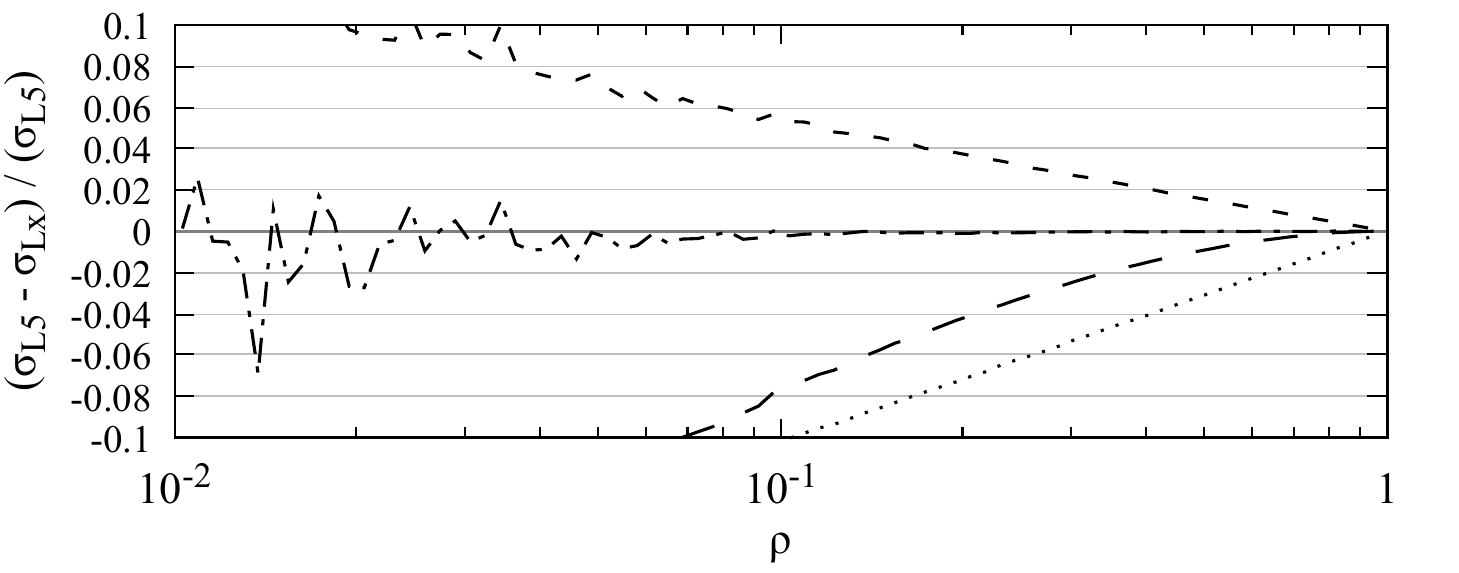}
\end{subfigure}
\caption{The veto cross-section as a function of veto scale for $Z \to q \bar{q}$. Solid: Full colour (L5), Dash-dotted: \LC\ + FCR (L4), Long-dashed: \LC\ + LCR + singlets (L3), Dotted: \LC\ + LCR (L2), Short-dashed: strict LC (L1). The different coloured curves correspond to different multiplicities (0 up to 6 emissions) and the black curves are the total cross-section. The lower residual plot is for the total cross-section.}
\label{fig:qq-multiplicity}
\end{figure}

Fig.~\ref{fig:qq-multiplicity} shows the veto cross-section broken down by multiplicity and colour accuracy. The line-types used in this plot will be used consistently throughout the remainder of the paper for processes involving any number of quarks or anti-quarks.  They are defined as follows.
\begin{itemize}
\item  Short-dashed: the strictly leading colour result. We will refer to this as our L1 result.
\item Dotted: the \LC\ result with the leading colour approximation in the real emissions. We will refer to this as our L2 result.
\item Long-dashed: the \LC\ result but with the real emissions computed by also including singlet emissions. This approximation should be closest to that of a standard parton shower algorithm. We will refer to this as our L3 result.
\item Dash-dotted:  the \LC\ result with a full colour treatment of the real emissions. This is the first approximation to include dipoles, rings and strings. It is our L4 result.
\item Solid: the full colour result. Here we go beyond the \LC\ approximation for the virtual corrections, i.e. we go beyond $d=0$ in Eq.~\eqref{eq:simon}. This involves substantially greater computational effort than the other approximations and is our L5 result.
\end{itemize}
A striking observation here is that the L4 and L5 results agree at the percent level. This is in accord with the observation in \cite{Hatta:2013iba} if we assume that our L4 result is approximately equal to the mean field approximation of \cite{Hatta:2013iba}. A corollary to this is that we can safely use the $d=0$ \LC\ approximation in the expansion of the virtual corrections. In other words, $\Nc/2 \to C_F$ in the diagonal part of the virtual corrections combined with a full colour treatment of the real emissions is sufficient to capture sub-leading colour effects in $Z \to q \bar{q}$. As we shall see, this feature does not extend to other processes. Note also that the zero emission curves for L2--L4 are exactly degenerate since the subsequent enhancements only impact upon the real emissions.
 
\begin{figure}[htbp!]
\centering
\begin{subfigure}[t]{0.48\textwidth}
\centering
\includegraphics[width=1.0\textwidth]{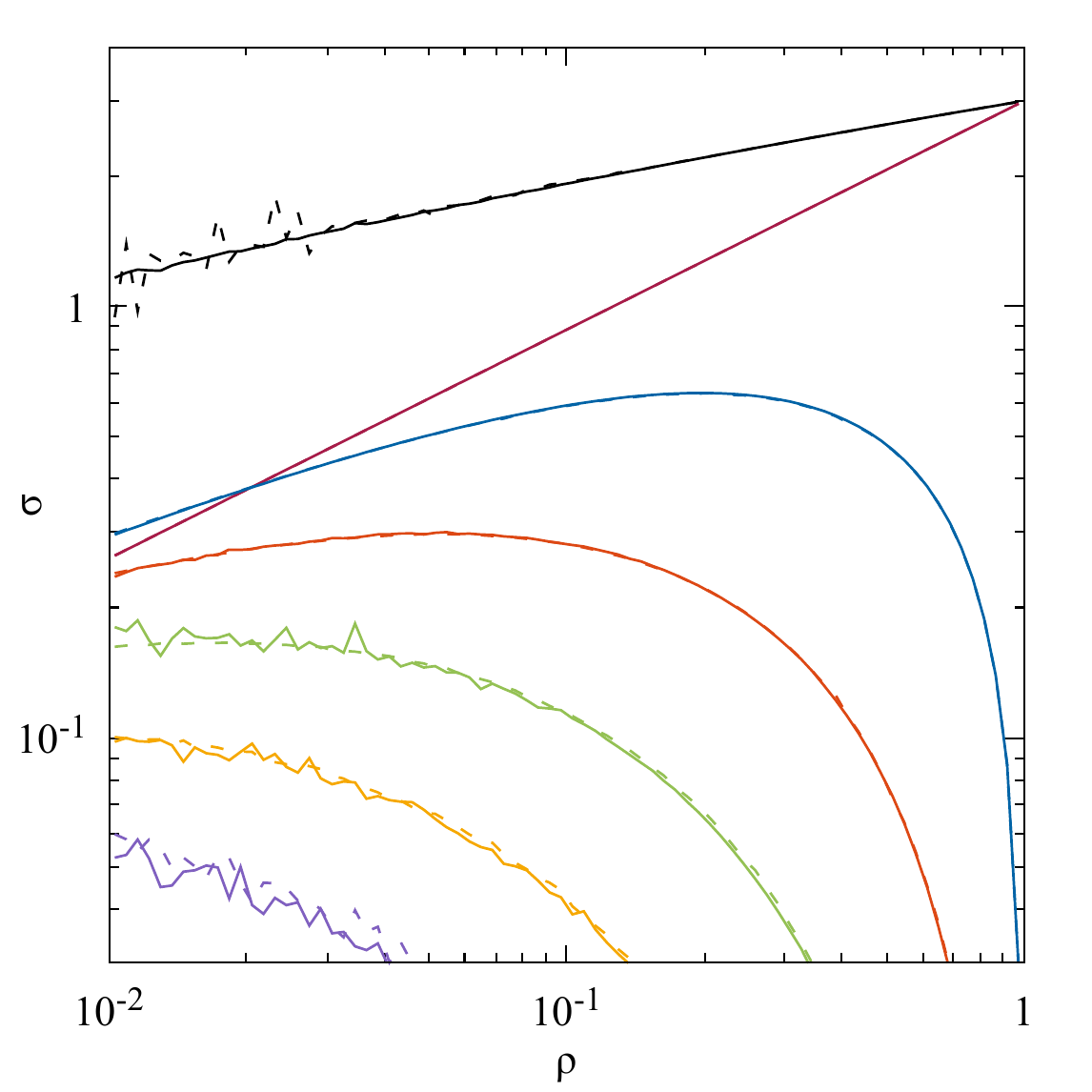}
\caption{Testing Lorentz invariance. The solid curves are for the standard kinematic configuration whilst the dashed curves correspond to a boosted configuration where the dijets have an opening angle of $\pi/4$. The interjet region and collinear cutoff are correspondingly boosted.}
\label{fig:controlLorentz}
\end{subfigure} \hfill
\begin{subfigure}[t]{0.48\textwidth}
\centering
\includegraphics[width=1.0\textwidth]{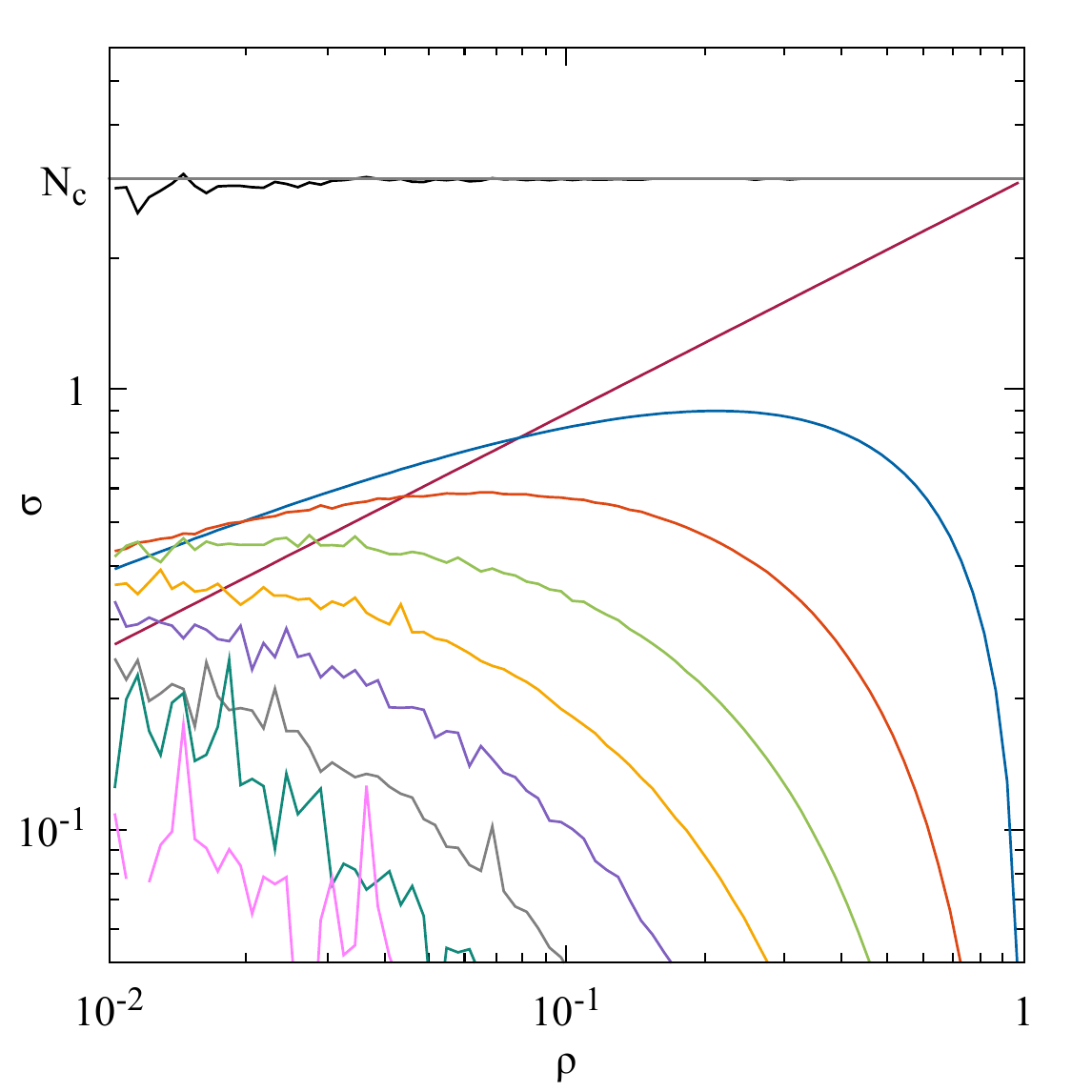}
\caption{Testing unitarity. The gap is eliminated so that real emissions are not vetoed and the fixed multiplicity curves sum up to a constant total ($\Nc$).}
\label{fig:controlUnitarity}
\end{subfigure} \\
\begin{subfigure}[c]{0.48\textwidth}
\centering
\includegraphics[width=1.0\textwidth]{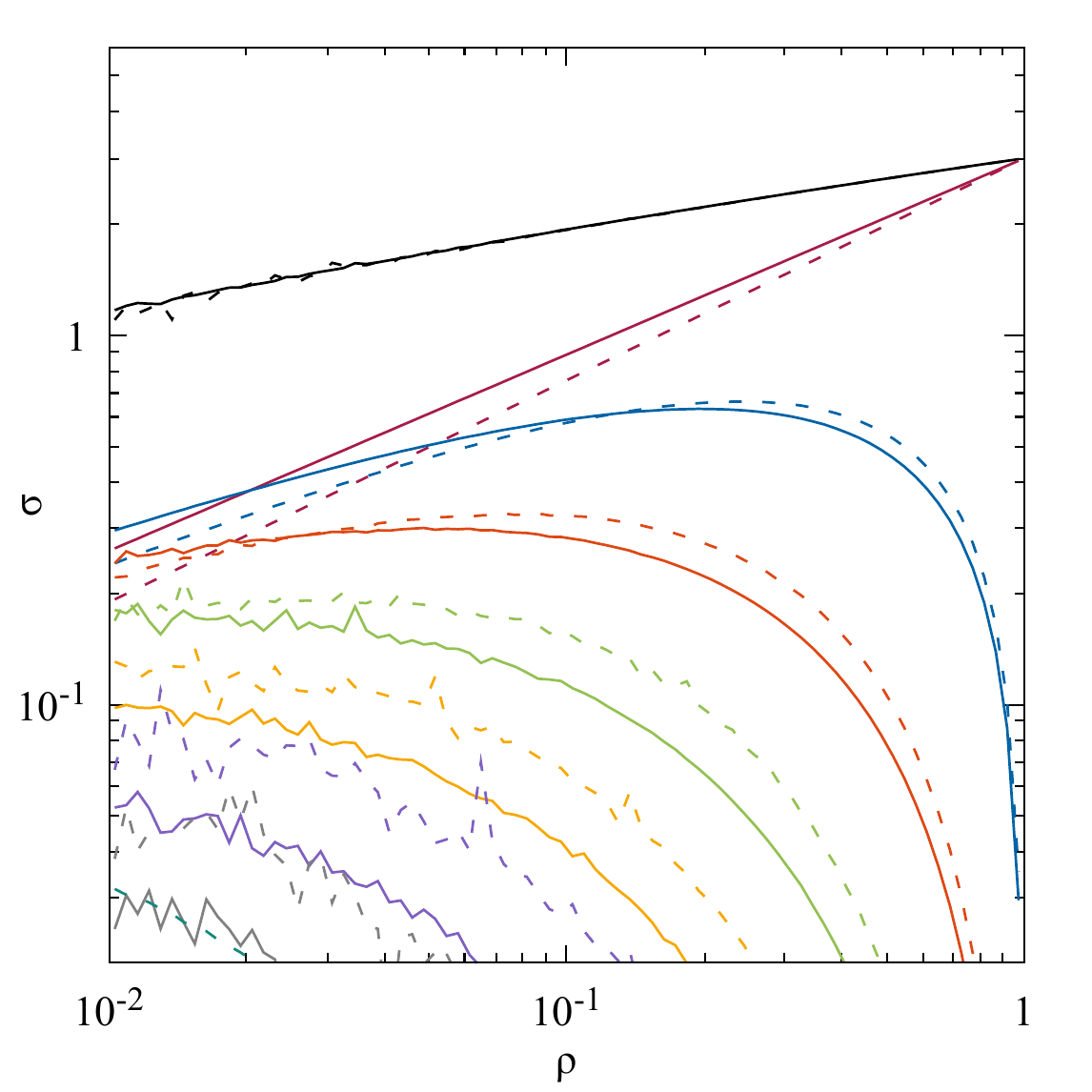}
\caption{Testing cutoff independence. The solid curves correspond to $\lambda = 0.01$ and the dashed curves to $\lambda = 0.005$. Though the results are very different per multiplicity, the total is independent of $\lambda$.}
\label{fig:controlCollinear}
\end{subfigure}
\caption{Control plots for $Z \to q \bar{q}$. In all cases, the upper curves are for the total and the lower curves are the contributions from different multiplicities (from 0 to 5 emissions).}
\label{fig:control}
\end{figure}

Fig.~\ref{fig:control} shows three control plots to illustrate that \CVolver\ is working well. The first plot compares the veto cross-section obtained in the standard configuration with that obtained by boosting the two hard partons such that their opening angle is $\pi/4$. The same boost is applied to the definition of the interjet region and to the collinear cutoff, i.e. $\lambda = 0.01$ for the standard configuration and $\lambda = 0.00146$ for the boosted configuration, such that the ratio $n_i \cdot n_j /\lambda$ is constant. The second plot shows that when we remove the gap the individual multiplicities give contributions that sum up to $N_c$. This is a highly non-trivial check of the algorithm. As is the third plot, which checks that our results do not depend on the choice of collinear cutoff over the range in $\rho$ that we explore. This is despite the fact that reducing the collinear cutoff enhances the effect of higher multiplicities, e.g. the dashed-grey 5 emission contribution, corresponding to the smaller cutoff, is similar in size to the solid-purple 6 emissions contribution for the larger cutoff. 

\begin{figure}[htbp!]
\centering
\begin{subfigure}[t]{0.9\textwidth}
\centering
\includegraphics[width=1.0\textwidth]{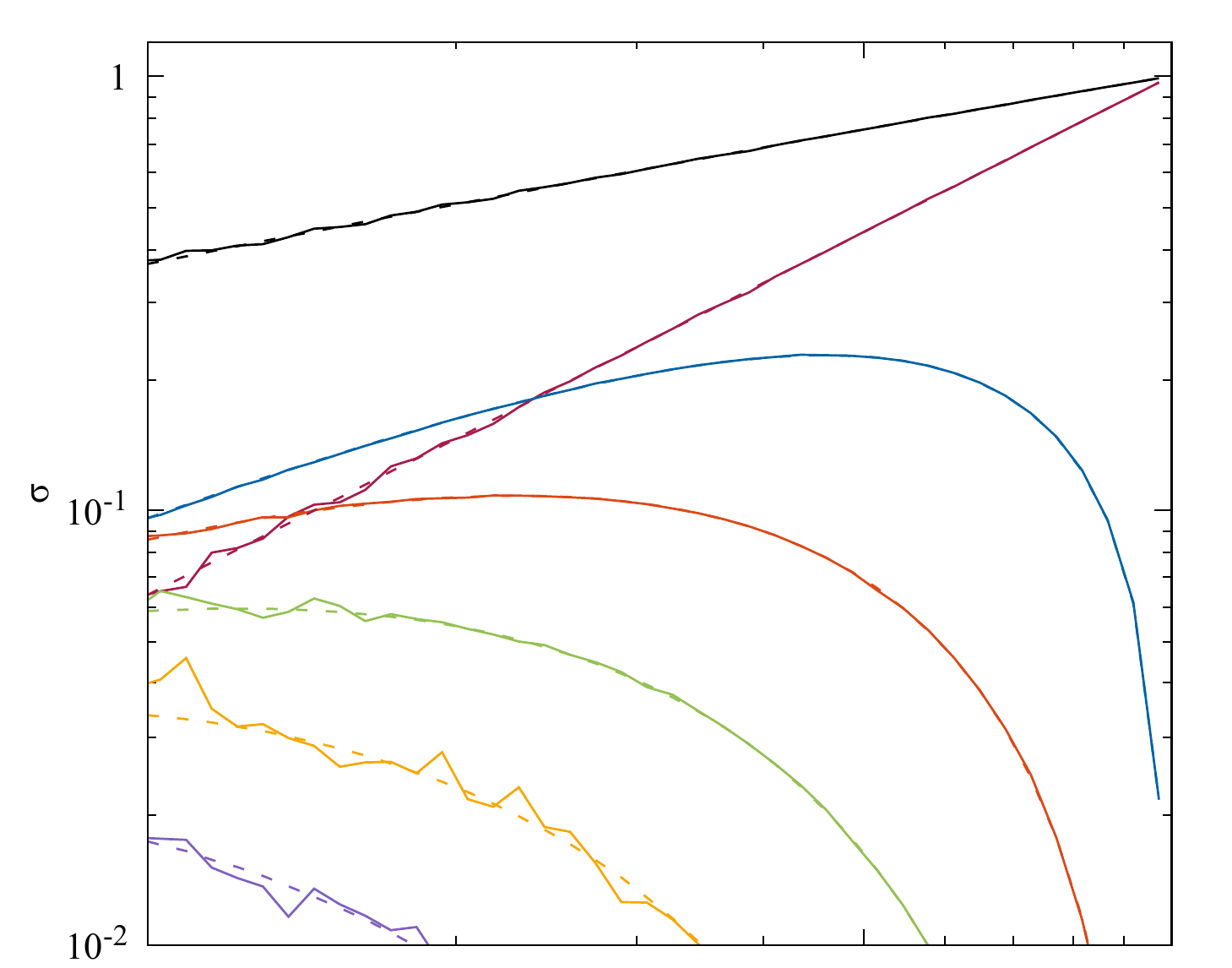}
\end{subfigure} \hfill
\begin{subfigure}[t]{0.9\textwidth}
\centering
\includegraphics[width=1.0\textwidth]{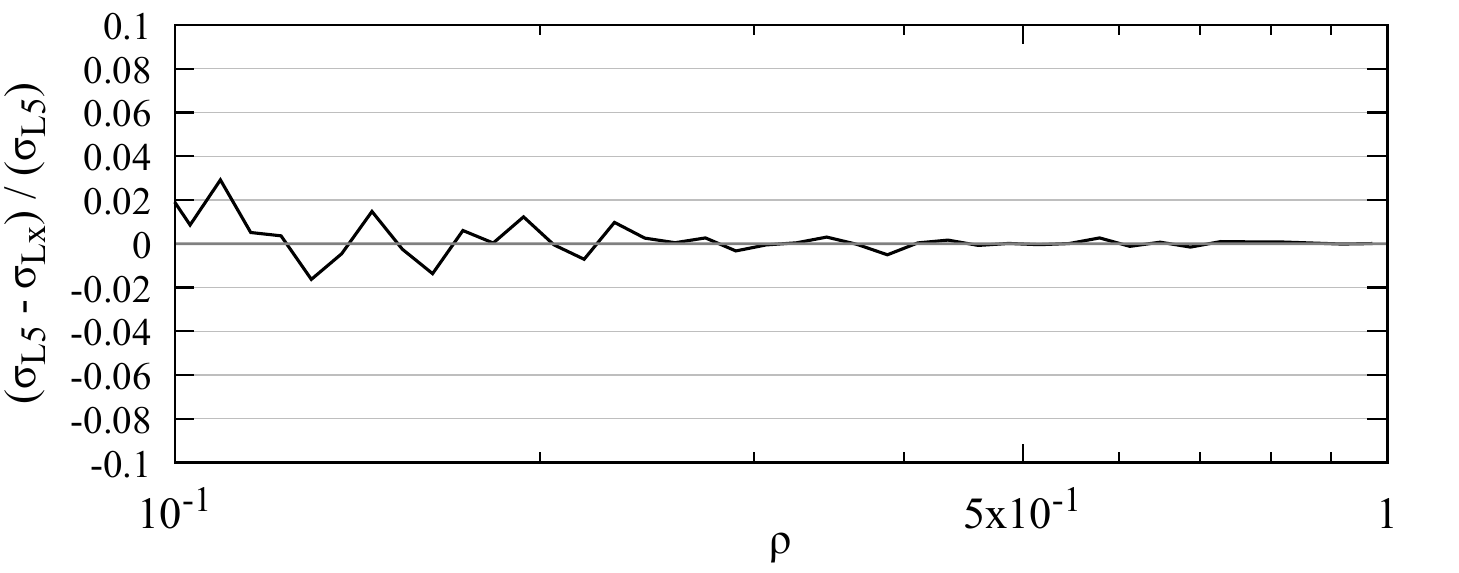}
\end{subfigure}
\caption{The veto cross-section as a function of veto scale for $H \to gg$. Solid: Full colour, Dashed: leading colour. The different coloured curves correspond to different multiplicities (0 up to 5 emissions) and the black curves are the total cross-section. The lower residual plot is for the total cross-section.}
\label{fig:ggHatta}
\end{figure}

\begin{figure}[htbp!]
\centering
\includegraphics[width=0.75\textwidth]{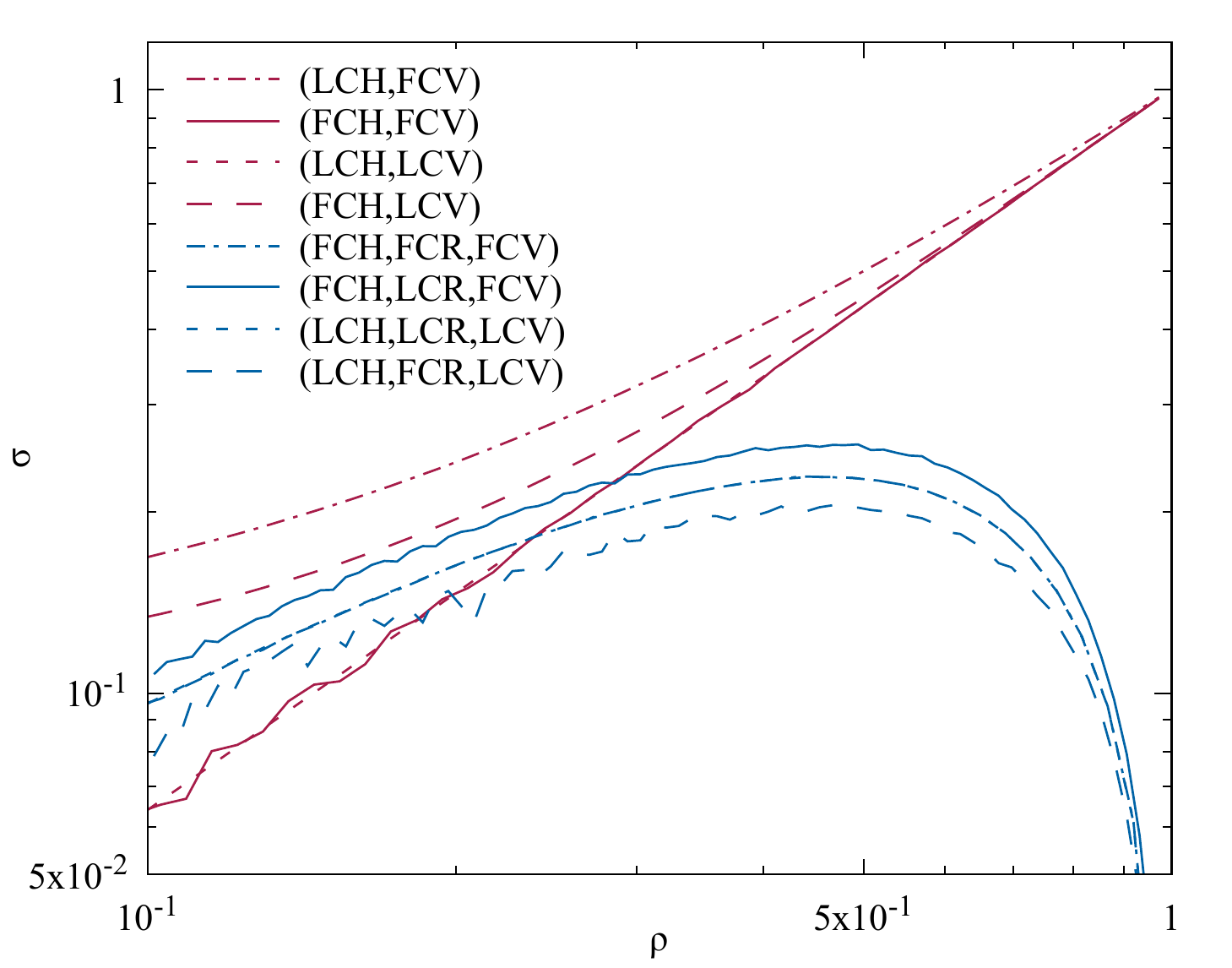}
\caption{The veto cross-section as a function of veto scale for $H \to gg$. The red curves correspond to 0 emissions, and the blue curves to 1 emission. }
\label{fig:ggHatta01}
\end{figure}

We now turn to consider colour singlet production of a pair of gluons, e.g. $H \to gg$. In this case the hard process density matrix is
\begin{align}
H & = |10\>\<10| -\frac{1}{\Nc}|10\>\<01| -\frac{1}{\Nc}|01\>\<10| + \frac{1}{\Nc^2}|01\>\<01|,
\end{align}
which is represented by the matrix whose elements are $(\mathbf{H})_{ij} = \<i|H|j\>$ where
\begin{align}
\mathbf{H} & = 
\begin{pmatrix}
1/\Nc^2 & -1/\Nc \\
-1/\Nc & 1 
\end{pmatrix}
\end{align}
and the scalar product matrix is
$S = \unitop = \sum_i |i]\<i|$,
which is represented by the matrix whose elements are $(\mathbf{S})_{ij} = [i|S|j] = \< i|j\>$ where
\begin{align}
\mathbf{S} & = 
\begin{pmatrix}
\Nc^2 & \Nc \\
\Nc & \Nc^2
\end{pmatrix}.
\end{align}
The total cross-section is $\sigma_0 = \text{Tr}(\mathbf{H}\mathbf{S}) = \Nc^2 - 1.$
The veto cross section is presented in Fig.~\ref{fig:ggHatta} and it exhibits the remarkable property first noted in \cite{Hatta:2020wre} whereby the full colour result is correctly described by the leading colour approximation up to a factor of $(\Nc^2-1)/\Nc^2$. We cannot explain this result but it is noteworthy that three of the four contributions to $\mathbf{H}$ do not evolve at all (the three terms involving singlet gluons). In other words, the singlet gluons do not emit and their colour evolution is trivial. However, it is surprising that even the multi-gluon emission cross-section with full colour evolution still only differs from the leading colour result by a factor $(\Nc^2-1)/\Nc^2$. Note that the approximations equivalent to L1, L2 and L3 in the quark case (but with LC rather than \LC) are all equivalent in the gluon case. Approximation L4 differs because of the way that real emissions from rings and strings can move the density matrix away from the leading colour form. As a result, for purely gluonic processes, we shall refer to the LCR (leading colour real) approximation, which excludes rings and strings, and FCR (full colour real). We shall also refer to LCH (leading colour in the hard scatter matrix) and FCH (full colour in the hard scatter matrix), and to LCV (leading colour virtuals) and FCV (full colour virtuals). In this language, Fig.~\ref{fig:ggHatta} illustrates that (FCH,FCR,FCV) is equal to (LCH,LCR,LCV) up to a factor $(\Nc^2-1)/\Nc^2$. This means there is a conspiracy between the real and virtual evolution that leads to a cancellation of the sub-leading colour.

\begin{figure}[htbp!]
\centering
\begin{subfigure}[t]{0.48\textwidth}
\centering
\includegraphics[width=1.0\textwidth]{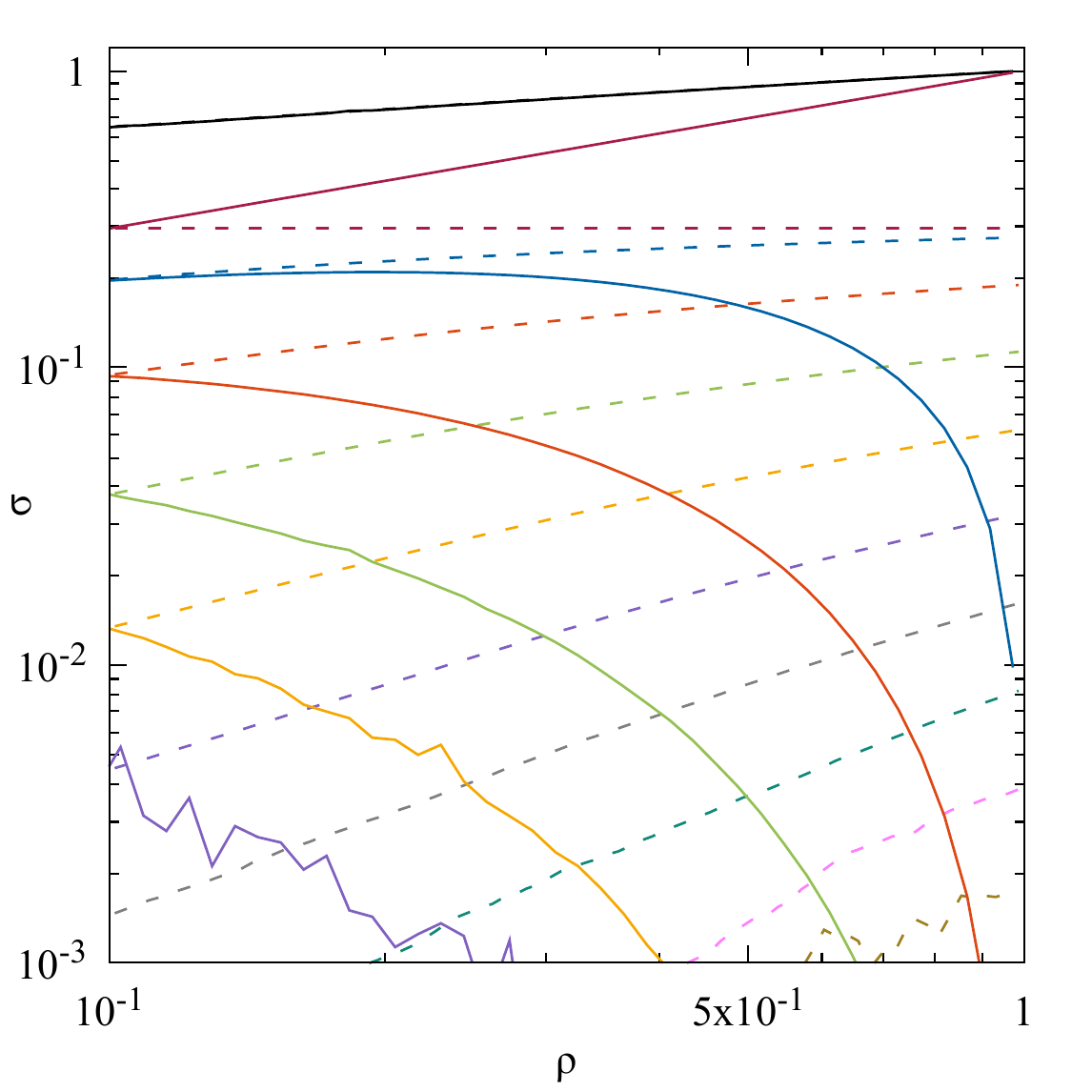}
\caption{The veto cross section for $Z \to q\bar{q}$.}
\label{fig:qqEventGen}
\end{subfigure} \hfill
\begin{subfigure}[t]{0.48\textwidth}
\centering
\includegraphics[width=1.0\textwidth]{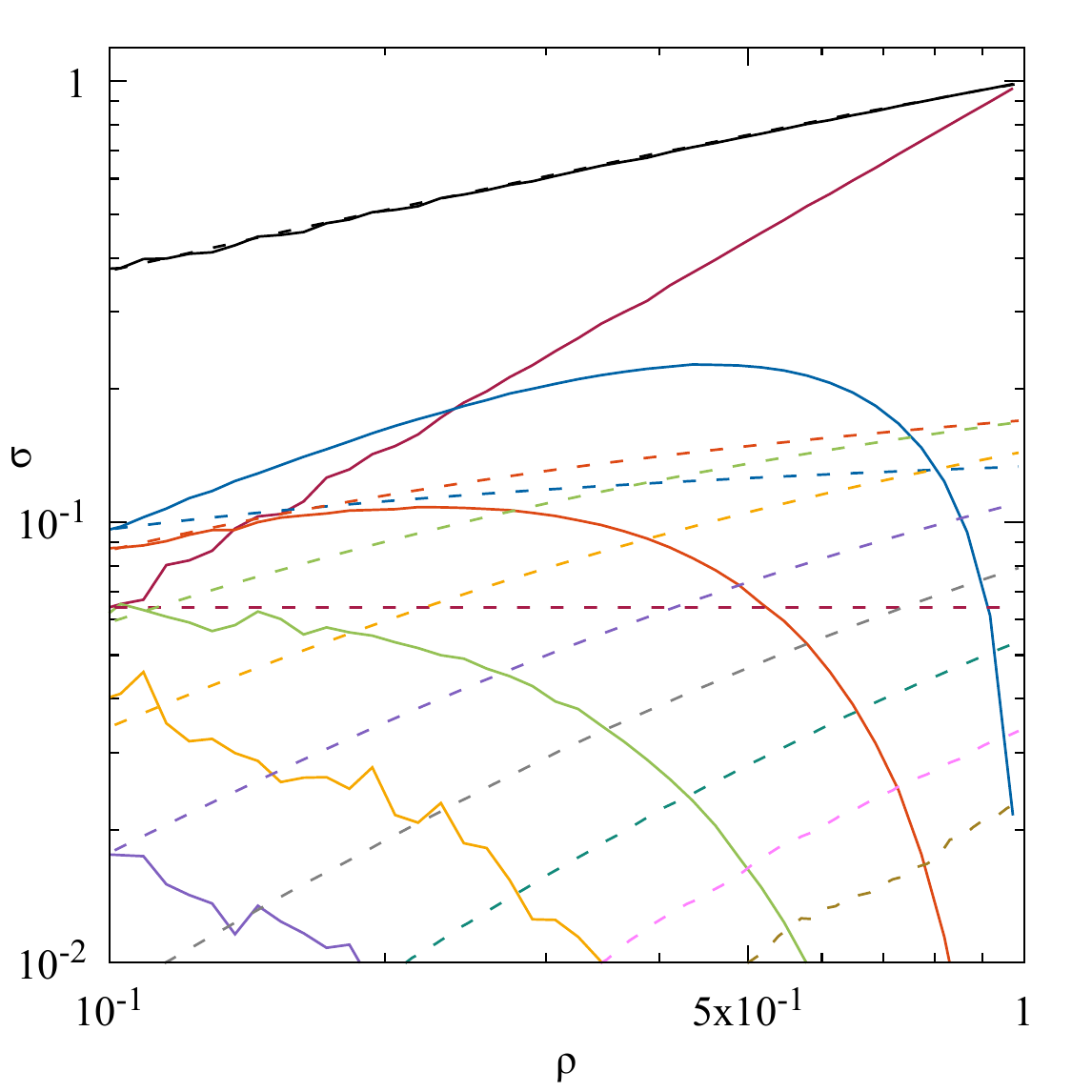}
\caption{The veto cross section for $H \to gg$.}
\label{fig:ggEventGen}
\end{subfigure}
\caption{Comparison of the dedicated (solid curves) and event generator (dashed curves) modes of \CVolver\ broken down by multiplicity. The total cross sections agree perfectly whilst those for individual multiplicities are very different.}
\label{fig:EventGen}
\end{figure}

This intriguing result is explored further in Fig.~\ref{fig:ggHatta01}, which shows results for zero and one gluon emission. In the zero emission case, the two new curves correspond to (LCH,FCV) evolution (the upper dot-dashed curve) and (FCH,LCV) (long dashed). Both of these are very different from the (FCH,FCV) and (LCH,LCV) curves (solid and short dashed). Remember that we always re-scale the LCH curves to match the total cross-section, i.e. to ensure agreement with FCH at $\rho=1$. The message for zero emissions is clear: it is all or nothing, in that one must evolve the complete hard scatter matrix with full colour evolution or the leading colour hard scatter matrix with leading colour evolution. The way in which the conspiracy works is illustrated by the one gluon emission curves in the figure. We now show the (FCH,LCR,FCV) (solid) and (LCH,FCR,LCV) (long dashed) curves. Remarkably, the (FCH,LCR,FCV) curve is larger than the correct result ((FCH,FCR,FCV)$=$(LCH,LCR,LCV)) by a factor $\Nc^2/(\Nc^2-1)$ and the (LCH,FCR,LCV) result is smaller than the correct result by a factor $(\Nc^2-1)/\Nc^2$. This scaling persists at least up to 5 emissions and indicates that the (very simple) hard process normalisation is somehow encoded in the (very complicated) real emissions.

We conclude this section by showcasing the operation of \CVolver\ as an event generator. Fig.~\ref{fig:EventGen} shows the veto cross-section for $Z \to q \bar{q}$ and $H \to gg$ broken down by multiplicity. Notice that there is perfect agreement between the two modes for the total cross-section, which is built up from a highly non-trivial sum over multiplicities. The dashed curves correspond to the veto cross-sections for a fixed number of emissions and they exhibit the correct physical behaviour, i.e. flat for zero emissions and falling as the veto increases in severity (i.e. as $\rho$ falls) for higher multiplicities. In this case, emissions may be in the angular region where the veto operates provided they have energy less than $Q_0$. This is in stark contrast to the dedicated mode where all emissions are necessarily emitted at angles outside of the veto region and $Q_0$ is fixed by the energy of the first emission into the veto region. A detailed study of colour evolution using the dedicated event generator mode will be presented in an accompanying paper. For the remainder of this paper, we shall focus on results obtained using the dedicated mode.  

\section{Two-to-Two processes}     
\label{sec:2to2}

\begin{figure}[htbp!]
\centering
\includegraphics[width=0.55\textwidth]{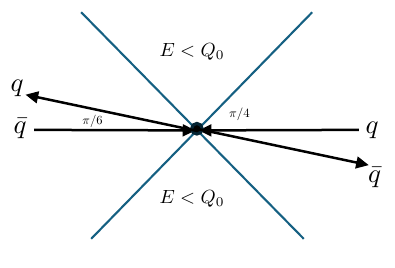}
\caption{Two outgoing partons are produced back-to-back and at an angle of $\pi/6$ relative to the axis defined by the incoming particles. Radiation is vetoed in the central region.}
\label{fig:GBJ-2to2}
\end{figure}

\begin{figure}[htbp!]
\centering
\begin{align*}
\mathcal{M}_{q\bar{q} \rightarrow q\bar{q}} \quad & = \quad \mathcal{M}_s \mspace{4mu} \raisebox{1.5mm}{\adjincludegraphics[valign=c,width=0.29\textwidth]{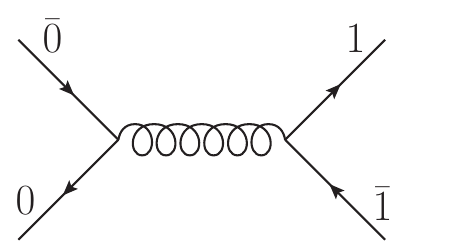}}
\mspace{3mu} + \mathcal{M}_t \adjincludegraphics[valign=c,width=0.25\textwidth]{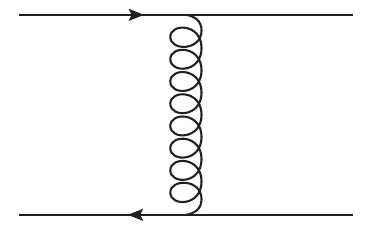}
\\ & \hspace{1mm} + \quad \mathcal{A}_s \hspace{2mm} \adjincludegraphics[valign=c,width=0.25\textwidth]{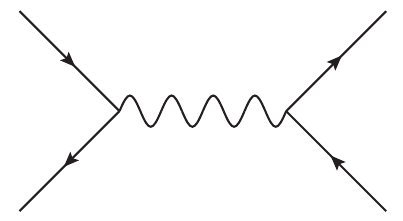}
 \quad \mspace{15mu} +\mathcal{A}_t \mspace{5mu} \adjincludegraphics[valign=c,width=0.25\textwidth]{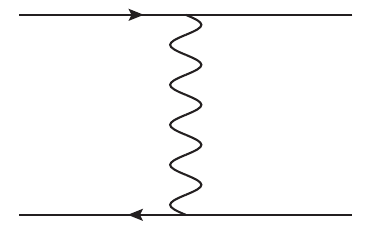}
\end{align*}
\caption{The diagrams contributing to the amplitude for $q\bar{q} \to q\bar{q}$.}
\label{fig:qq2qqFeynman}
\end{figure}

\begin{figure}[htbp!]
\centering
\begin{subfigure}[t]{0.75\textwidth}
\centering
\begin{align*}
\mspace{8mu} \adjincludegraphics[valign=c,width=0.32\textwidth]{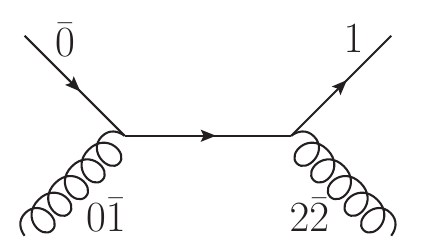}\mspace{-8mu} & = |021\> - \frac{1}{N_c}|120\> - \frac{1}{N_c}|012\> + \frac{1}{N_c^2}|102\>
\\ \adjincludegraphics[valign=c,width=0.30\textwidth]{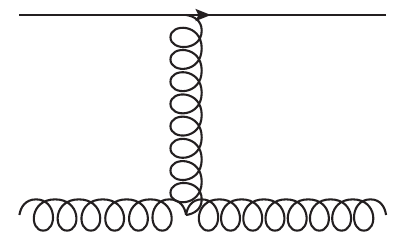} & = |210\> -|021\>
\\ \adjincludegraphics[valign=c,width=0.30\textwidth]{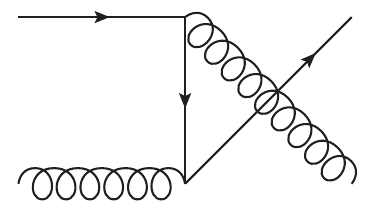} & = |210\> - \frac{1}{N_c}|120\> - \frac{1}{N_c}|012\> + \frac{1}{N_c^2}|102\>
\end{align*}
\caption{$qg \to qg$}
\label{fig:qg2qgFeynman}
\end{subfigure}	
\begin{subfigure}[t!]{0.75\textwidth}
\centering
\begin{align*}
\adjincludegraphics[valign=c,width=0.30\textwidth]{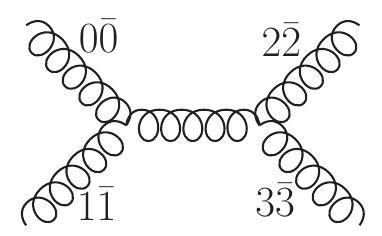} & = |3012\> + |1230\> - |1302\> - |2031\>
\\ \adjincludegraphics[valign=c,width=0.30\textwidth]{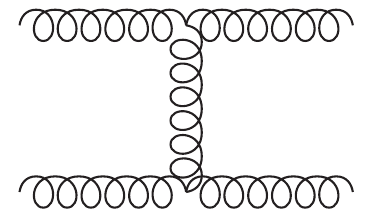} & = |3201\> + |2310\> - |1302\> - |2031\>
\\ \adjincludegraphics[valign=c,width=0.30\textwidth]{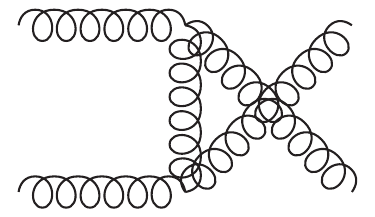} & = |3201\> + |2310\> - |3012\> - |1230\>
\end{align*}
\caption{$gg \to gg$}
\label{fig:gg2ggFeynman}
\end{subfigure}
\caption{The diagrams and colour flows for the $qg$ and $gg$ scattering processes.}
\end{figure}

We now turn our attention to two-to-two hard scattering processes. The goal is to explore the dynamics of sub-leading colour in a controlled and systematic way. To this end we will analyse $q\bar{q} \to q\bar{q}$, $qg \to qg$ and $gg \to gg$. We will consider a veto region defined as in Fig.~\ref{fig:GBJ-2to2}. This is the same configuration as that studied in \cite{Hatta:2020wre} and it allows us to compare with the results in that paper. We will also consider a more asymmetric configuration. For this paper, we do not include the $i \pi$ terms due to Coulomb gluon exchanges in the virtual evolution operator. This is because the cross-section is divergent in the limit $\lambda \to 0$ in the soft approximation due to the presence of super-leading logarithms \cite{Forshaw:2006fk,Forshaw:2008cq,Becher:2021zkk,Becher:2023mtx,Boer:2024hzh,Becher:2024nqc}. We do see this as a logarithmic dependence upon the collinear cutoff for all two-to-two scattering processes. This dependence cancels in the processes we considered in the previous section, since the Coulomb gluon contribution precisely cancels in those cases. We aim to study Coulomb exchanges in more detail in a future analysis. 

\subsection{\boldmath$q\bar{q} \to q\bar{q}$}
\label{sec:qq2qq}

The matrix element for $q\bar{q} \to q \bar{q}$ is illustrated in Fig.~\ref{fig:qq2qqFeynman} and, in the colour flow basis, it can be written
\begin{align}
\mathcal{M}_{q\bar{q}\to q\bar{q}} = \left( \frac{\mathcal{M}_s}{2} -	\frac{\mathcal{M}_t}{2\Nc} + \mathcal{A}_t \right) |10\> + \left( \frac{\mathcal{M}_t}{2} -	\frac{\mathcal{M}_s}{2\Nc} + \mathcal{A}_s \right) |01\>.
\end{align}
We shall not concern ourselves with the kinematic factors ($\mathcal{M}_{s,t}$ and $\mathcal{A}_t$), preferring instead to focus on the contributions from specific terms in the density matrix. The hard scatter matrix includes contributions from $|01 \> \< 01|$, $|10 \> \< 10|$ and $|01 \> \< 10|$ and we shall now investigate the veto cross-section arising from each of these three possibilities.   

\begin{figure}[t]
\centering
\begin{subfigure}[t]{0.9\textwidth}
\centering
\includegraphics[width=1.0\textwidth]{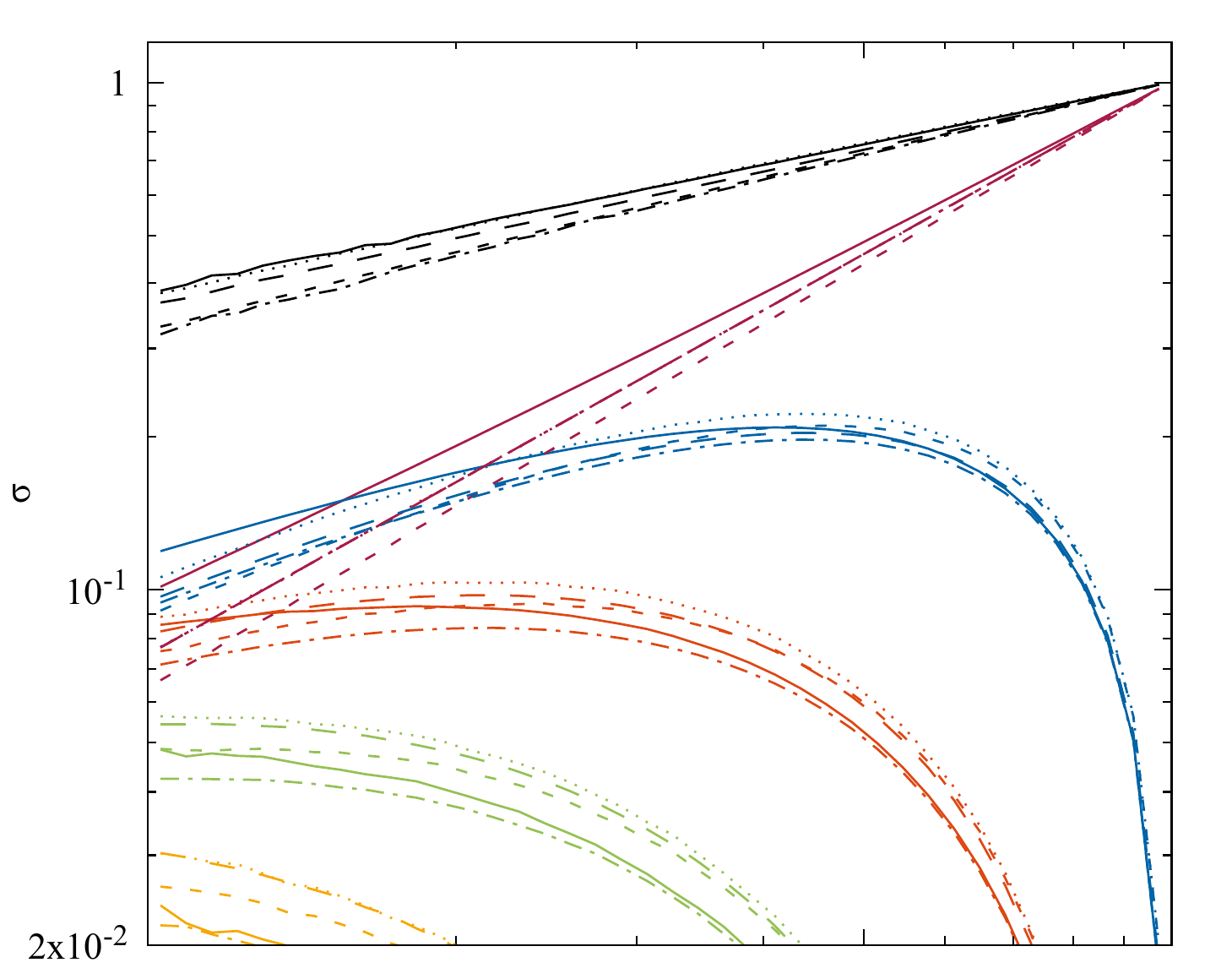}
\end{subfigure} \\
\begin{subfigure}[t]{0.9\textwidth}
\centering
\includegraphics[width=1.0\textwidth]{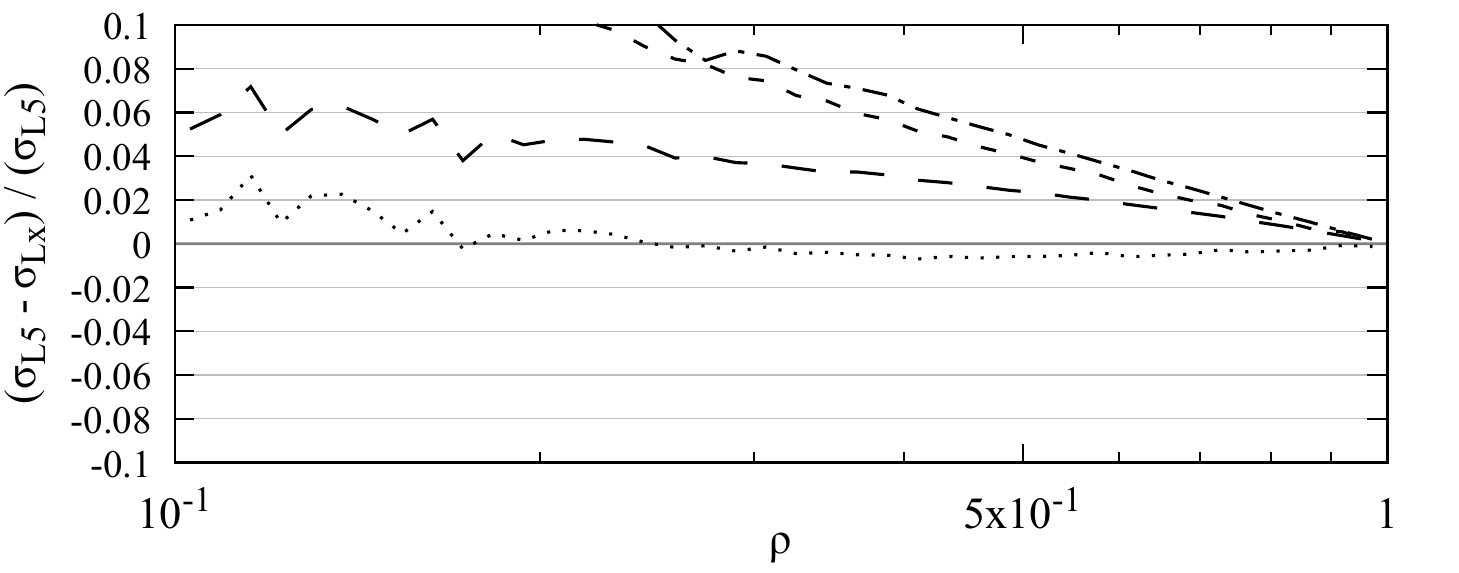}
\end{subfigure} \\
\caption{The veto cross-section for the $|01\>\<01|$ contribution to $q\bar{q} \to q\bar{q}$. Solid: Full colour (L5), Dash-dotted: \LC\ + FCR (L4), Long-dashed: \LC\ + LCR + singlets (L3), Dotted: \LC\ + LCR (L2), Short-dashed: strict LC (L1). }
\label{fig:qq2qq0101}
\end{figure}

In Fig.~\ref{fig:qq2qq0101} we show the L1 to L5 curves for the case $H = |01\>\<01|$. This is the configuration where the outgoing particles are colour connected and the incoming particles are colour connected. At leading colour in the hard process, this corresponds to $t$-channel gluon exchange (or $s$-channel $\gamma/Z$ exchange). The lower pane shows the residuals for the total rates and indicates that leading colour \LC\ (L2) agrees best with the full colour result. This is in contrast to the $Z \to q\bar{q}$ case, where L4 provided the best approximation. Notice that the agreement between L1 and L2 is only for the total. Even at zero emissions there is a big difference between red L5 and red L4=L3=L2, which implies large effects from virtual swaps. The differences between L5 and L4 show that virtual swaps are also involved in the cancellations beyond L2.

\begin{figure}[t]
\centering
\begin{subfigure}[t]{0.9\textwidth}
\centering
\includegraphics[width=1.0\textwidth]{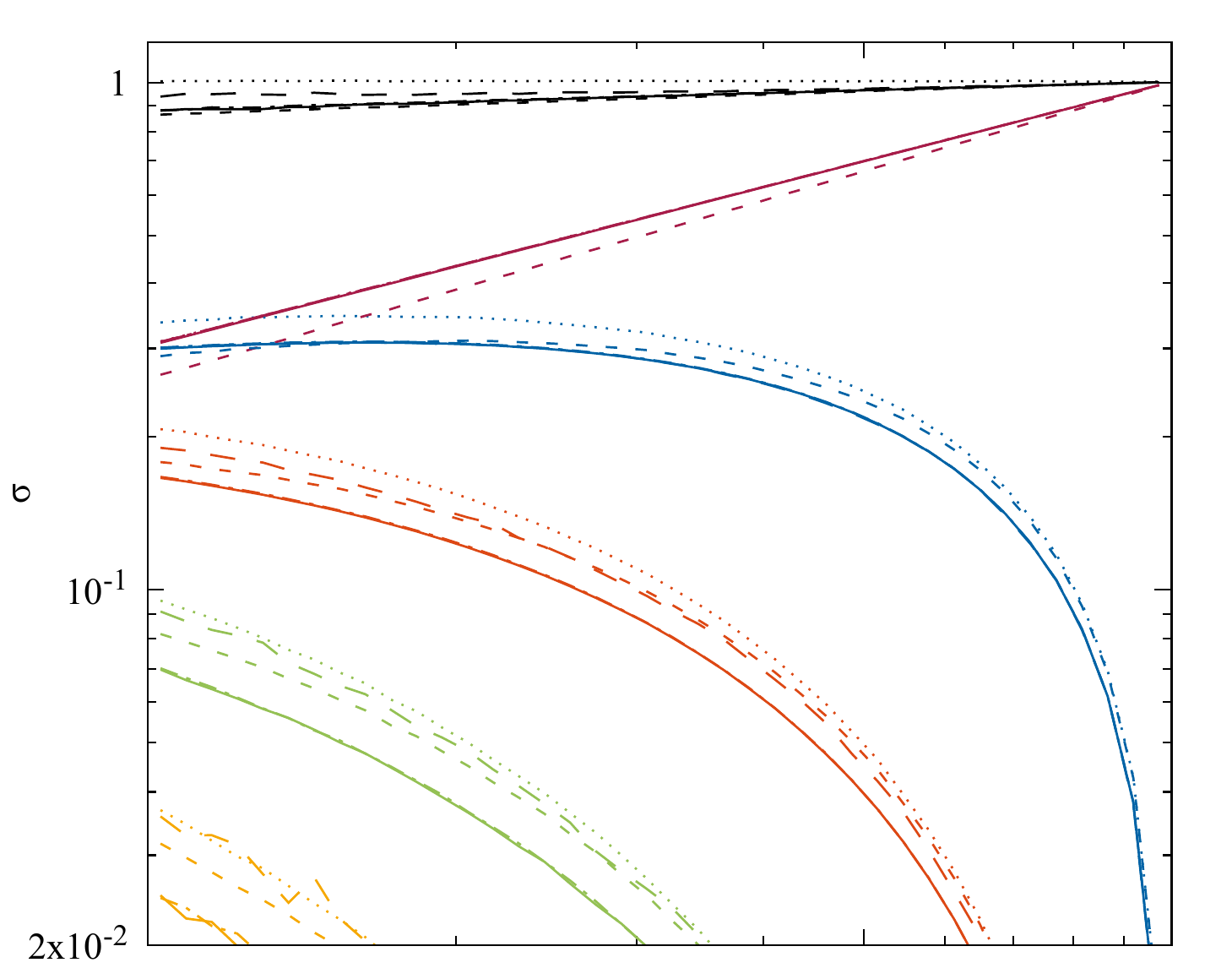}
\end{subfigure} \hfill
\begin{subfigure}[t]{0.9\textwidth}
\centering
\includegraphics[width=1.0\textwidth]{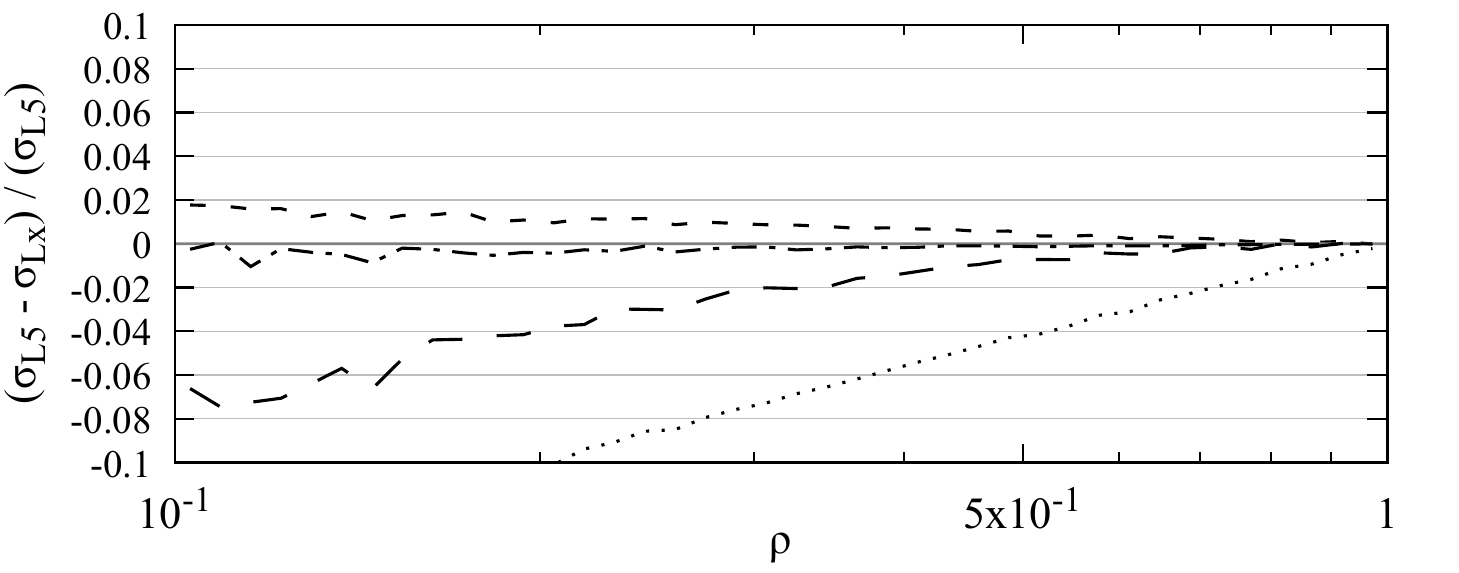}
\end{subfigure}
\caption{The veto cross-section for the $|10\>\<10|$ contribution to $q\bar{q} \to q\bar{q}$. Solid: Full colour (L5), Dash-dotted: \LC\ + FCR (L4), Long-dashed: \LC\ + LCR + singlets (L3), Dotted: \LC\ + LCR (L2), Short-dashed: strict LC (L1). }
\label{fig:qq2qq1010}
\end{figure}

In Fig.~\ref{fig:qq2qq1010} we show the L1 to L5 curves for the case $H = |10\>\<10|$. This is the configuration where the incoming quark is colour connected to the outgoing quark and similarly for the incoming and outgoing antiquarks. At leading colour in the hard process, this corresponds to $s$-channel gluon exchange (or $t$-channel $\gamma/Z$ exchange). In contrast to the previous case, it is L4 that agrees best with the full colour result and the leading colour \LC\ approximation fails. This conclusion appears to be in agreement with \cite{Hatta:2020wre} who study $qq \to qq$ with $t$-channel electroweak boson exchange and find that the mean field approximation agrees best with full colour\footnote{Though \cite{Hatta:2020wre} say that their result may be unreliable due to lattice artifacts.}. Contrary to Fig.~\ref{fig:qq2qq0101}, this time the agreement L4=L5 works at the level of each multiplicity, which shows there is little effect from virtual swaps, i.e. it seems the evolution does not favour swapping to the other colour configuration.

We now turn to the interference contribution arising from $H = |10\>\<01|$, which is illustrated in Fig.~\ref{fig:qq2qq1001}. Only L3, L4 and L5 have any emissions. That is because the L1 and L2 approximations require that emission be off dipoles and in this case there are no dipoles to emit from. Hence the L1 and L2 total cross-section is equal to the zero emission cross-section. The L3, one-emission curve is the blue, long-dashed curve towards the bottom of the plot. It is negative and we plot the absolute value. As before, the zero emission curves for L2, L3 and L4 are all coincident. Clearly, none of the \LC\ approximations is good with differences in all cases exceeding 50\% at $\rho = 0.1$.

\begin{figure}[t]
\centering
\begin{subfigure}[t]{0.9\textwidth}
\centering
\includegraphics[width=1.0\textwidth]{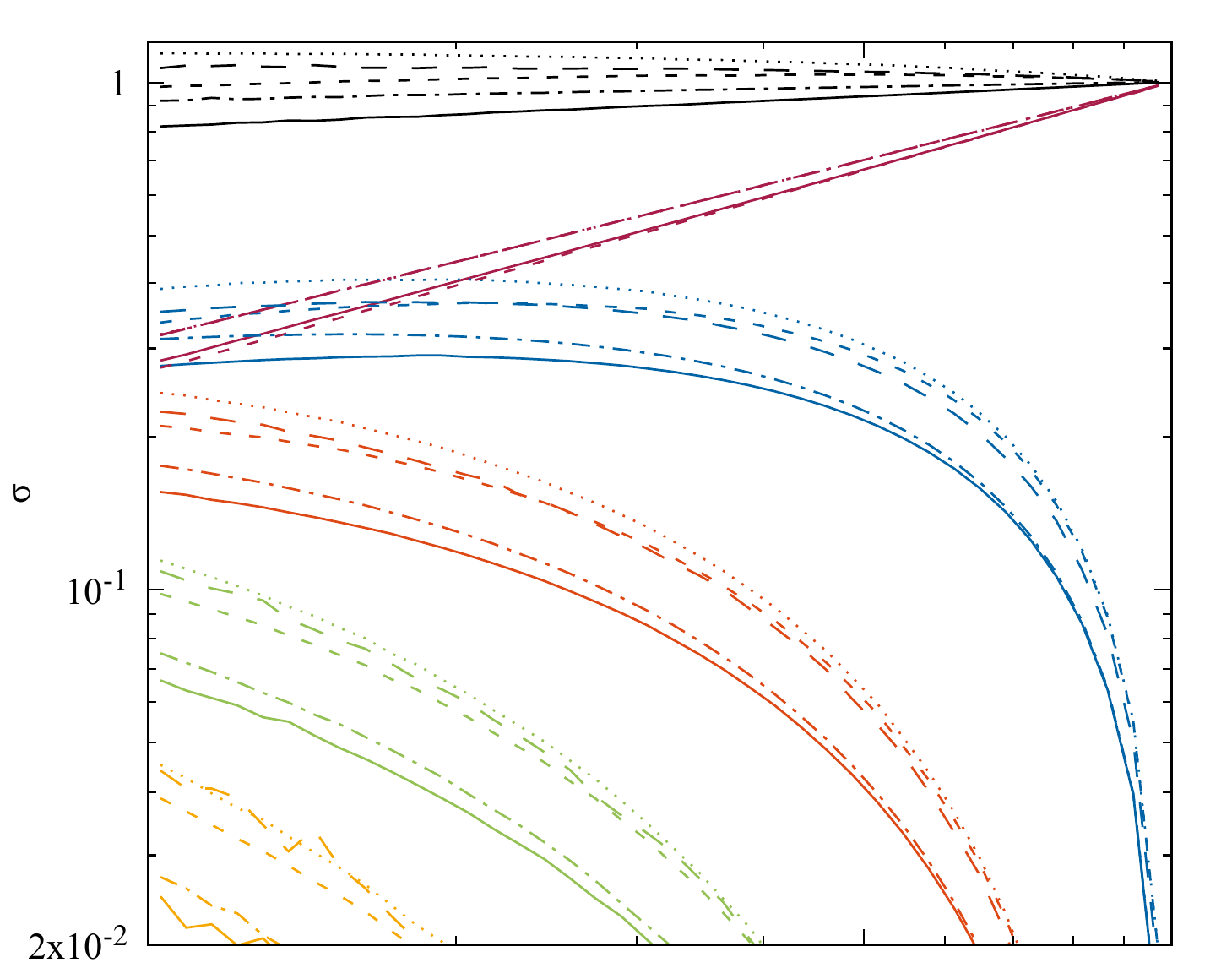}
\end{subfigure} \hfill
\begin{subfigure}[t]{0.9\textwidth}
\centering
\includegraphics[width=1.0\textwidth]{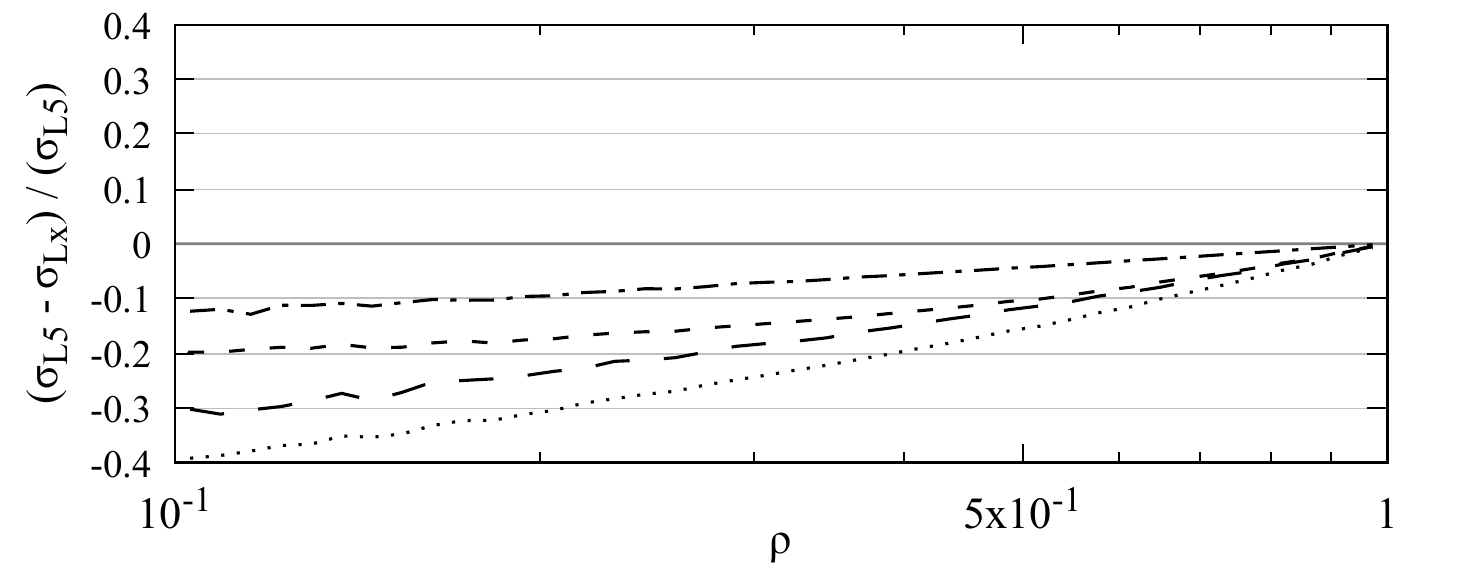}
\end{subfigure}
\caption{The veto cross-section for the $s$-channel gluon exchange contribution to $q\bar{q} \to q\bar{q}$. Solid: Full colour (L5), Dash-dotted: \LC\ + FCR (L4), Long-dashed: \LC\ + LCR + singlets (L3), Dotted: \LC\ + LCR (L2), Short-dashed: strict LC (L1).}
\label{fig:qq2qqs}
\end{figure}

\begin{figure}[t]
\centering
\begin{subfigure}[t]{0.9\textwidth}
\centering
\includegraphics[width=1.0\textwidth]{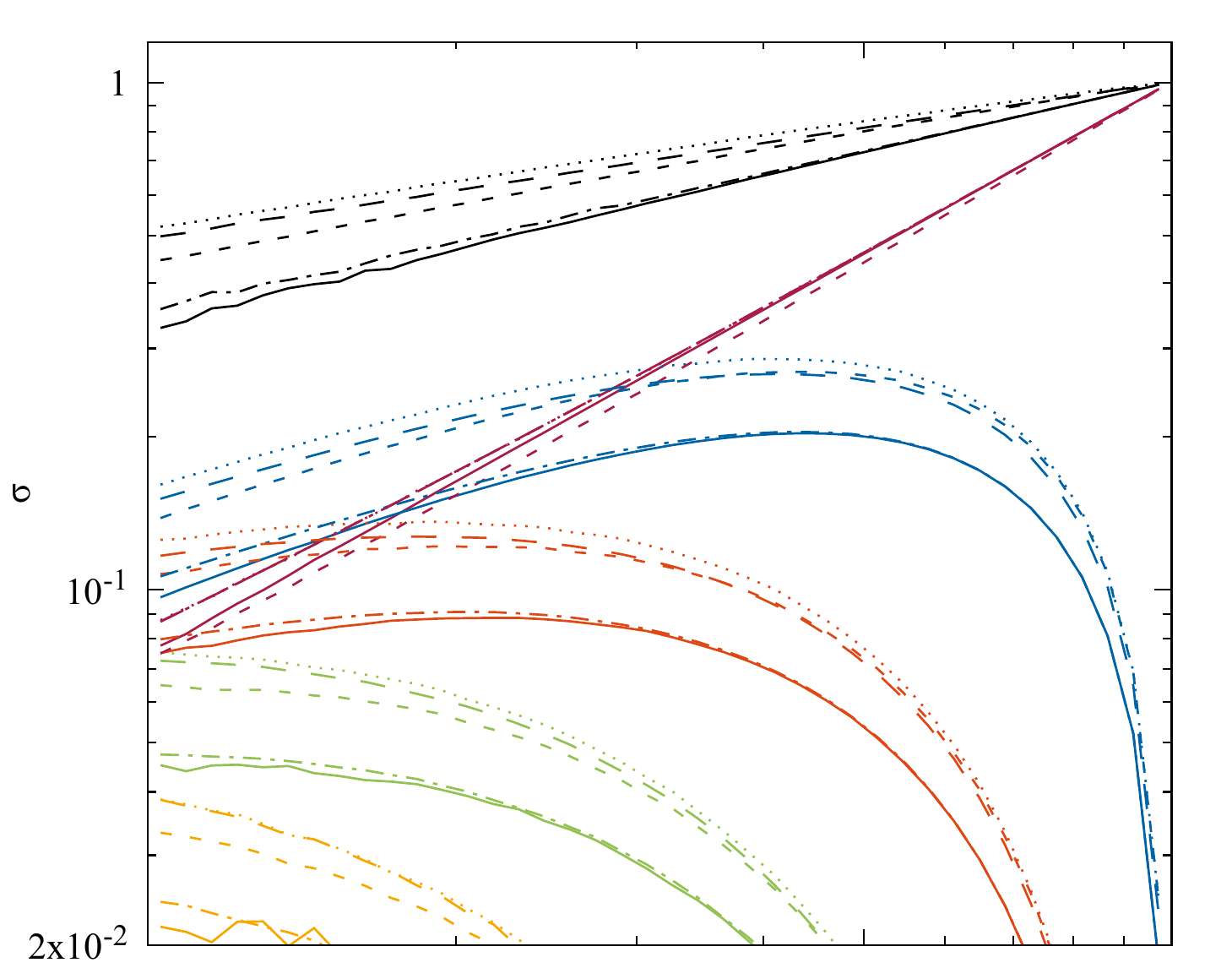}
\end{subfigure} \hfill
\begin{subfigure}[t]{0.9\textwidth}
\centering
\includegraphics[width=1.0\textwidth]{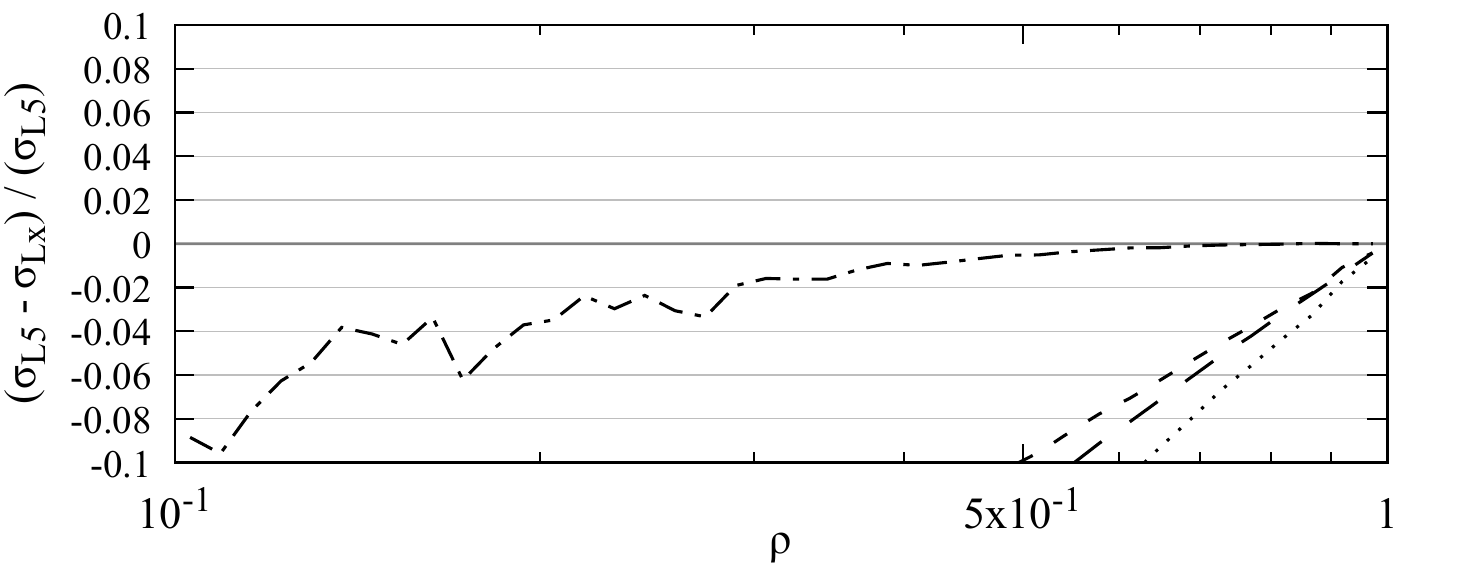}
\end{subfigure}
\caption{The veto cross-section for the $t$-channel gluon exchange contribution to $q\bar{q} \to q\bar{q}$. Solid: Full colour (L5), Dash-dotted: \LC\ + FCR (L4), Long-dashed: \LC\ + LCR + singlets (L3), Dotted: \LC\ + LCR (L2), Short-dashed: strict LC (L1).}
\label{fig:qq2qqt}
\end{figure}

We can also explore the veto cross-section for $s$-channel and $t$-channel gluon exchanges, which is achieved by combining the previous results. The corresponding hard scatter matrices are proportional to
\begin{align}
\mathbf{H}_s = 
\begin{pmatrix}
1/\Nc^2 & -1/\Nc \\
-1/\Nc & 1 
\end{pmatrix} \text{ and }
\mathbf{H}_t  = 
\begin{pmatrix}
1 & -1/\Nc \\
-1/\Nc & 1/\Nc^2 
\end{pmatrix}. \label{eq:fchqqqq}
\end{align}
The $s$-channel contribution is the complete lowest-order QCD contribution to $q \bar{q} \to Q \bar{Q}$. The corresponding veto cross-section is shown in Fig.~\ref{fig:qq2qqs}. Again we see that, though the approximations do steadily improve, none agrees well with the full colour result. In fact in all cases, the difference exceeds 10\% at $\rho=0.1$. The equivalent curves in the case of $t$-channel gluon exchange are in Fig.~\ref{fig:qq2qqt}. This would be the complete lowest-order QCD contribution to $q \bar{Q} \to q \bar{Q}$. Again none of the L1--L3 approximations is good and the L4 approximation is breaking down at low $\rho$.

\begin{figure}[t]
\centering
\begin{subfigure}[t]{0.9\textwidth}
\centering
\includegraphics[width=1.0\textwidth]{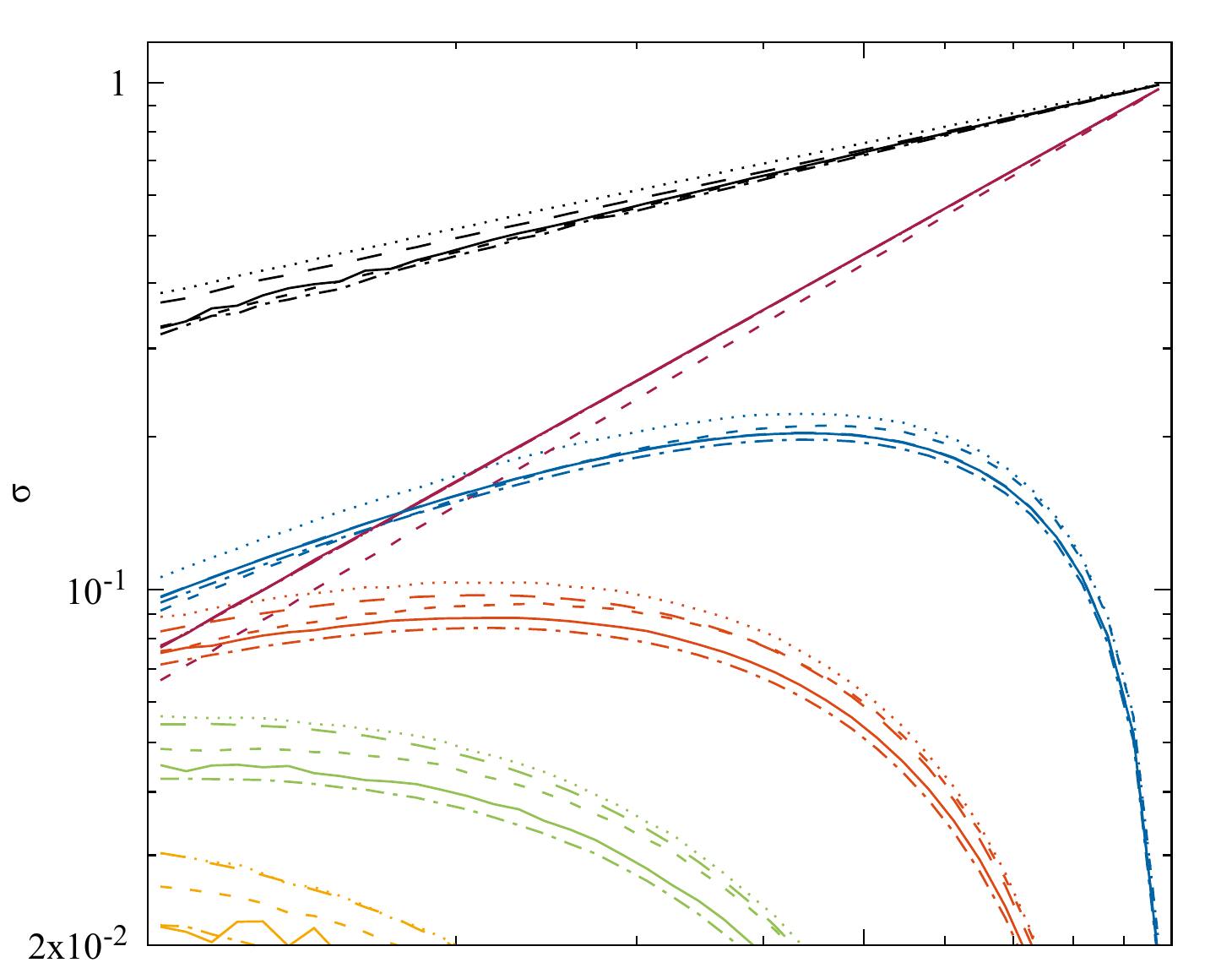}
\end{subfigure} \hfill
\begin{subfigure}[t]{0.9\textwidth}
\centering
\includegraphics[width=1.0\textwidth]{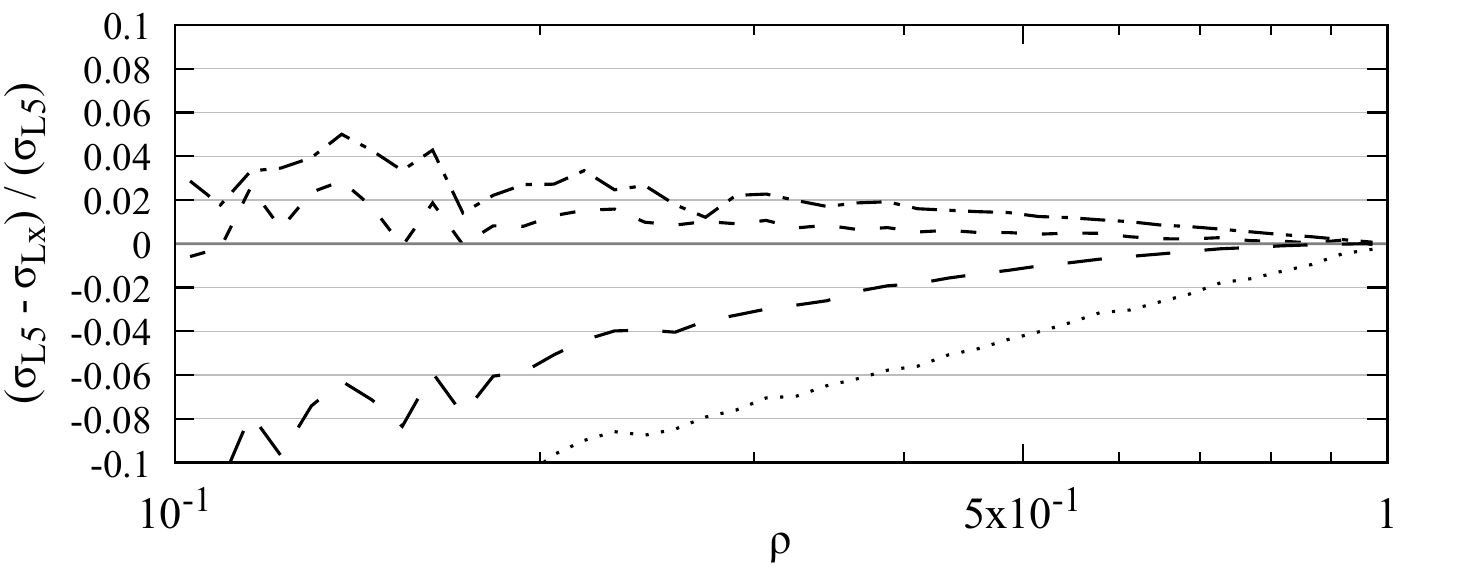}
\end{subfigure}
\caption{The veto cross-section for the $t$-channel gluon exchange contribution to $q\bar{q} \to q\bar{q}$.  Solid: Full colour (L5), Dash-dotted: \LC\ + FCR (L4), Long-dashed: \LC\ + LCR + singlets (L3), Dotted: \LC\ + LCR (L2), Short-dashed: strict LC (L1). For the L1--L4 curves we start the evolution using the leading-colour approximation to the hard-scatter matrix.}
\label{fig:qq2qqt-LCH}
\end{figure}

\begin{figure}[t]
\centering
\begin{subfigure}[t]{0.9\textwidth}
\centering
\includegraphics[width=1.0\textwidth]{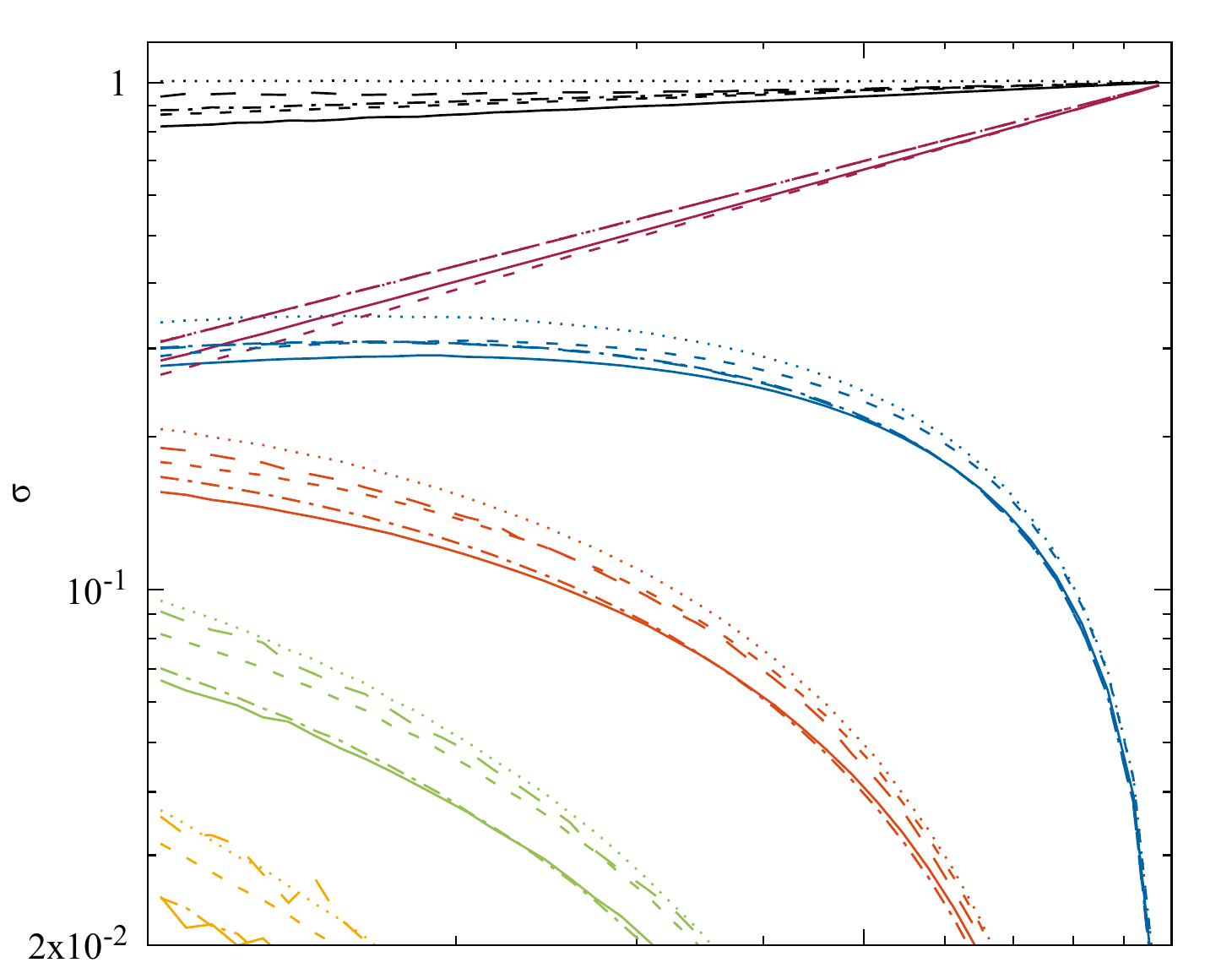}
\end{subfigure} \hfill
\begin{subfigure}[t]{0.9\textwidth}
\centering
\includegraphics[width=1.0\textwidth]{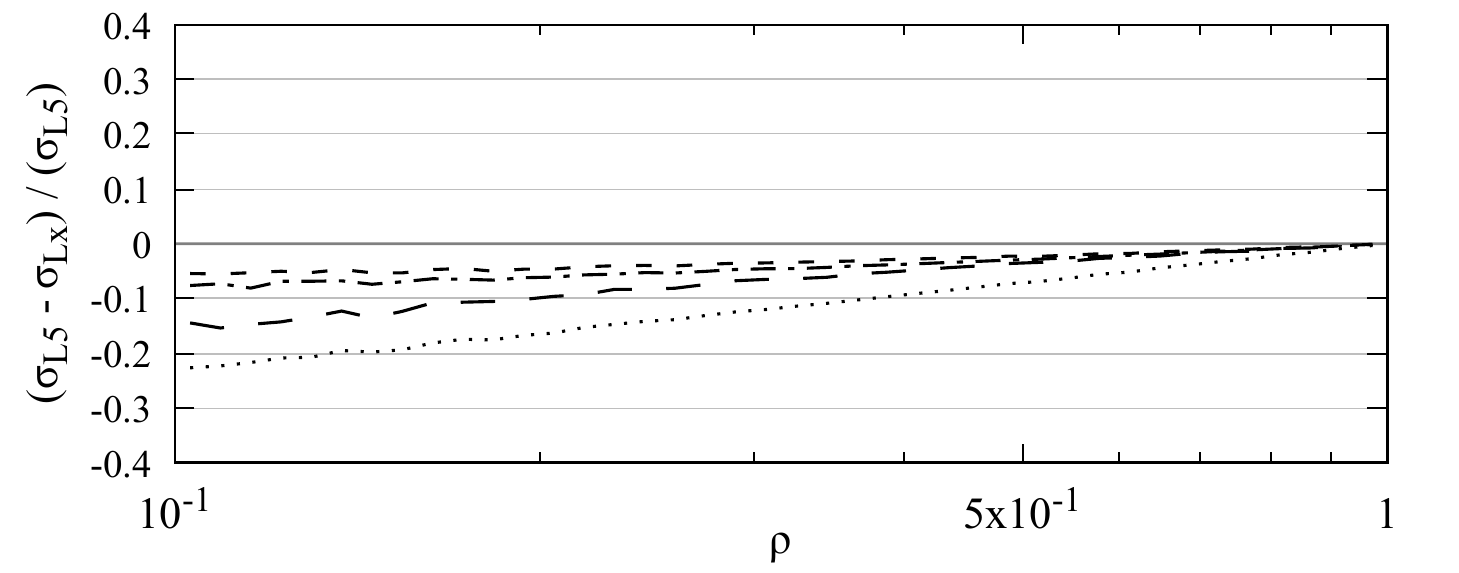}
\end{subfigure}
\caption{The veto cross-section for the $s$-channel gluon exchange contribution to $q\bar{q} \to q\bar{q}$. Solid: Full colour (L5), Dash-dotted: \LC\ + FCR (L4), Long-dashed: \LC\ + LCR + singlets (L3), Dotted: \LC\ + LCR (L2), Short-dashed: strict LC (L1). For the L1--L4 curves we start the evolution using the leading-colour approximation to the hard-scatter matrix.}
\label{fig:qq2qqs-LCH}
\end{figure}

The results we just presented for $s$ and $t$-channel gluon exchange evolved from the full-colour hard-scatter matrices in Eq.~\eqref{eq:fchqqqq}. Starting instead from the leading-colour hard-scatter matrices for the L1--L4 approximations (i.e. we still use the full-colour hard-scatter matrix for L5) we find that the L1--L4 approximations fare significantly better than before\footnote{The L1--L4 curves on Fig.~\ref{fig:qq2qqt-LCH} are identical to those on Fig.~\ref{fig:qq2qq0101} and those on Fig.~\ref{fig:qq2qqs-LCH} are identical to those on Fig.~\ref{fig:qq2qq1010}. Only the solid (L5) curves differ.}. In fact, the $t$-channel contribution shown in Fig.~\ref{fig:qq2qqt-LCH} again confirms a result first found in \cite{Hatta:2020wre} that the full-colour evolution of the full-colour hard process is very well approximated by the leading-colour evolution of the leading-colour hard process (i.e. the strictly leading colour approximation). This is the same result we found for $H \to gg$, though in this case is perhaps even more surprising since the agreement is found by taking $C_F \to \Nc/2$ even in the diagonal part of the virtual evolution. Indeed, this change shifts the slope of the zero emission curve in such a way to compensate shifts in the higher multiplicity contributions, which is in contrast to Fig.~\ref{fig:ggHatta01} where the agreement occurs for each multiplicity. The success of the strictly leading colour approximation does not hold for $s$-channel gluon exchange, as illustrated in Fig.~\ref{fig:qq2qqs-LCH}.

\begin{figure}[t]
\centering
\begin{subfigure}[t]{0.9\textwidth}
\centering
\includegraphics[width=1.0\textwidth]{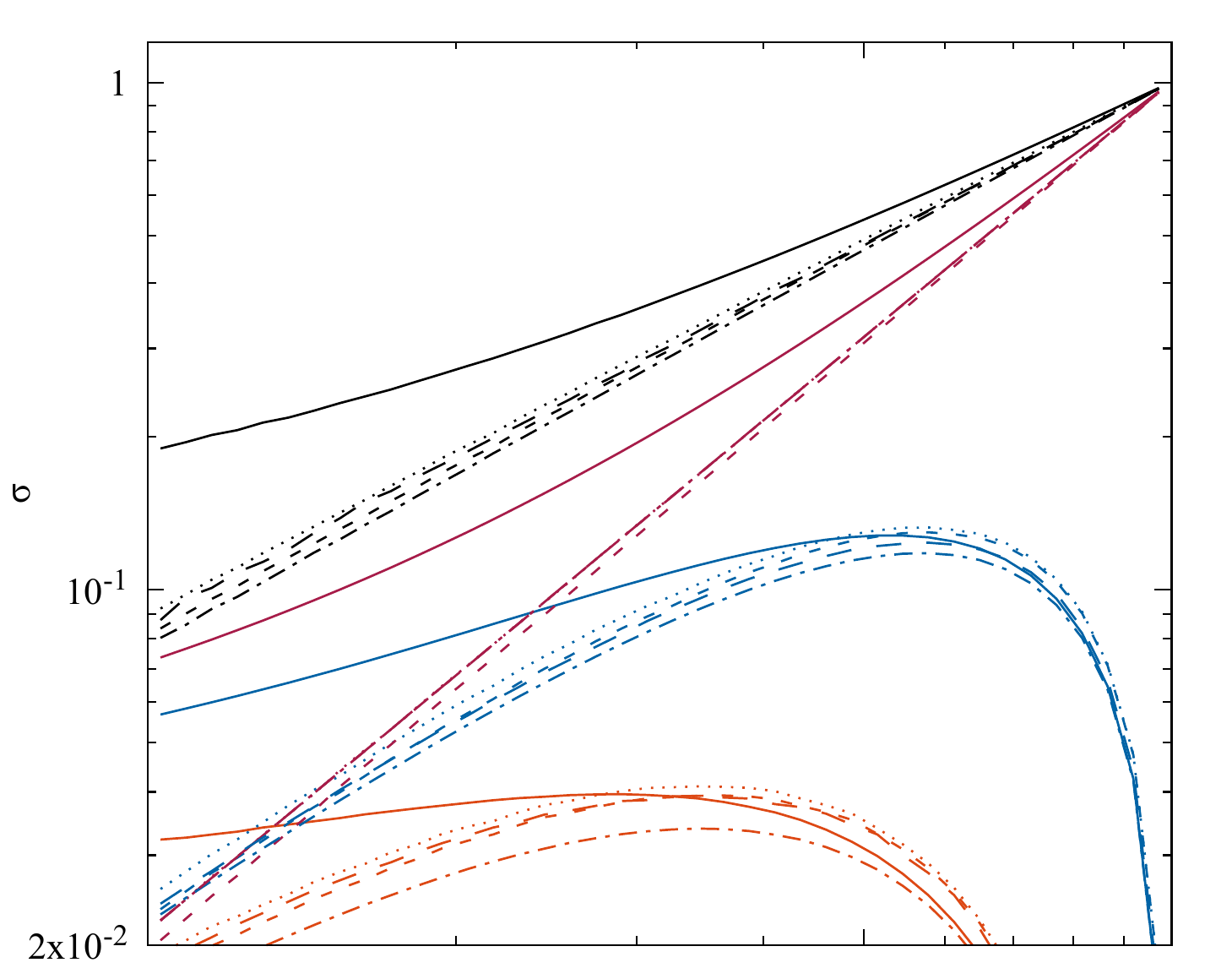}
\end{subfigure} \hfill
\begin{subfigure}[t]{0.9\textwidth}
\centering
\includegraphics[width=1.0\textwidth]{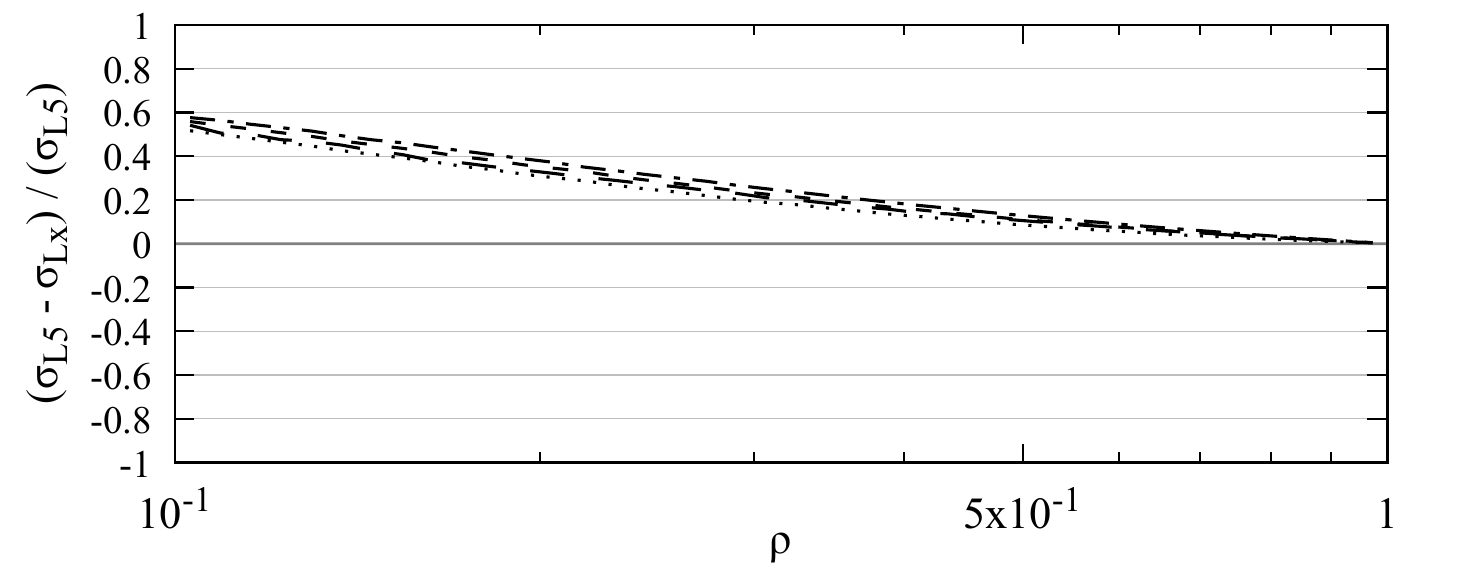}
\end{subfigure}
\caption{The veto cross-section for the $|01\>\<01|$ contribution to $q\bar{q} \to q\bar{q}$ in the asymmetric configuration. Solid: Full colour (L5), Dash-dotted: \LC\ + FCR (L4), Long-dashed: \LC\ + LCR + singlets (L3), Dotted: \LC\ + LCR (L2), Short-dashed: strict LC (L1). }
\label{fig:qq2qq0101a}
\end{figure}

\begin{figure}[t]
\centering
\begin{subfigure}[t]{0.9\textwidth}
\centering
\includegraphics[width=1.0\textwidth]{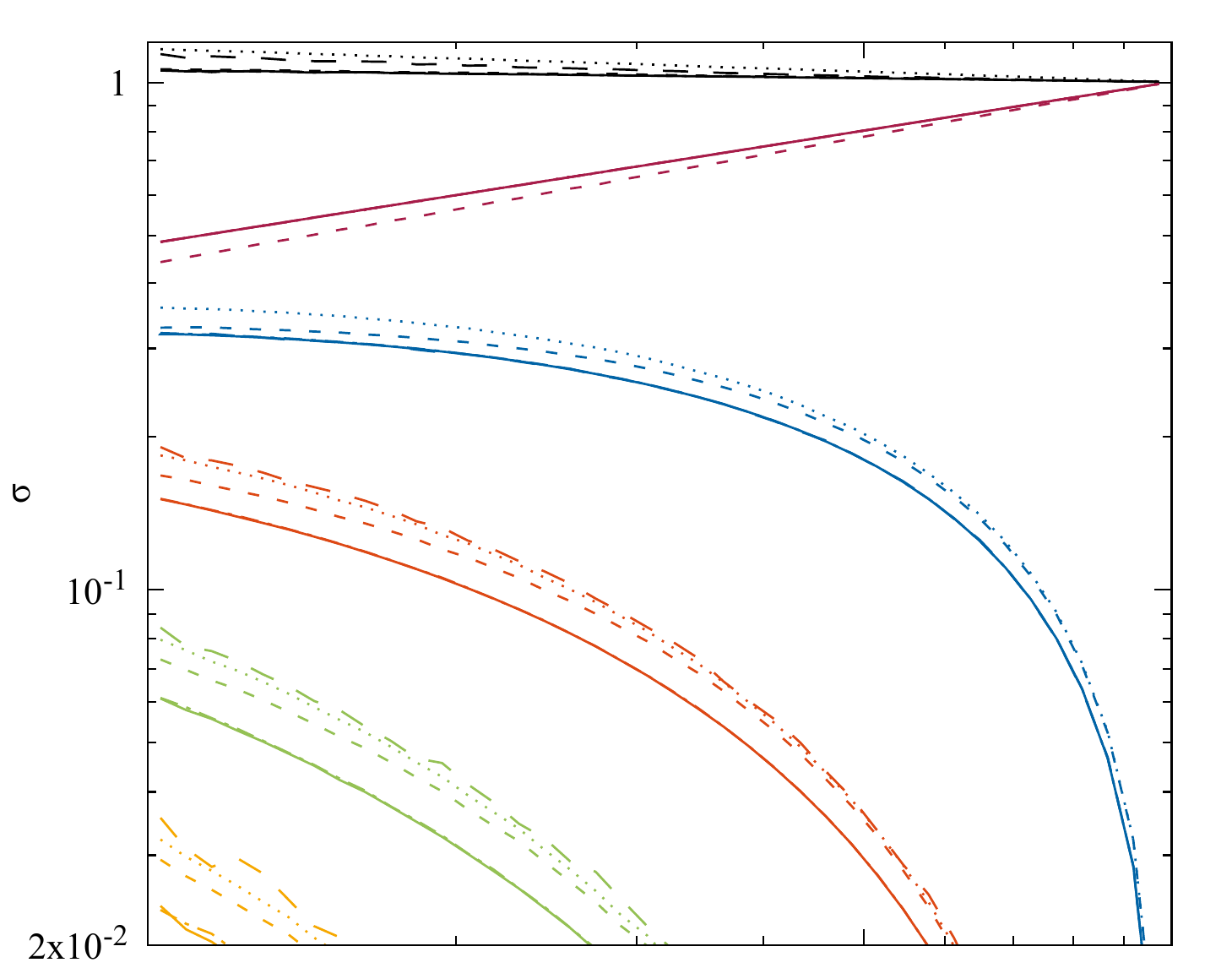}
\end{subfigure} \hfill
\begin{subfigure}[t]{0.9\textwidth}
\centering
\includegraphics[width=1.0\textwidth]{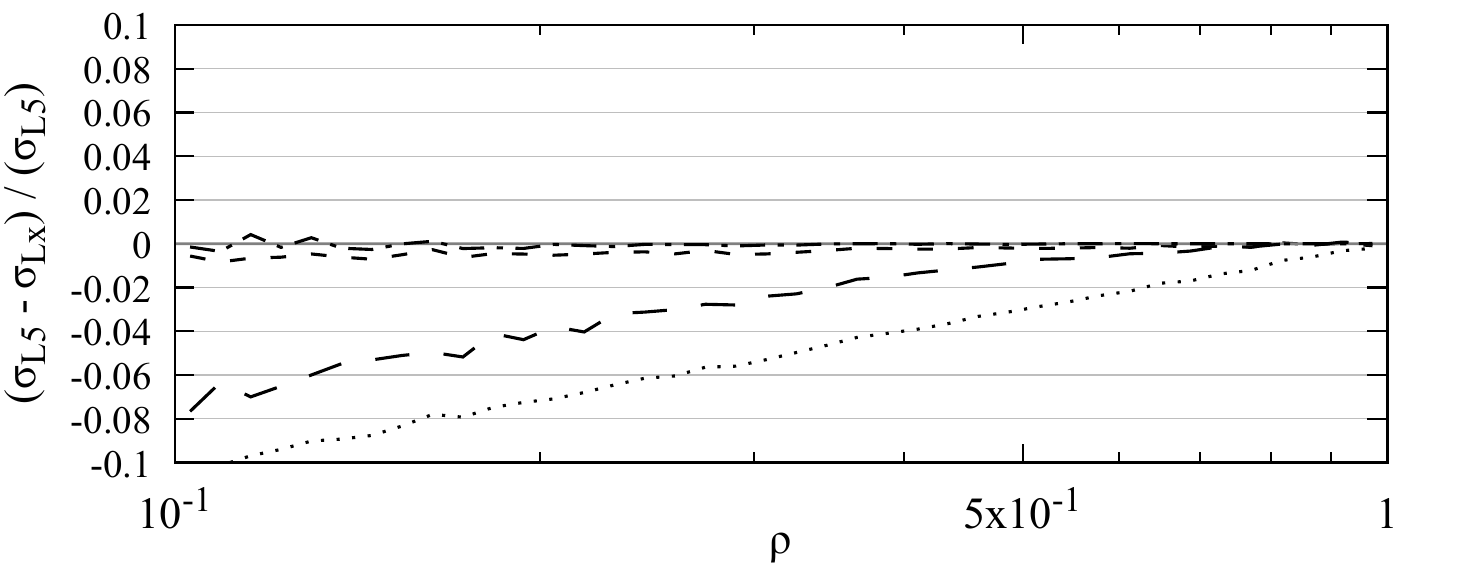}
\end{subfigure}
\caption{The veto cross-section for the $|10\>\<10|$ contribution to $q\bar{q} \to q\bar{q}$ in the asymmetric configuration. Solid: Full colour (L5), Dash-dotted: \LC\ + FCR (L4), Long-dashed: \LC\ + LCR + singlets (L3), Dotted: \LC\ + LCR (L2), Short-dashed: strict LC (L1). }
\label{fig:qq2qq1010a}
\end{figure}

Let us now change the kinematics and put the final state quark at rapidity $y=4$ and the final state antiquark at $y=-2$. The veto region is the interjet region $-1.3 < y < 3.3$. This asymmetric, low-angle scenario leads to very different results from the previous configuration. The $|10\>\<10|$ configuration is much less likely to radiate into the veto region than the $|01\>\<01|$\footnote{To a lesser extent this was also true for the previous kinematics.} and this immediately renders all of the \LC\ approximations inadequate whenever the $|01\>\<01|$ contribution is important since none are able to flip the colour of the $|01\>\<01|$ configuration, via a virtual exchange, into the $|10\>\<10|$ state. We expect that this effect would be even more dramatic had we included Coulomb exchanges. Note that in a hadron-hadron collision, the $|01\>\<01|$ contribution would be leading since $t$-channel gluon exchange dominates for low-angle scattering. Conversely, the L4 approximation is very good for the evolution of the $|10\>\<10|$ configuration. These features are all illustrated in Fig.~\ref{fig:qq2qq0101a}--Fig.~\ref{fig:qq2qq1001a}. Fig.~\ref{fig:qq2qq1010a} also shows that the L1 (strictly leading colour) approximation is a good approximation for the evolution of the $|10\>\<10|$ configuration.

\begin{figure}[t]
\centering
\begin{subfigure}[t]{0.9\textwidth}
\centering
\includegraphics[width=1.0\textwidth]{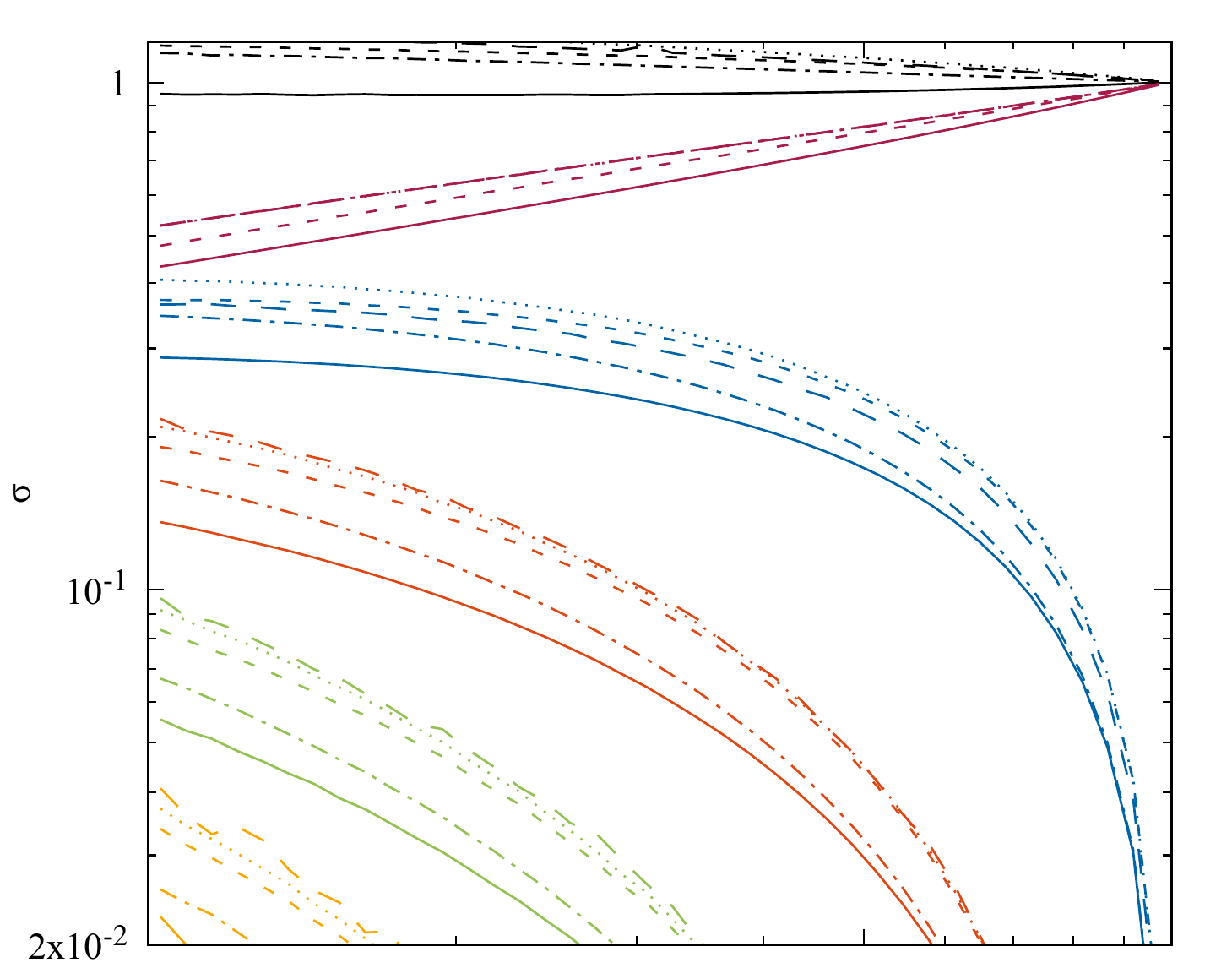}
\end{subfigure} \hfill
\begin{subfigure}[t]{0.9\textwidth}
\centering
\includegraphics[width=1.0\textwidth]{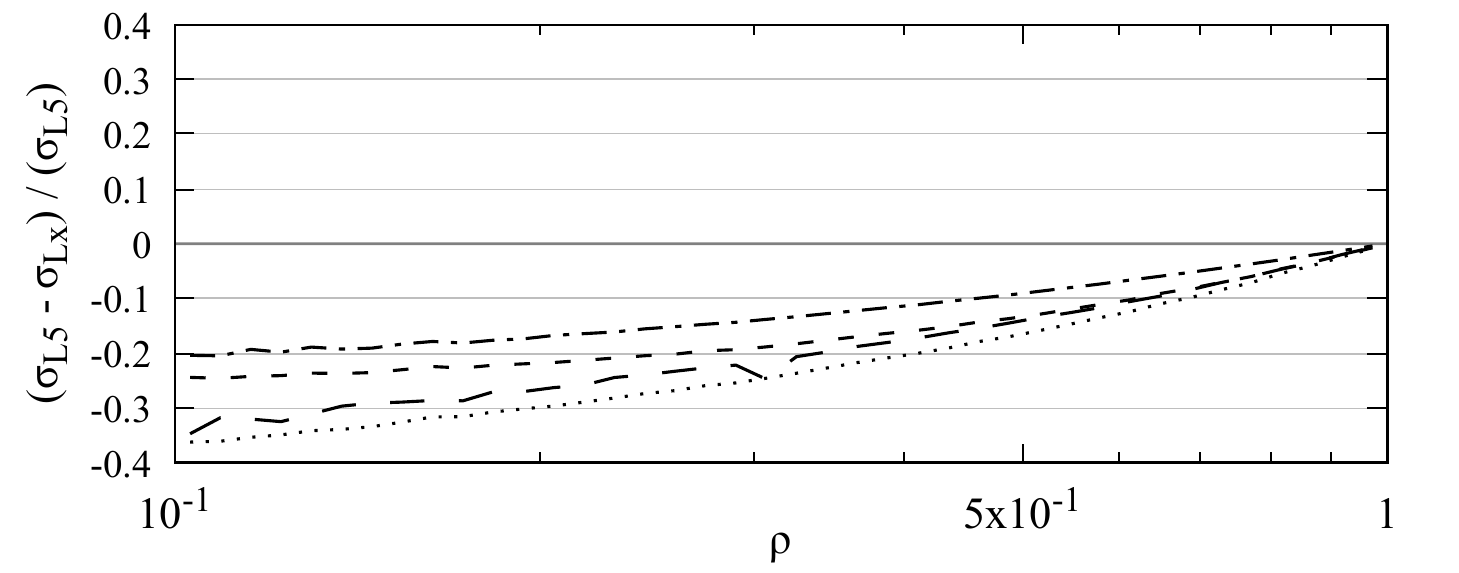}
\end{subfigure}
\caption{The veto cross-section for the $s$-channel gluon exchange contribution to $q\bar{q} \to q\bar{q}$ in the asymmetric configuration. Solid: Full colour (L5), Dash-dotted: \LC\ + FCR (L4), Long-dashed: \LC\ + LCR + singlets (L3), Dotted: \LC\ + LCR (L2), Short-dashed: strict LC (L1).}
\label{fig:qq2qqsa}
\end{figure}

\begin{figure}[t]
\centering
\begin{subfigure}[t]{0.9\textwidth}
\centering
\includegraphics[width=1.0\textwidth]{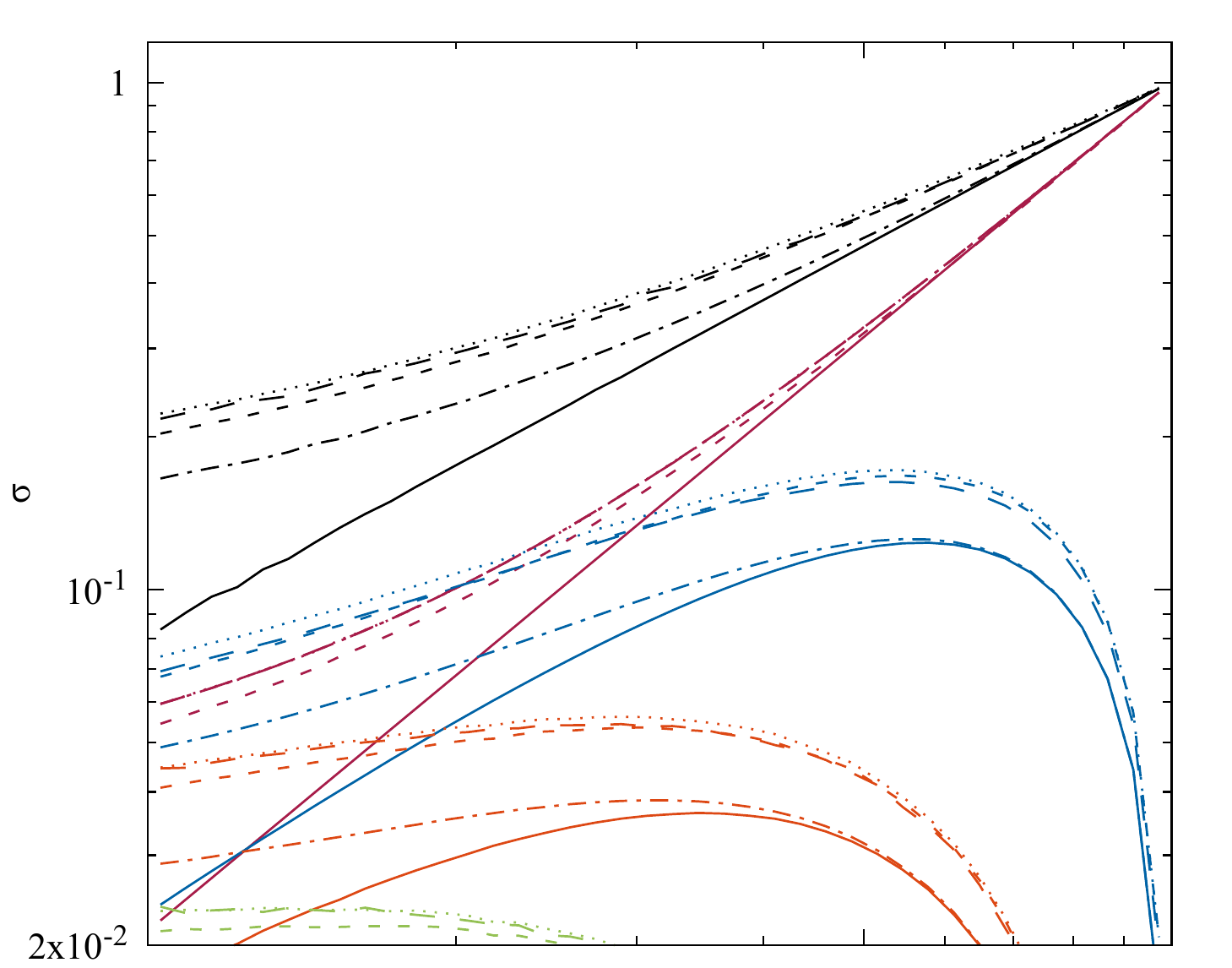}
\end{subfigure} \hfill
\begin{subfigure}[t]{0.9\textwidth}
\centering
\includegraphics[width=1.0\textwidth]{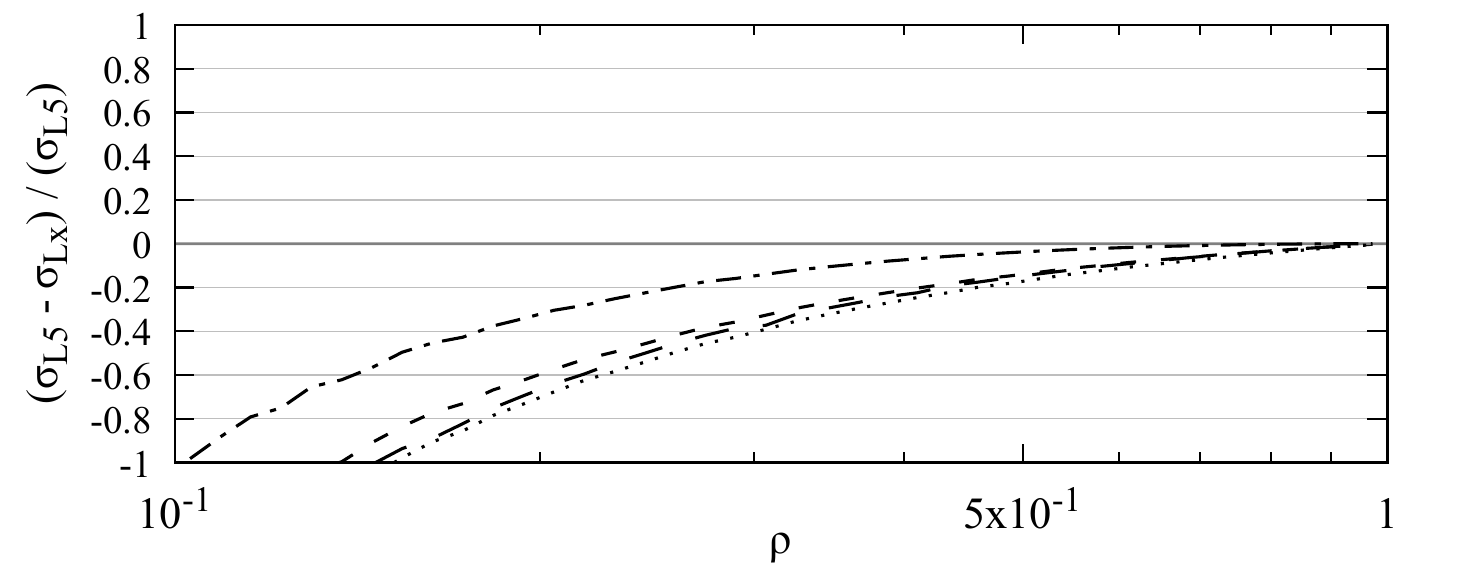}
\end{subfigure}
\caption{The veto cross-section for the $t$-channel gluon exchange contribution to $q\bar{q} \to q\bar{q}$ in the asymmetric configuration. Solid: Full colour (L5), Dash-dotted: \LC\ + FCR (L4), Long-dashed: \LC\ + LCR + singlets (L3), Dotted: \LC\ + LCR (L2), Short-dashed: strict LC (L1).}
\label{fig:qq2qqta}
\end{figure}

\begin{figure}[t]
\centering
\begin{subfigure}[t]{0.9\textwidth}
\centering
\includegraphics[width=1.0\textwidth]{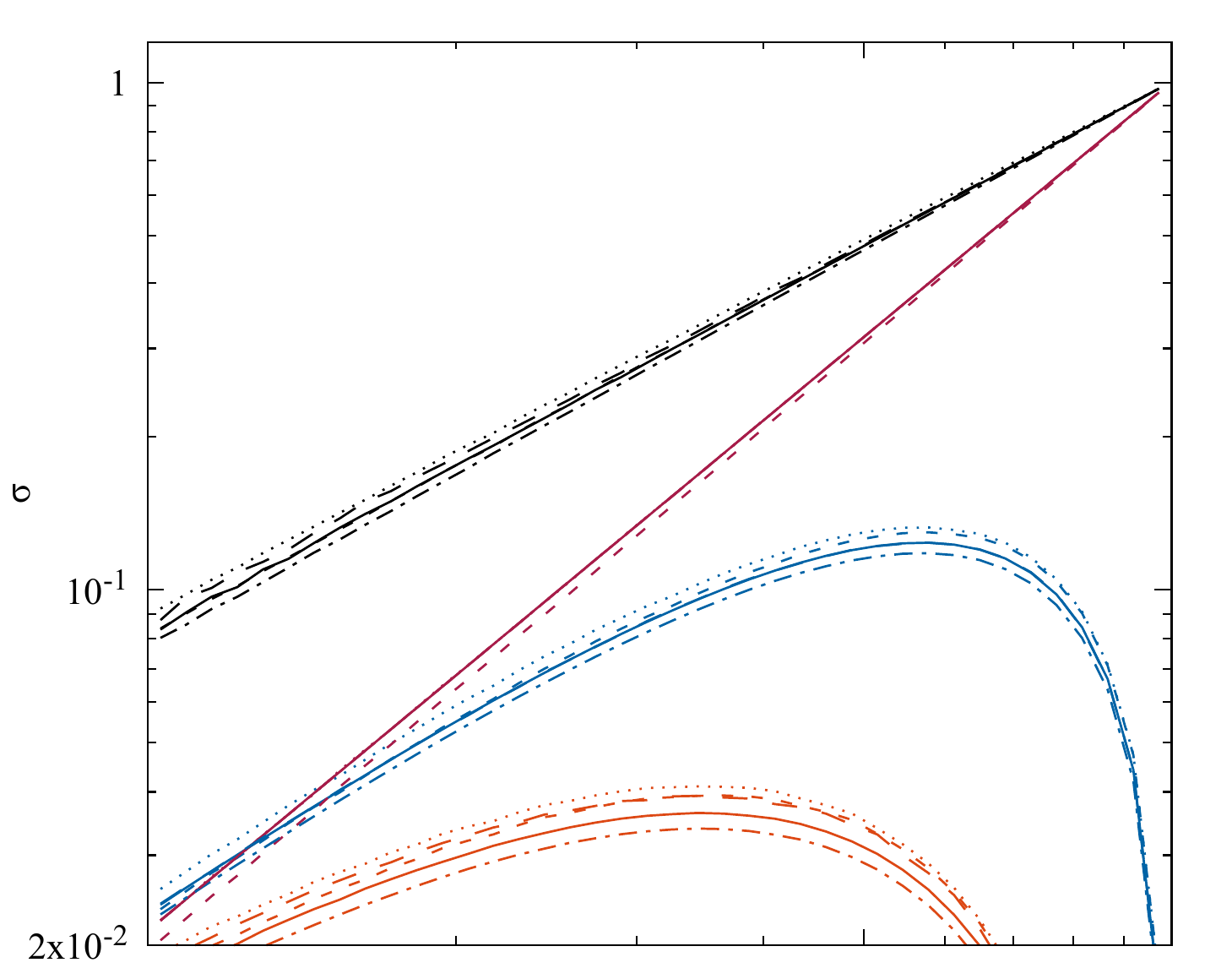}
\end{subfigure} \hfill
\begin{subfigure}[t]{0.9\textwidth}
\centering
\includegraphics[width=1.0\textwidth]{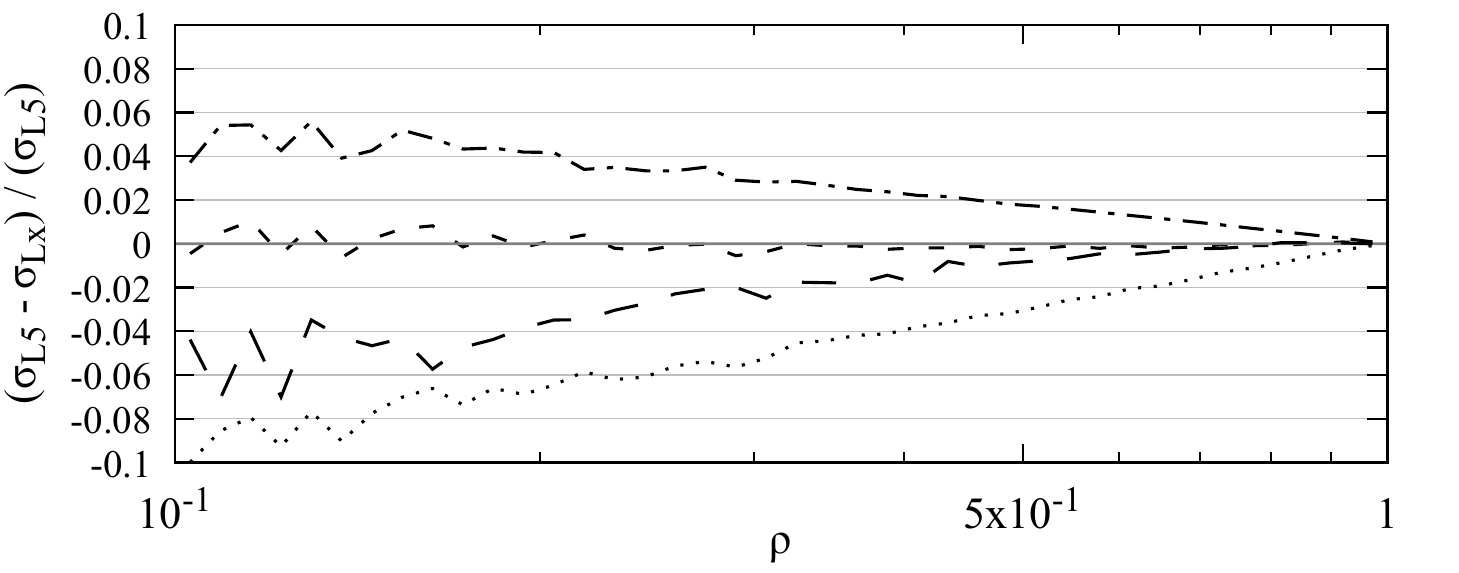}
\end{subfigure}
\caption{The veto cross-section for the $t$-channel gluon exchange contribution to $q\bar{q} \to q\bar{q}$ in the asymmetric configuration.  Solid: Full colour (L5), Dash-dotted: \LC\ + FCR (L4), Long-dashed: \LC\ + LCR + singlets (L3), Dotted: \LC\ + LCR (L2), Short-dashed: strict LC (L1). For the L1--L4 curves we start the evolution using the leading-colour approximation to the hard-scatter matrix.}
\label{fig:qq2qqta-LCH}
\end{figure}

\begin{figure}[t]
\centering
\begin{subfigure}[t]{0.9\textwidth}
\centering
\includegraphics[width=1.0\textwidth]{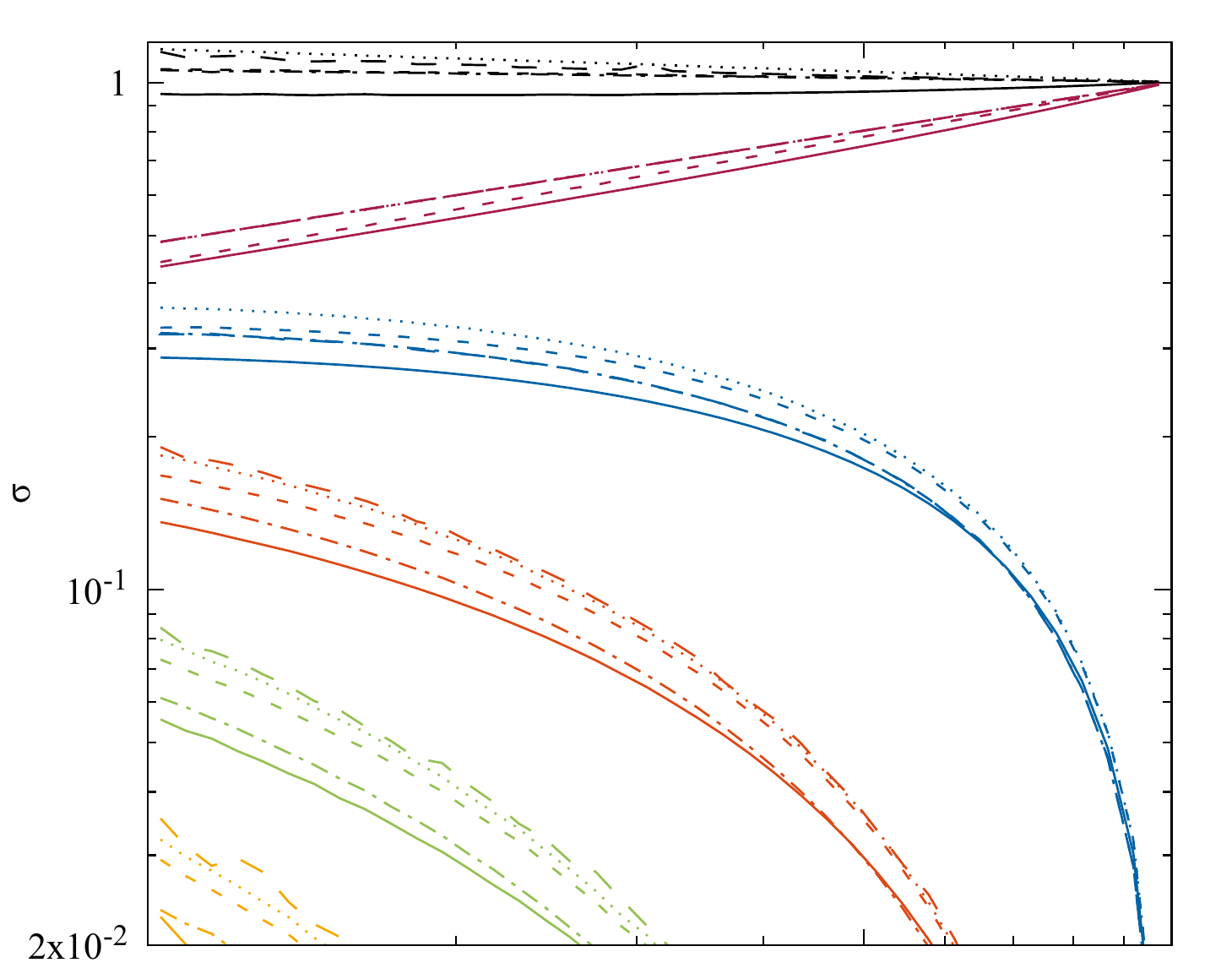}
\end{subfigure} \hfill
\begin{subfigure}[t]{0.9\textwidth}
\centering
\includegraphics[width=1.0\textwidth]{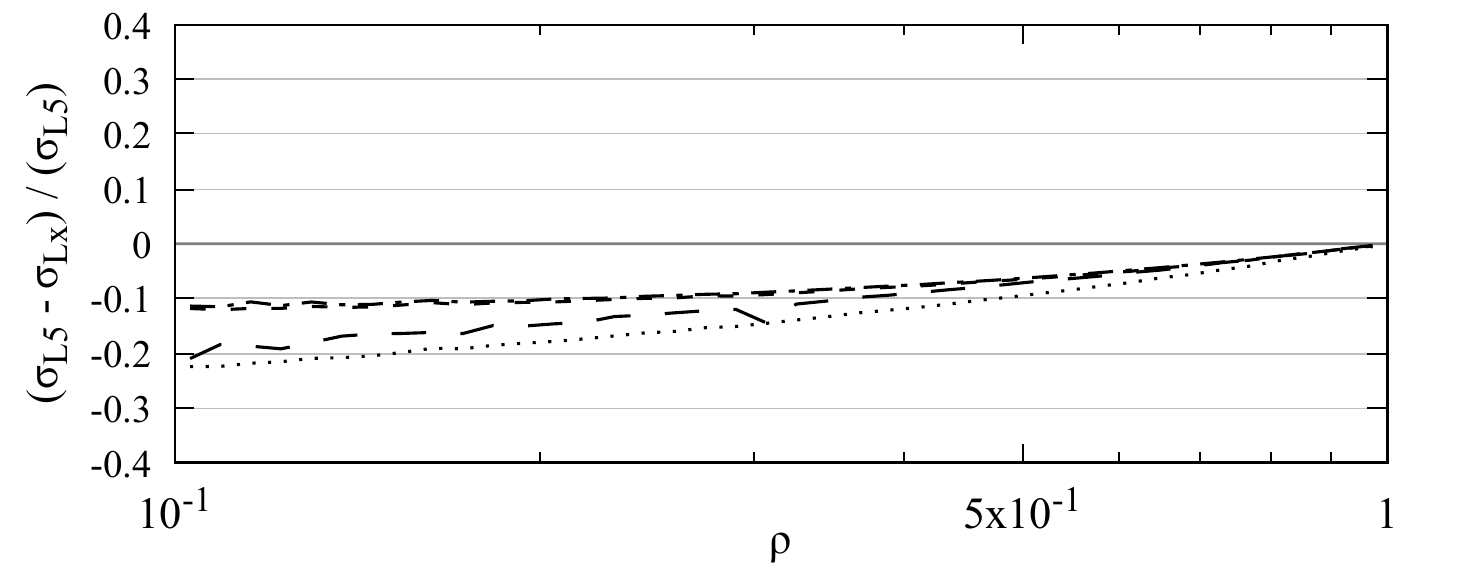}
\end{subfigure}
\caption{The veto cross-section for the $s$-channel gluon exchange contribution to $q\bar{q} \to q\bar{q}$ in the asymmetric configuration. Solid: Full colour (L5), Dash-dotted: \LC\ + FCR (L4), Long-dashed: \LC\ + LCR + singlets (L3), Dotted: \LC\ + LCR (L2), Short-dashed: strict LC (L1). For the L1--L4 curves we start the evolution using the leading-colour approximation to the hard-scatter matrix.}
\label{fig:qq2qqsa-LCH}
\end{figure}

For the evolution of the $s$-channel gluon exchange diagram starting from the full-colour hard scatter matrix ($H_s$), we find that all of the \LC\ approximations are at least 20\% above the full colour result at $\rho = 0.1$. For evolution of the dominant $t$-channel gluon exchange diagram, the situation is, as expected, even more extreme and none is correct within a factor of 2 at this value of $\rho$. Starting from the leading colour hard-scatter again improves matters and the L1 strictly leading colour curve once again agrees with the full colour result. These features are illustrated in Fig.~\ref{fig:qq2qqsa}--Fig.~\ref{fig:qq2qqsa-LCH}.

A word on the control that we have over these results. We have checked that the $d$-approximation is convergent (the $d=2$ approximation is sufficient for all of the plots presented so far) and we have checked that our results agree perfectly with independent analytic calculations for zero emissions and one emission. That the $d=2$ approximation is sufficient for the full-colour curves is at first a surprise since $d=4$ is formally required at this order in $1/\Nc$. However, the lower multiplicities have the largest relative contribution from larger $d$ for combinatoric reasons, i.e. there are many more possible $d=1$ contributions at high multiplicities, and for these we can check explicitly that $d=2$ is good. We have also checked that our results are precisely independent of the collinear cutoff. For the symmetric scenario we find that $\lambda = 0.01$ is sufficient and for the asymmetric scenario we need to take a smaller value, $\lambda = 0.001$.

\FloatBarrier 

\subsection{\boldmath$qg \to qg$}
\label{sec:qg2qg}

\begin{figure}[t]
\centering
\begin{subfigure}[t]{0.9\textwidth}
\centering
\includegraphics[width=1.0\textwidth]{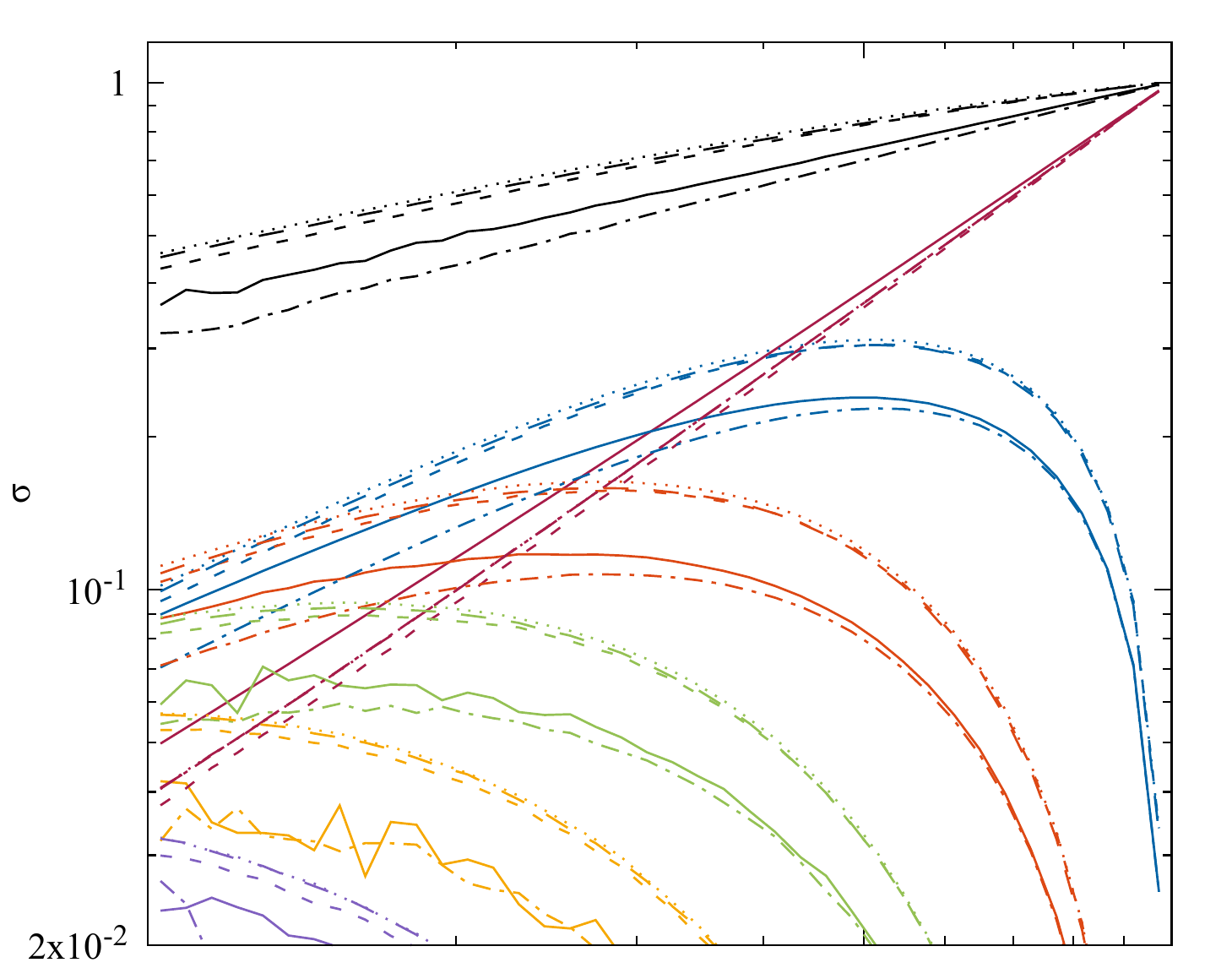}
\end{subfigure} \hfill
\begin{subfigure}[t]{0.9\textwidth}
\centering
\includegraphics[width=1.0\textwidth]{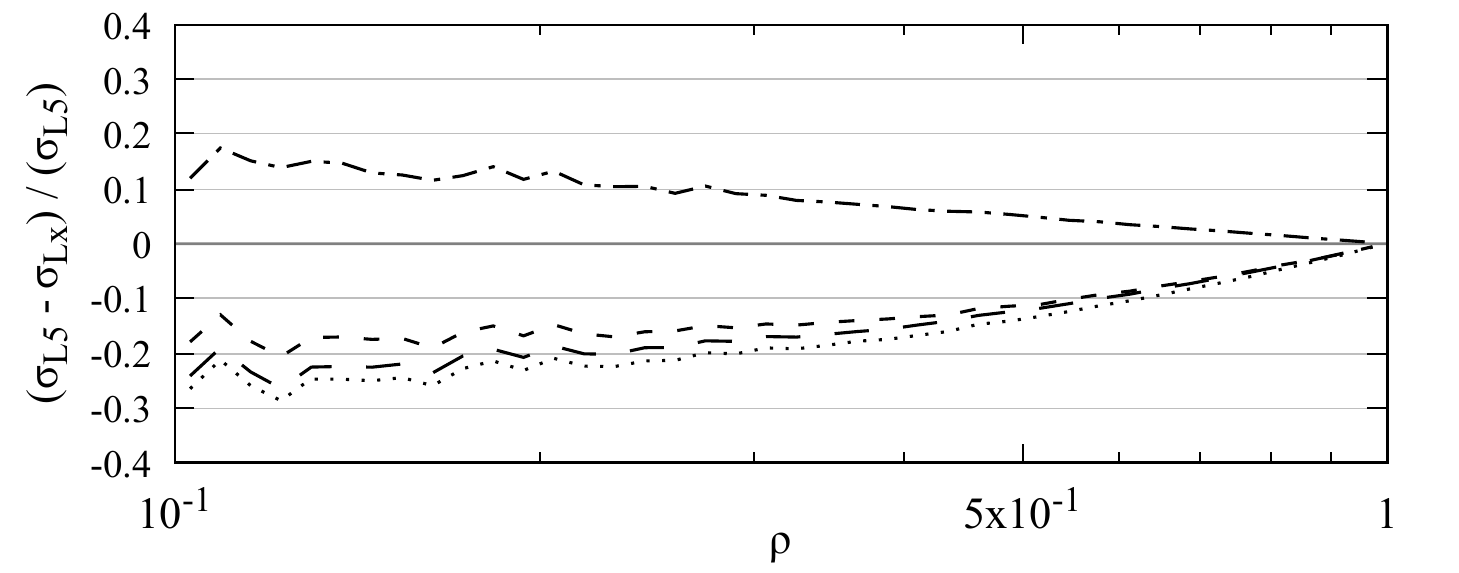}
\end{subfigure}
\caption{The veto cross-section for the $s$-channel quark exchange contribution to $qg \to qg$ in the symmetric configuration. Solid: Full colour (L5), Dash-dotted: \LC\ + FCR (L4), Long-dashed: \LC\ + LCR + singlets (L3), Dotted: \LC\ + LCR (L2), Short-dashed: strict LC (L1). Evolution starts from the full-colour hard-scatter matrix.}
\label{fig:qg2qg-s-FCHs}
\end{figure}

\begin{figure}[t]
\centering
\begin{subfigure}[t]{0.9\textwidth}
\centering
\includegraphics[width=1.0\textwidth]{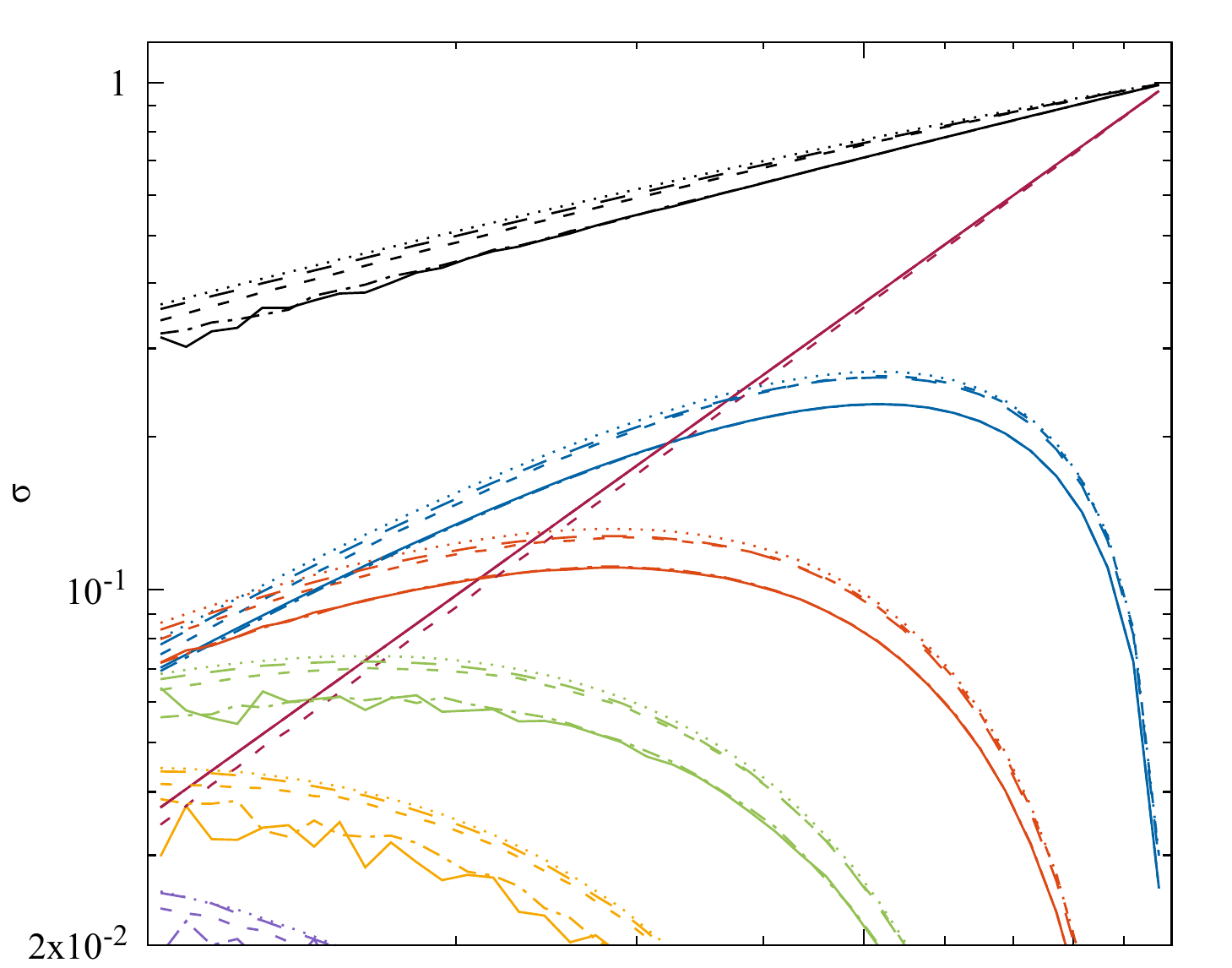}
\end{subfigure} \hfill
\begin{subfigure}[t]{0.9\textwidth}
\centering
\includegraphics[width=1.0\textwidth]{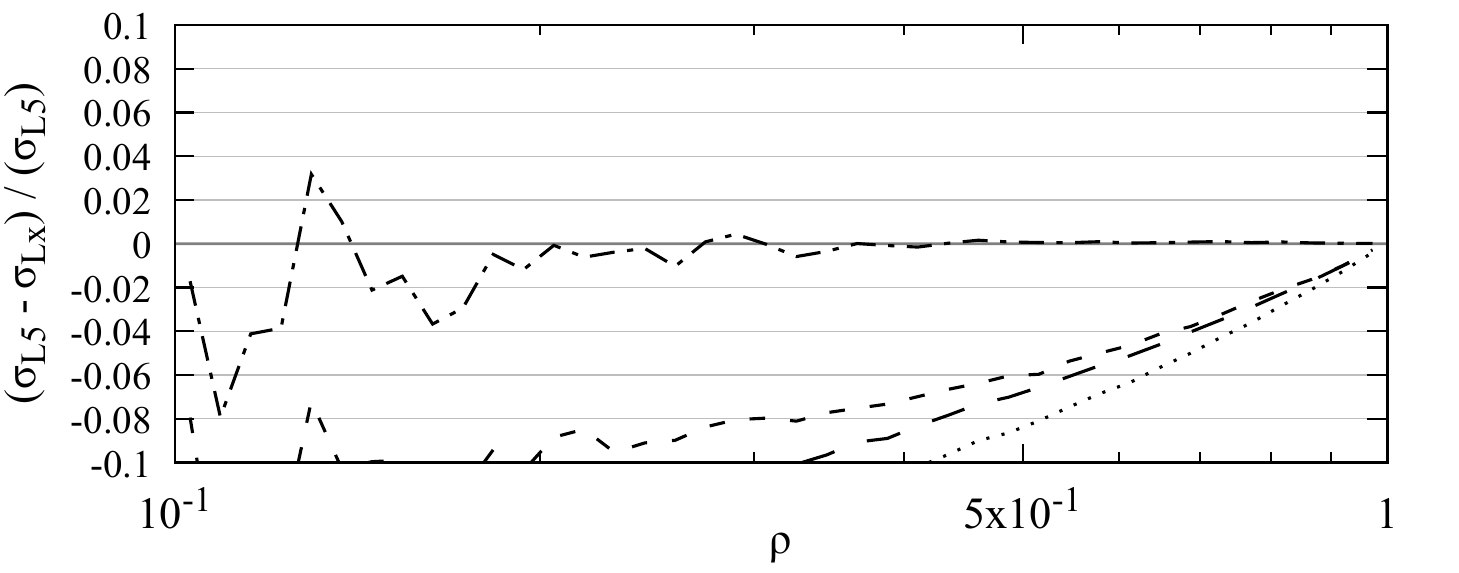}
\end{subfigure}
\caption{The veto cross-section for the $t$-channel gluon exchange contribution to $qg \to qg$ in the symmetric configuration. Solid: Full colour (L5), Dash-dotted: \LC\ + FCR (L4), Long-dashed: \LC\ + LCR + singlets (L3), Dotted: \LC\ + LCR (L2), Short-dashed: strict LC (L1). Evolution starts from the full-colour hard-scatter matrix.}
\label{fig:qg2qg-t-FCHs}
\end{figure}

\begin{figure}[t]
\centering
\begin{subfigure}[t]{0.9\textwidth}
\centering
\includegraphics[width=1.0\textwidth]{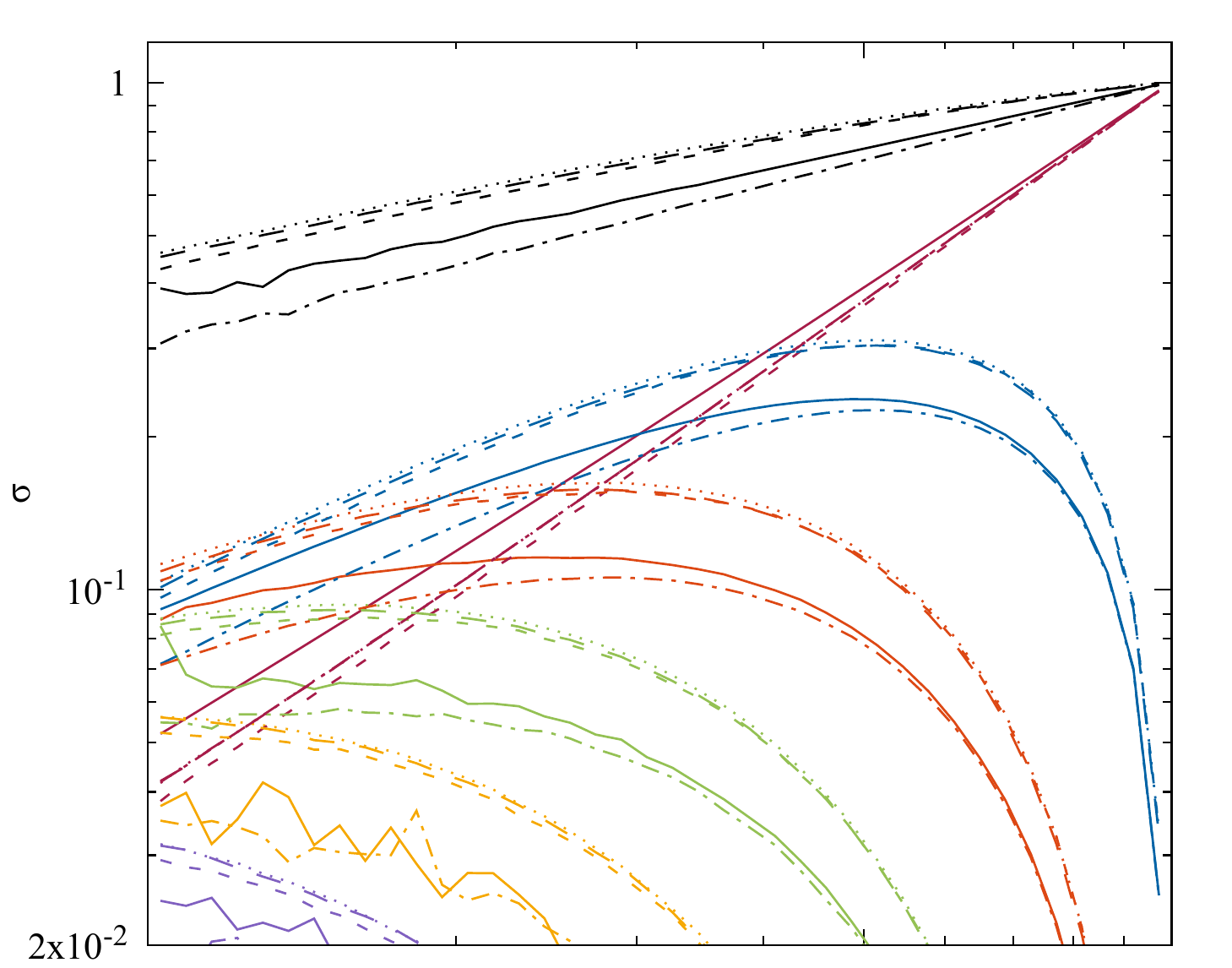}
\end{subfigure} \hfill
\begin{subfigure}[t]{0.9\textwidth}
\centering
\includegraphics[width=1.0\textwidth]{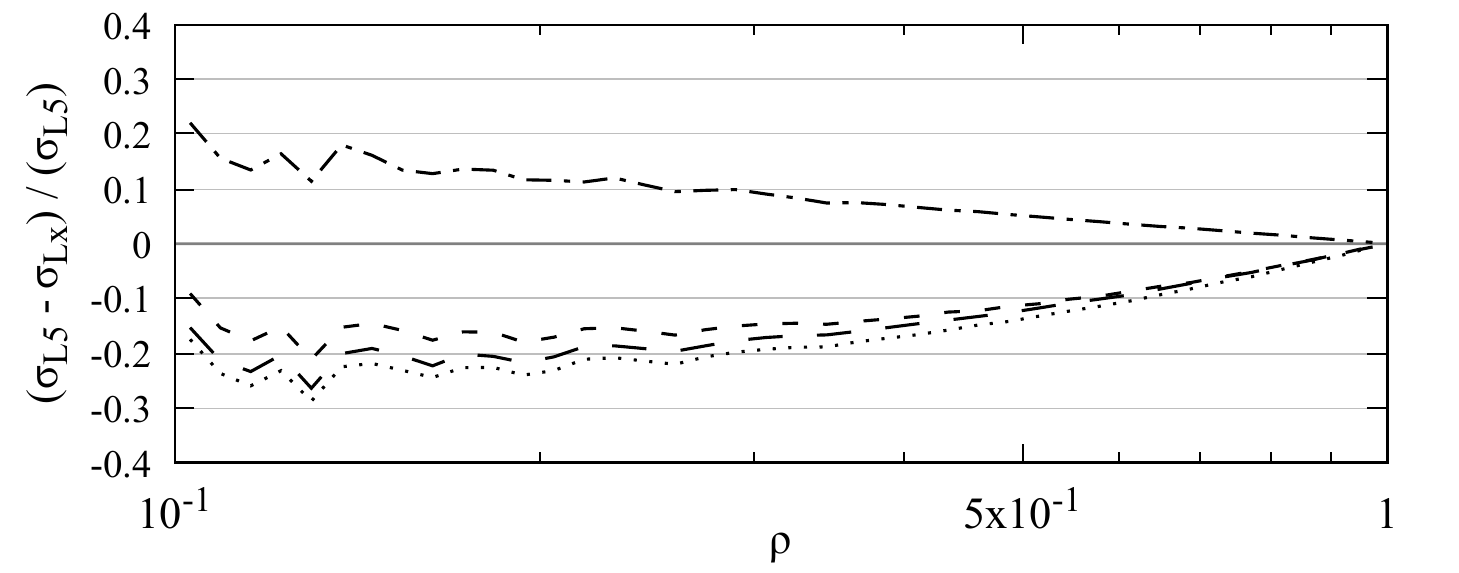}
\end{subfigure}
\caption{The veto cross-section for the $u$-channel quark exchange contribution to $qg \to qg$ in the symmetric configuration. Solid: Full colour (L5), Dash-dotted: \LC\ + FCR (L4), Long-dashed: \LC\ + LCR + singlets (L3), Dotted: \LC\ + LCR (L2), Short-dashed: strict LC (L1). Evolution starts from the full-colour hard-scatter matrix.}
\label{fig:qg2qg-u-FCHs}
\end{figure}

\begin{figure}[t]
\centering
\begin{subfigure}[t]{0.9\textwidth}
\centering
\includegraphics[width=1.0\textwidth]{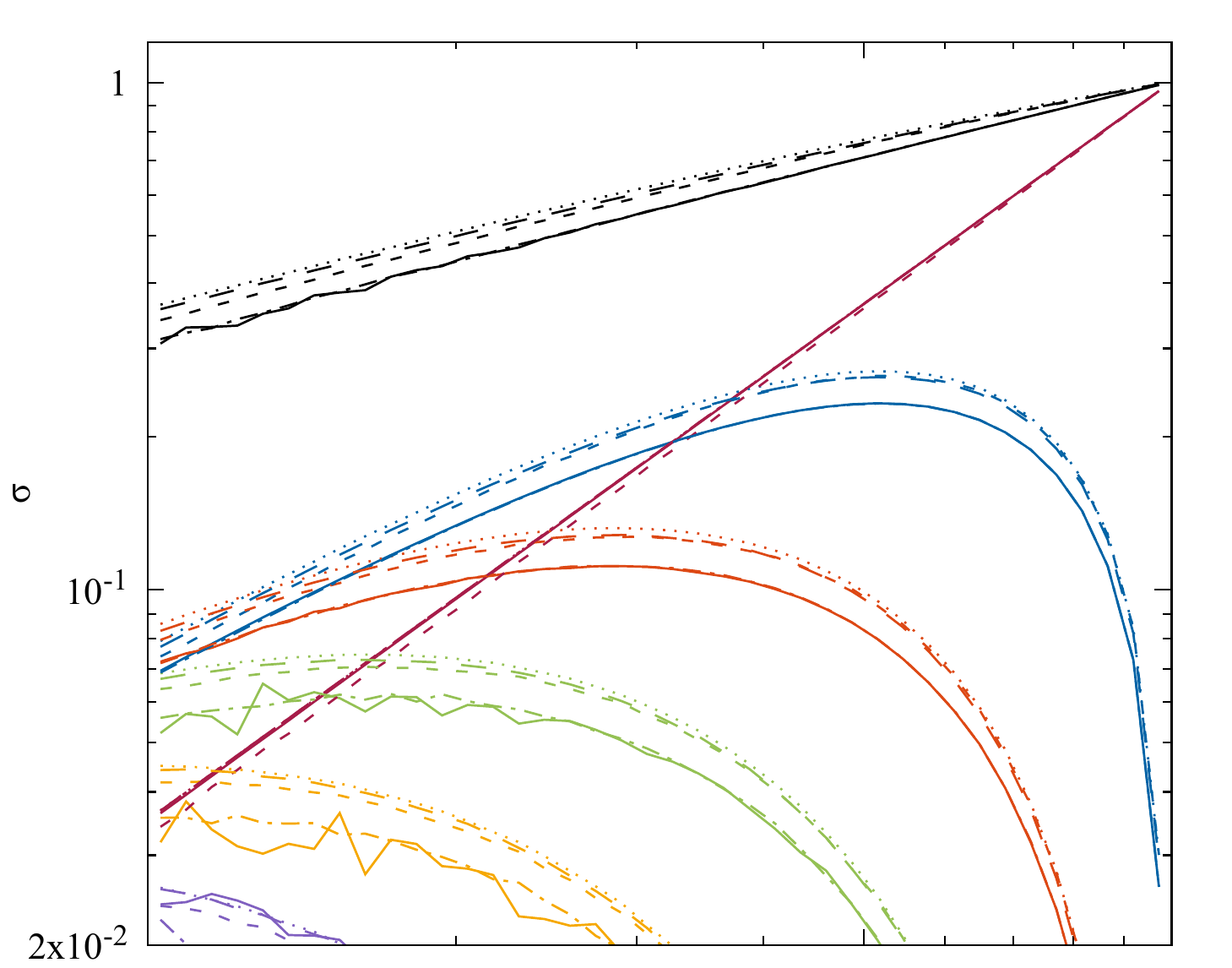}
\end{subfigure} \hfill
\begin{subfigure}[t]{0.9\textwidth}
\centering
\includegraphics[width=1.0\textwidth]{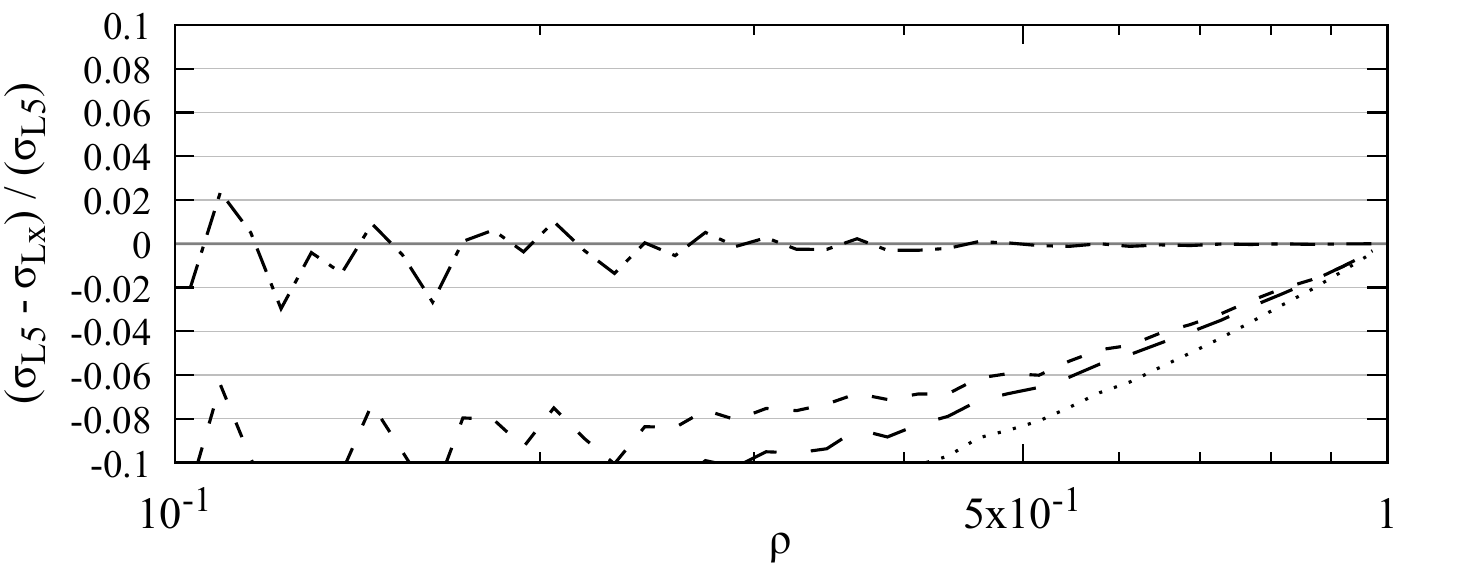}
\end{subfigure}
\caption{The veto cross-section for the $st$-channel interference contribution to $qg \to qg$ in the symmetric configuration. Solid: Full colour (L5), Dash-dotted: \LC\ + FCR (L4), Long-dashed: \LC\ + LCR + singlets (L3), Dotted: \LC\ + LCR (L2), Short-dashed: strict LC (L1). Evolution starts from the full-colour hard-scatter matrix.}
\label{fig:qg2qg-st-FCHs}
\end{figure}

\begin{figure}[t]
\centering
\begin{subfigure}[t]{0.9\textwidth}
\centering
\includegraphics[width=1.0\textwidth]{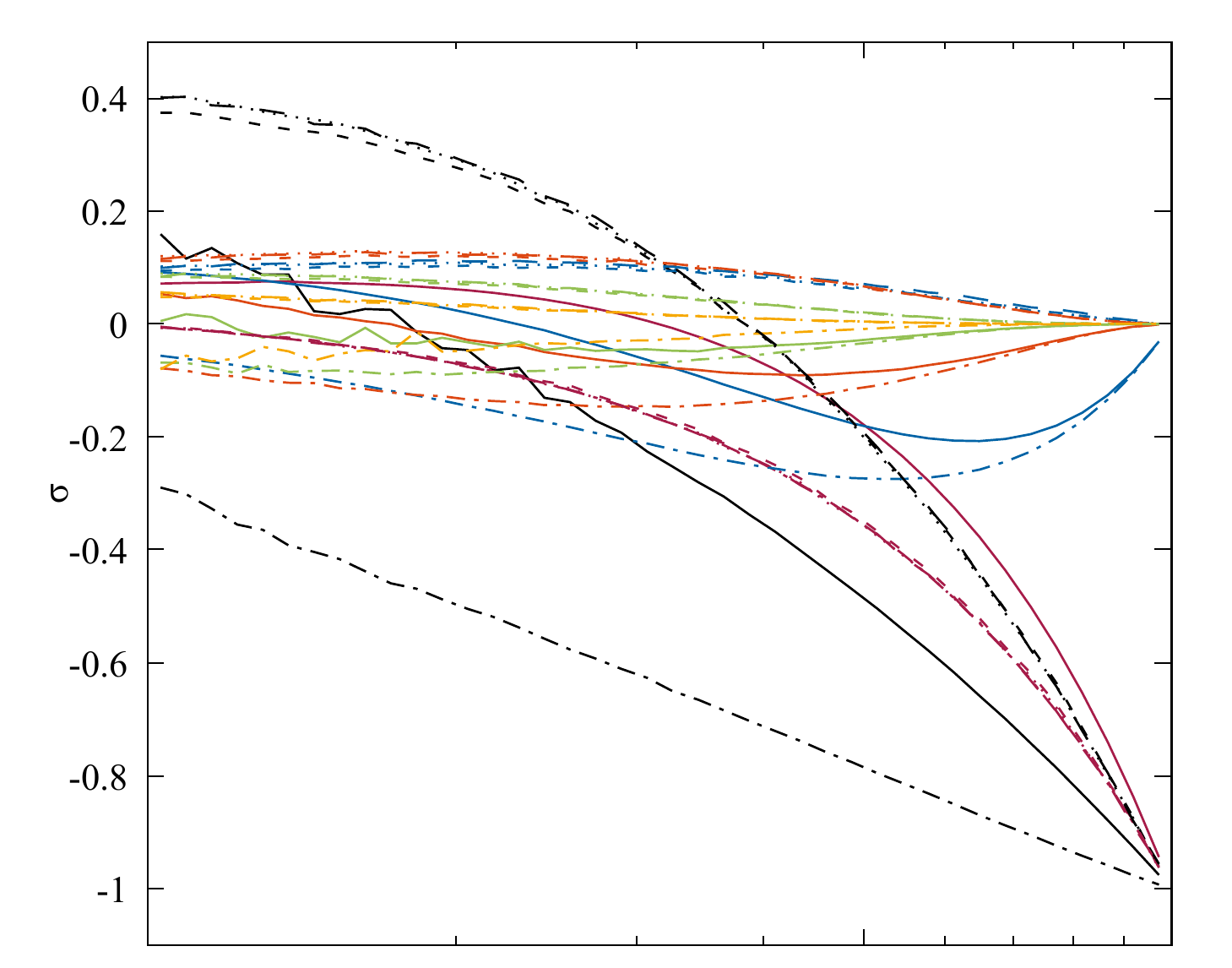}
\end{subfigure} \hfill
\begin{subfigure}[t]{0.9\textwidth}
\centering
\includegraphics[width=1.0\textwidth]{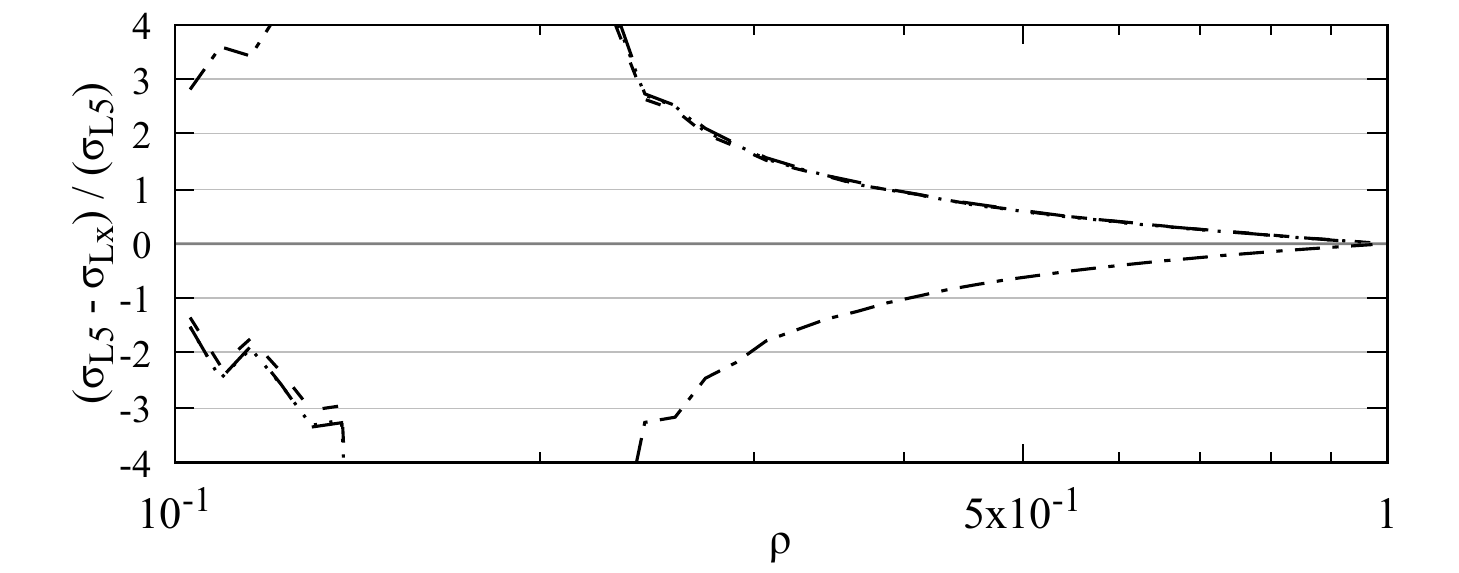}
\end{subfigure}
\caption{The veto cross-section for the $su$-channel interference contribution to $qg \to qg$ in the symmetric configuration. Solid: Full colour (L5), Dash-dotted: \LC\ + FCR (L4), Long-dashed: \LC\ + LCR + singlets (L3), Dotted: \LC\ + LCR (L2), Short-dashed: strict LC (L1). Evolution starts from the full-colour hard-scatter matrix.}
\label{fig:qg2qg-su-FCHs}
\end{figure}

\begin{figure}[t]
\centering
\begin{subfigure}[t]{0.9\textwidth}
\centering
\includegraphics[width=1.0\textwidth]{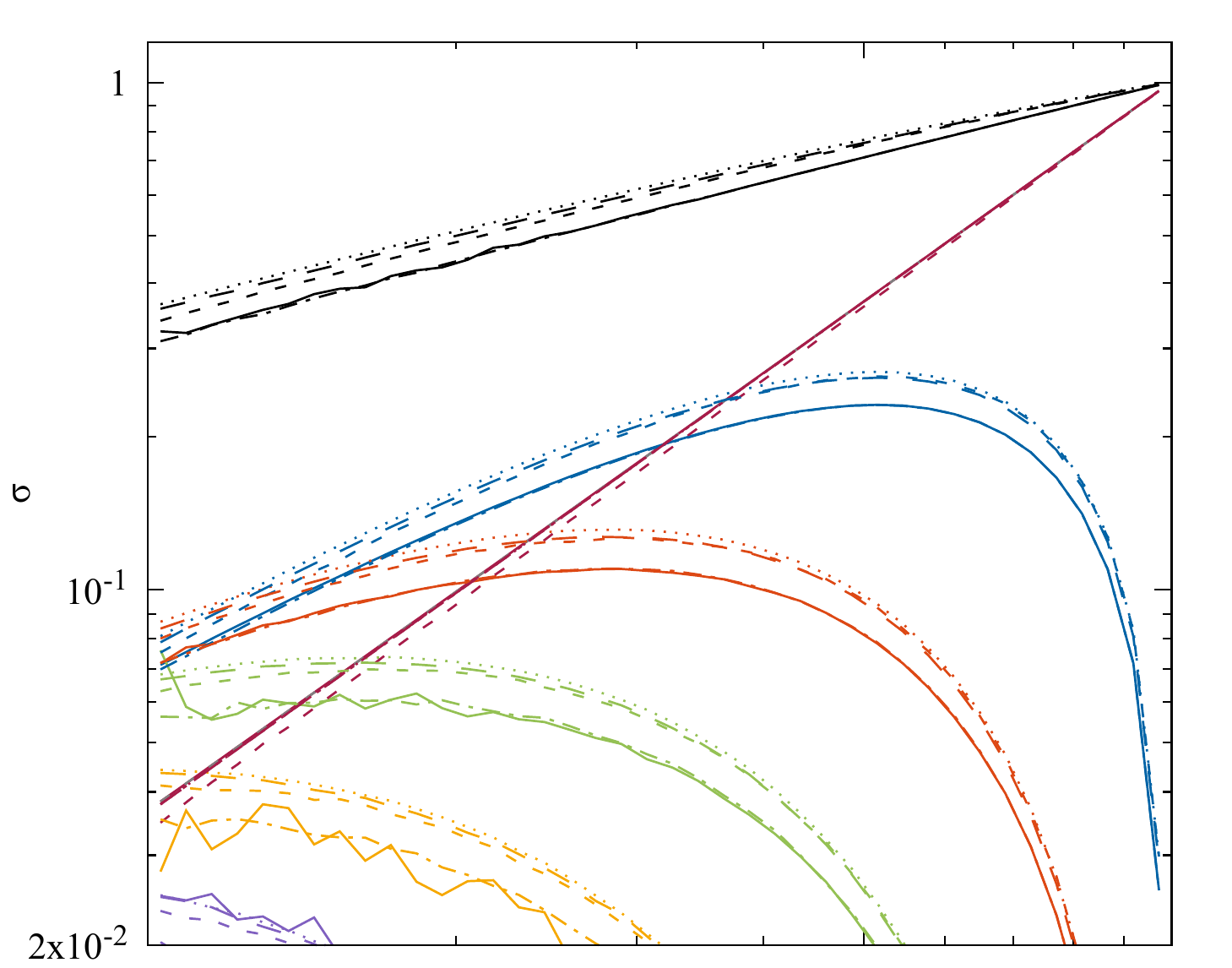}
\end{subfigure} \hfill
\begin{subfigure}[t]{0.9\textwidth}
\centering
\includegraphics[width=1.0\textwidth]{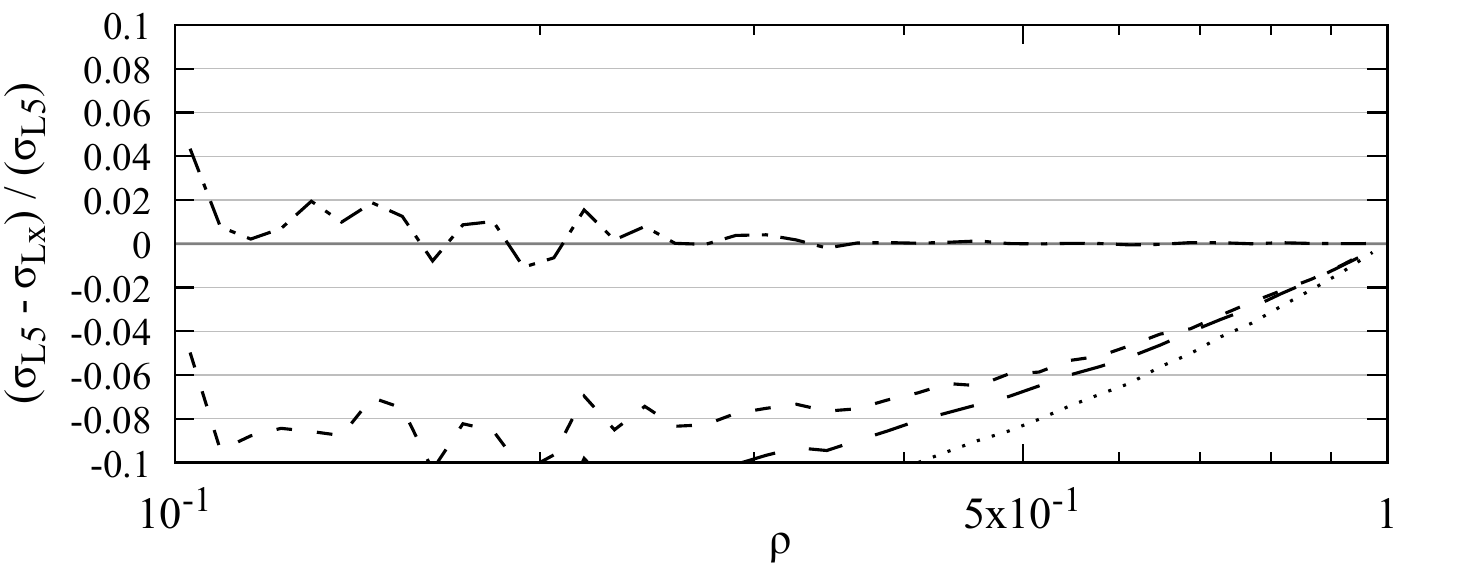}
\end{subfigure}
\caption{The veto cross-section for the $tu$-channel interference contribution to $qg \to qg$ in the symmetric configuration. Solid: Full colour (L5), Dash-dotted: \LC\ + FCR (L4), Long-dashed: \LC\ + LCR + singlets (L3), Dotted: \LC\ + LCR (L2), Short-dashed: strict LC (L1). Evolution starts from the full-colour hard-scatter matrix.}
\label{fig:qg2qg-tu-FCHs}
\end{figure}

\begin{figure}[t]
\centering
\begin{subfigure}[t]{0.9\textwidth}
\centering
\includegraphics[width=1.0\textwidth]{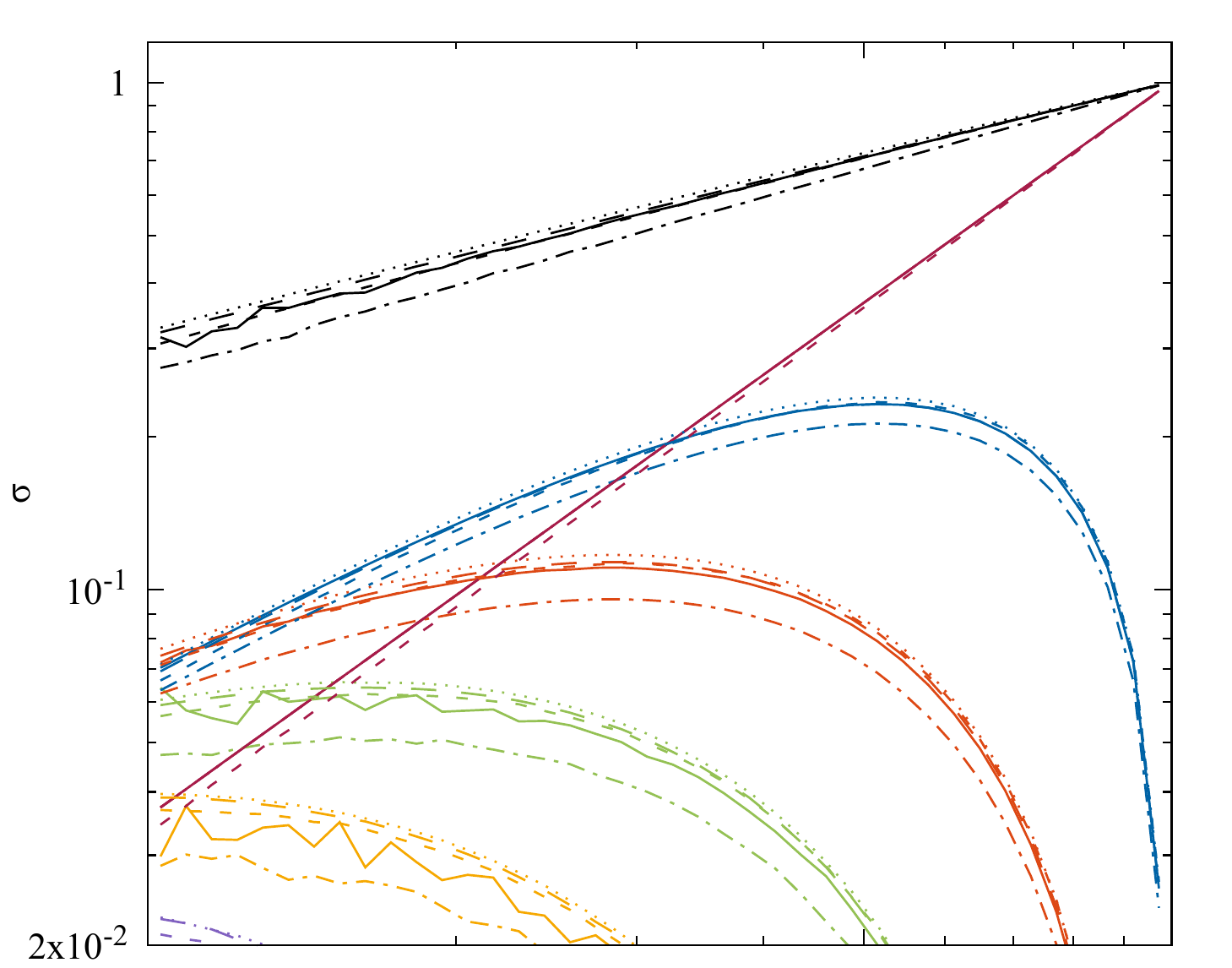}
\end{subfigure} \hfill
\begin{subfigure}[t]{0.9\textwidth}
\centering
\includegraphics[width=1.0\textwidth]{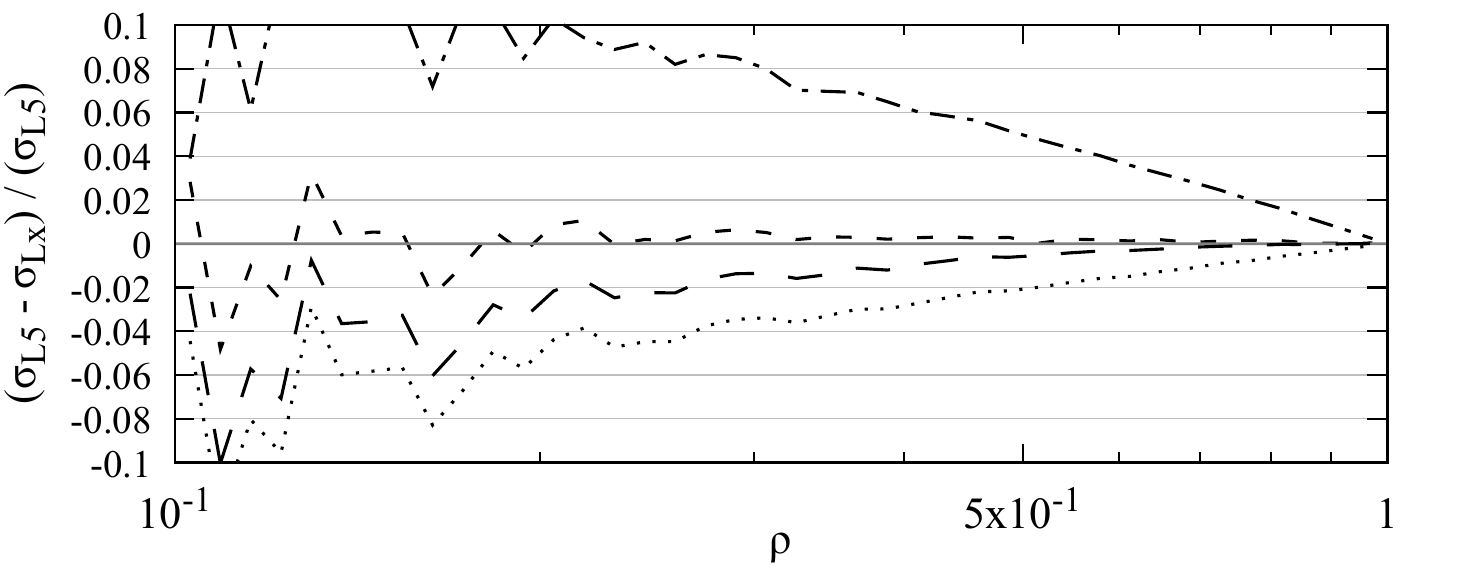}
\end{subfigure}
\caption{The veto cross-section for the $t$-channel gluon exchange contribution to $qg \to qg$ in the symmetric configuration. Solid: Full colour (L5), Dash-dotted: \LC\ + FCR (L4), Long-dashed: \LC\ + LCR + singlets (L3), Dotted: \LC\ + LCR (L2), Short-dashed: strict LC (L1). Evolution starts from the leading-colour hard-scatter matrix.}
\label{fig:qg2qg-t-LCHs}
\end{figure}

\begin{figure}[t]
\centering
\begin{subfigure}[t]{0.9\textwidth}
\centering
\includegraphics[width=1.0\textwidth]{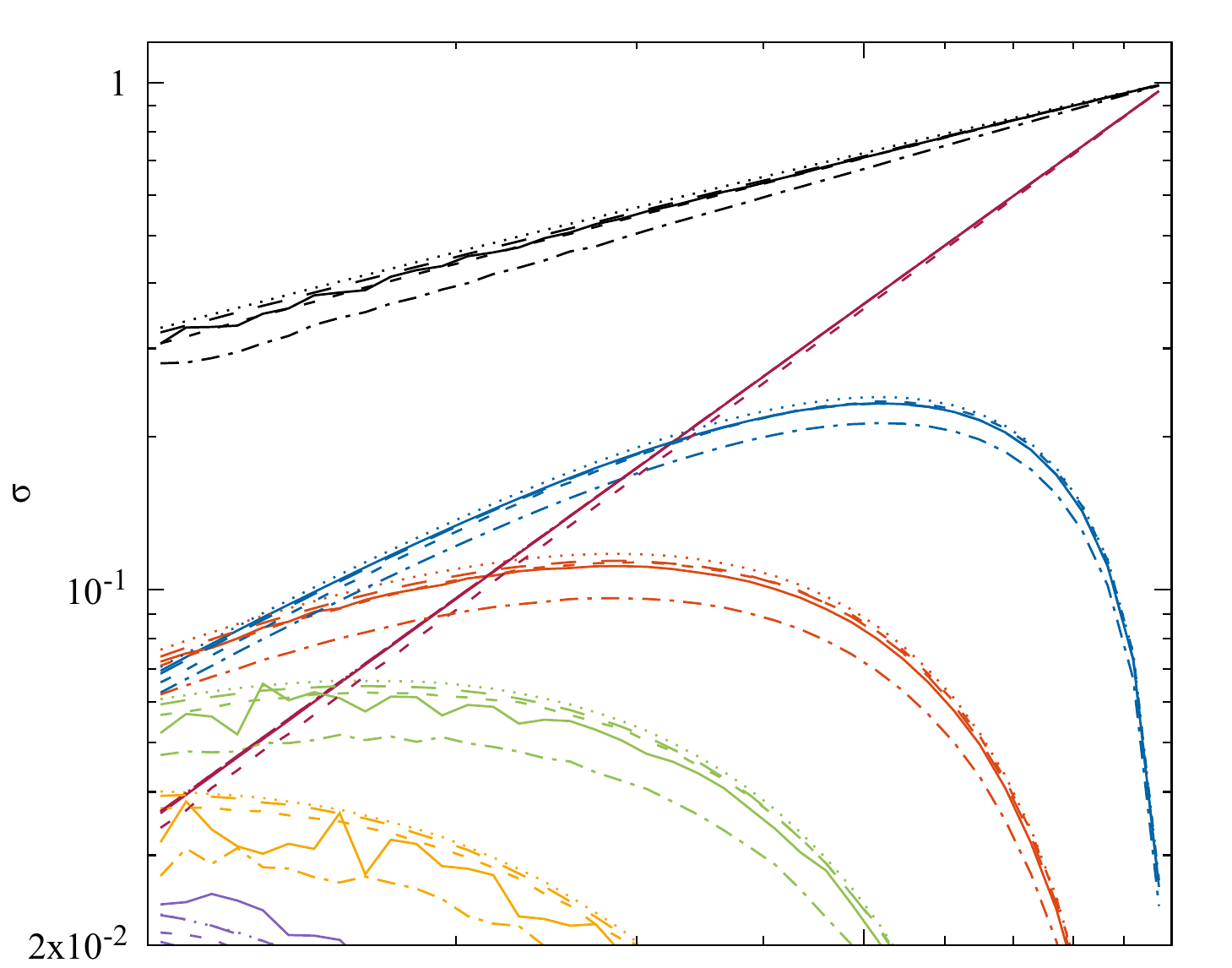}
\end{subfigure} \hfill
\begin{subfigure}[t]{0.9\textwidth}
\centering
\includegraphics[width=1.0\textwidth]{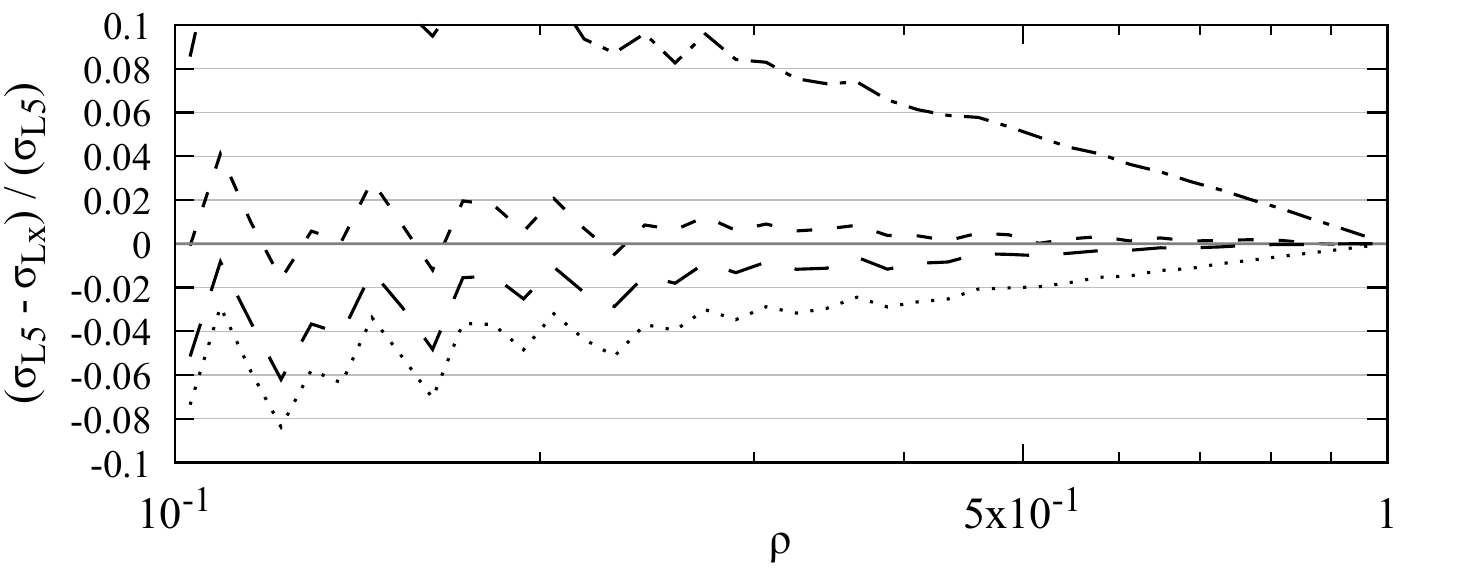}
\end{subfigure}
\caption{The veto cross-section for the $st$-channel interference contribution to $qg \to qg$ in the symmetric configuration. Solid: Full colour (L5), Dash-dotted: \LC\ + FCR (L4), Long-dashed: \LC\ + LCR + singlets (L3), Dotted: \LC\ + LCR (L2), Short-dashed: strict LC (L1). Evolution starts from the leading-colour hard-scatter matrix.}
\label{fig:qg2qg-st-LCHs}
\end{figure}

\begin{figure}[t]
\centering
\begin{subfigure}[t]{0.9\textwidth}
\centering
\includegraphics[width=1.0\textwidth]{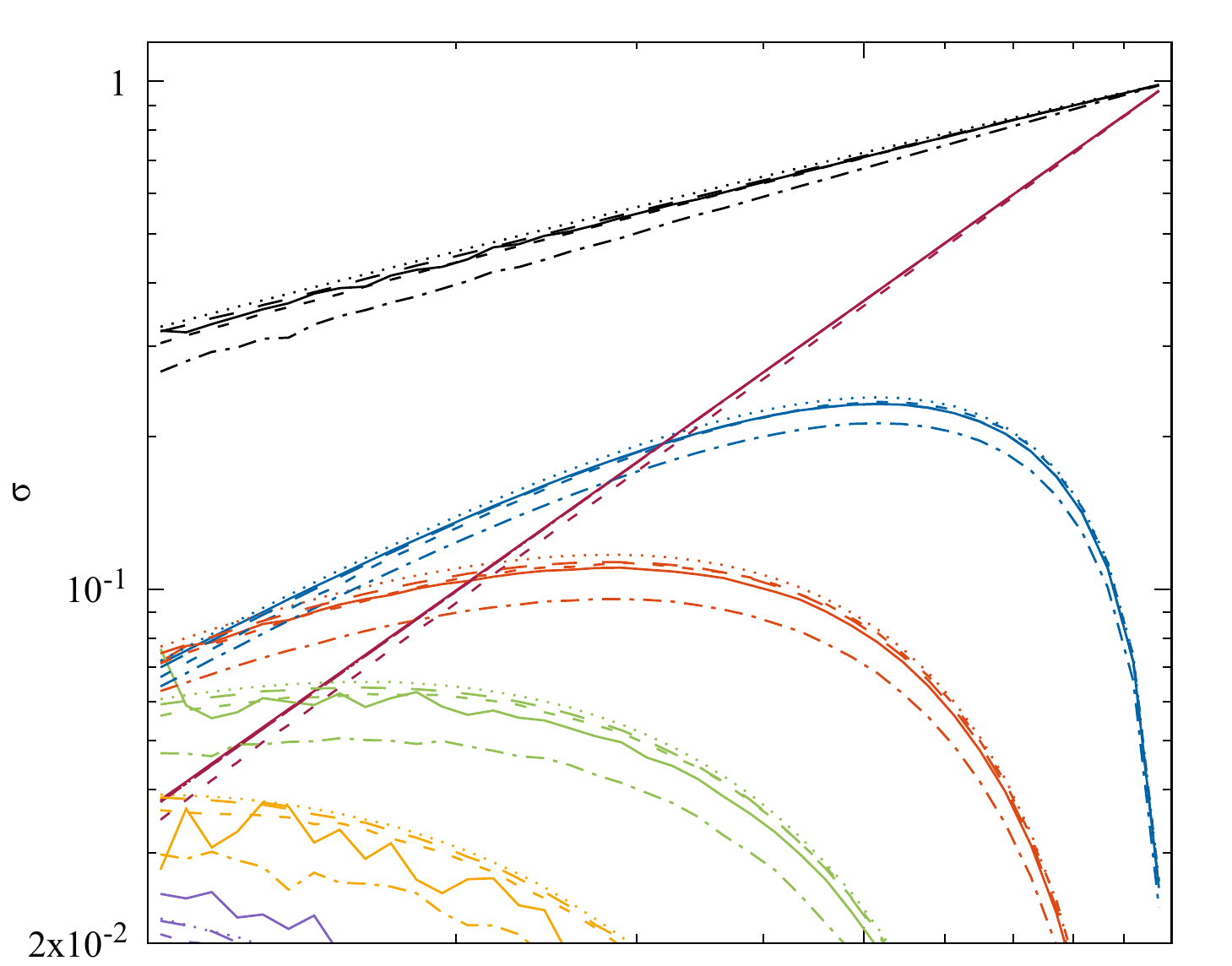}
\end{subfigure} \hfill
\begin{subfigure}[t]{0.9\textwidth}
\centering
\includegraphics[width=1.0\textwidth]{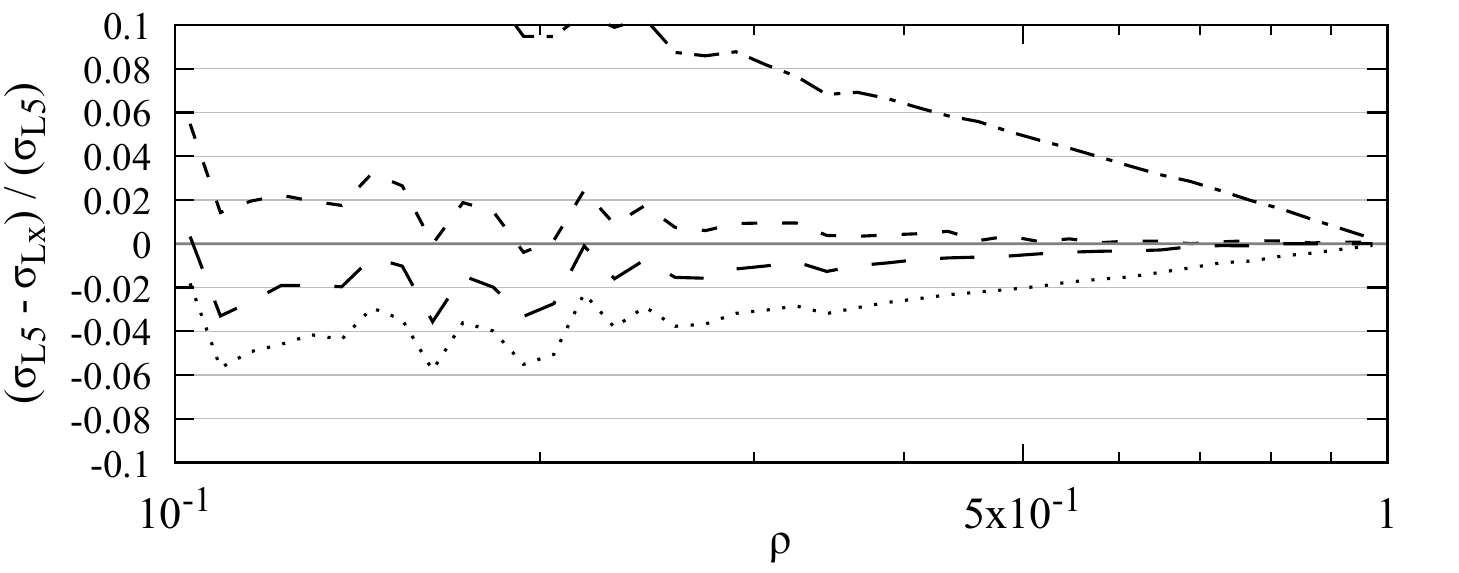}
\end{subfigure}
\caption{The veto cross-section for the $tu$-channel interference contribution to $qg \to qg$ in the symmetric configuration. Solid: Full colour (L5), Dash-dotted: \LC\ + FCR (L4), Long-dashed: \LC\ + LCR + singlets (L3), Dotted: \LC\ + LCR (L2), Short-dashed: strict LC (L1). Evolution starts from the leading-colour hard-scatter matrix.}
\label{fig:qg2qg-tu-LCHs}
\end{figure}

\begin{figure}[t]
\centering
\begin{subfigure}[t]{0.9\textwidth}
\centering
\includegraphics[width=1.0\textwidth]{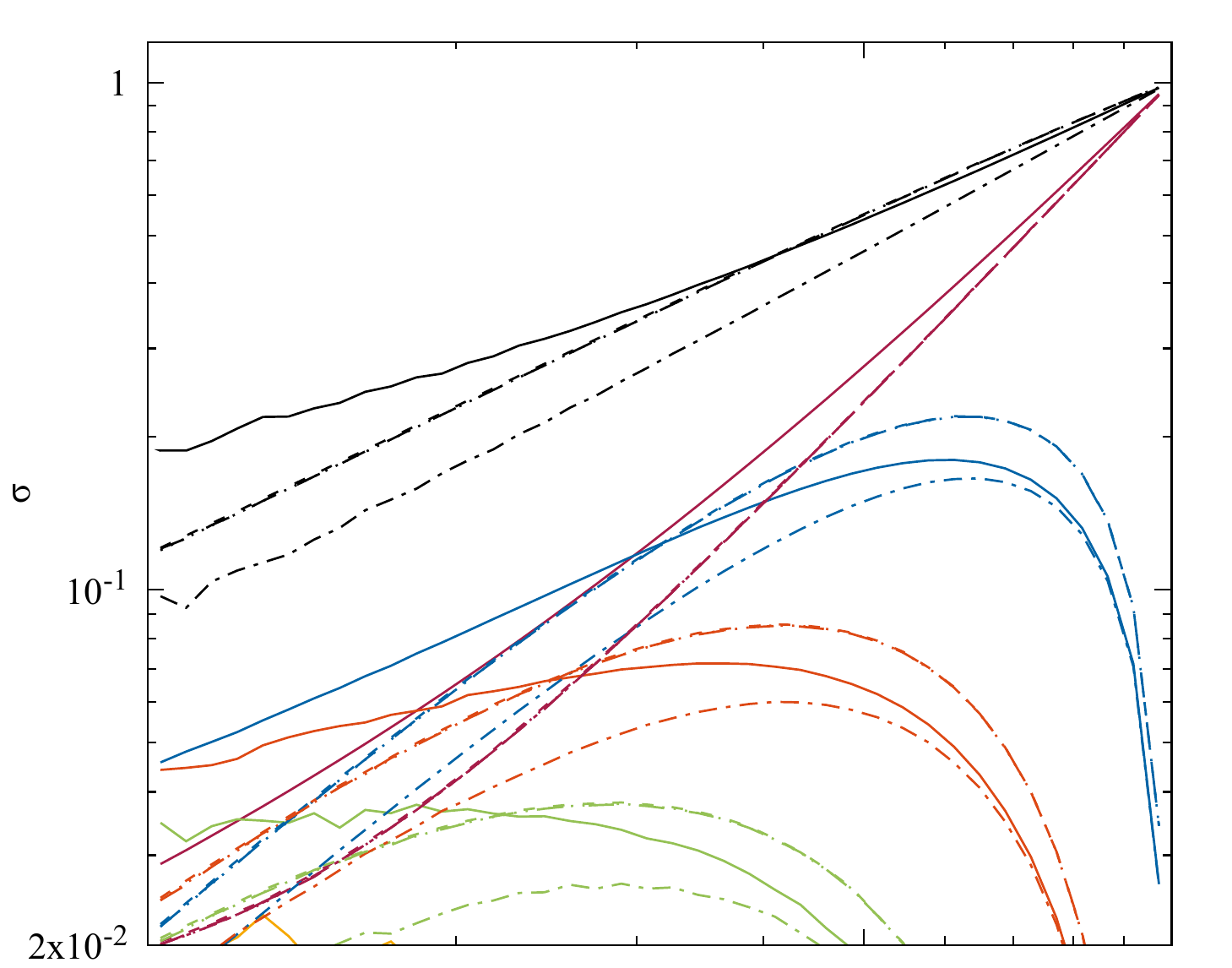}
\end{subfigure} \hfill
\begin{subfigure}[t]{0.9\textwidth}
\centering
\includegraphics[width=1.0\textwidth]{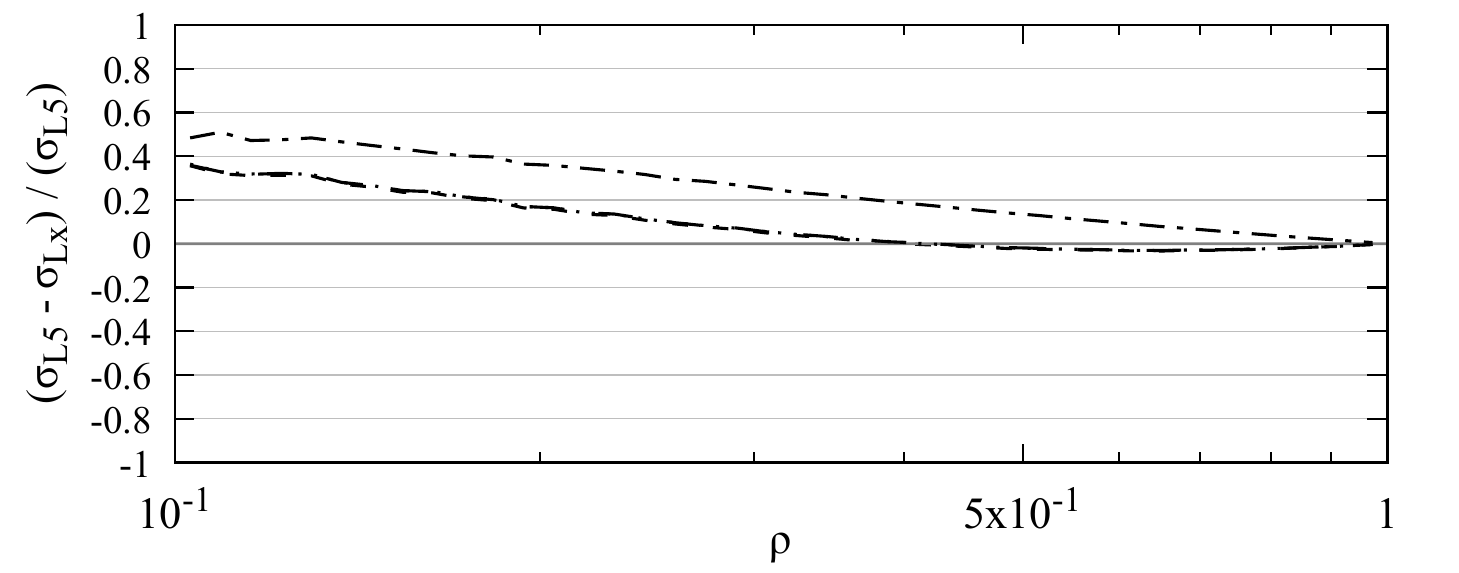}
\end{subfigure}
\caption{The veto cross-section for the $s$-channel quark exchange contribution to $qg \to qg$ in the asymmetric configuration. Solid: Full colour (L5), Dash-dotted: \LC\ + FCR (L4), Long-dashed: \LC\ + LCR + singlets (L3), Dotted: \LC\ + LCR (L2), Short-dashed: strict LC (L1). Evolution starts from the full-colour hard-scatter matrix.}
\label{fig:qg2qg-s-FCHa}
\end{figure}

\begin{figure}[t]
\centering
\begin{subfigure}[t]{0.9\textwidth}
\centering
\includegraphics[width=1.0\textwidth]{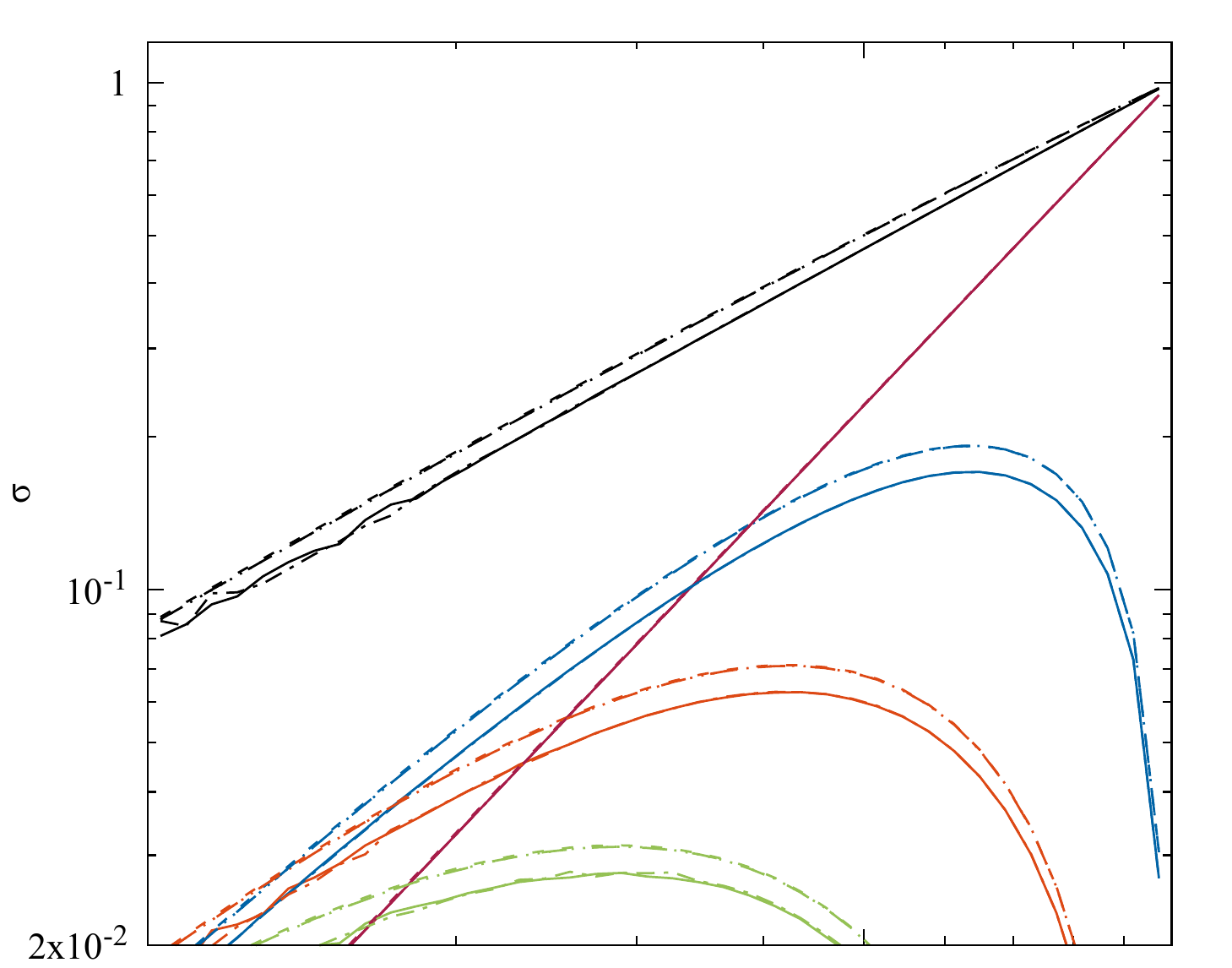}
\end{subfigure} \hfill
\begin{subfigure}[t]{0.9\textwidth}
\centering
\includegraphics[width=1.0\textwidth]{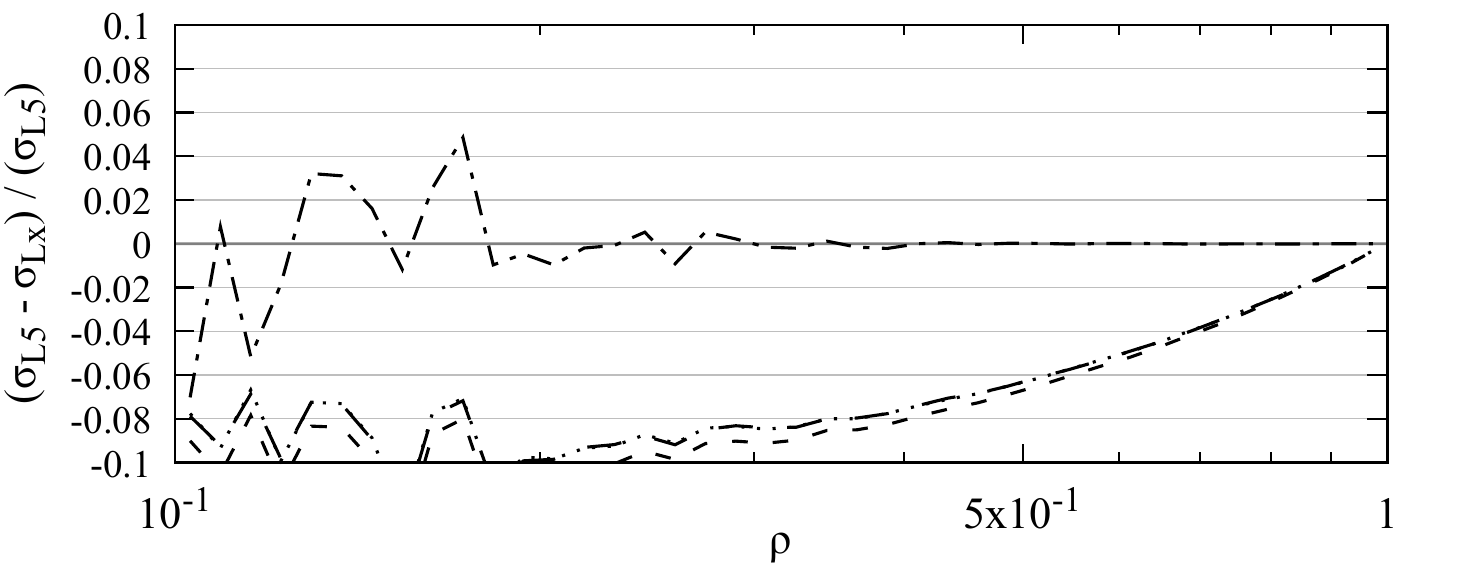}
\end{subfigure}
\caption{The veto cross-section for the $t$-channel gluon exchange contribution to $qg \to qg$ in the asymmetric configuration. Solid: Full colour (L5), Dash-dotted: \LC\ + FCR (L4), Long-dashed: \LC\ + LCR + singlets (L3), Dotted: \LC\ + LCR (L2), Short-dashed: strict LC (L1). Evolution starts from the full-colour hard-scatter matrix.}
\label{fig:qg2qg-t-FCHa}
\end{figure}

\begin{figure}[t]
\centering
\begin{subfigure}[t]{0.9\textwidth}
\centering
\includegraphics[width=1.0\textwidth]{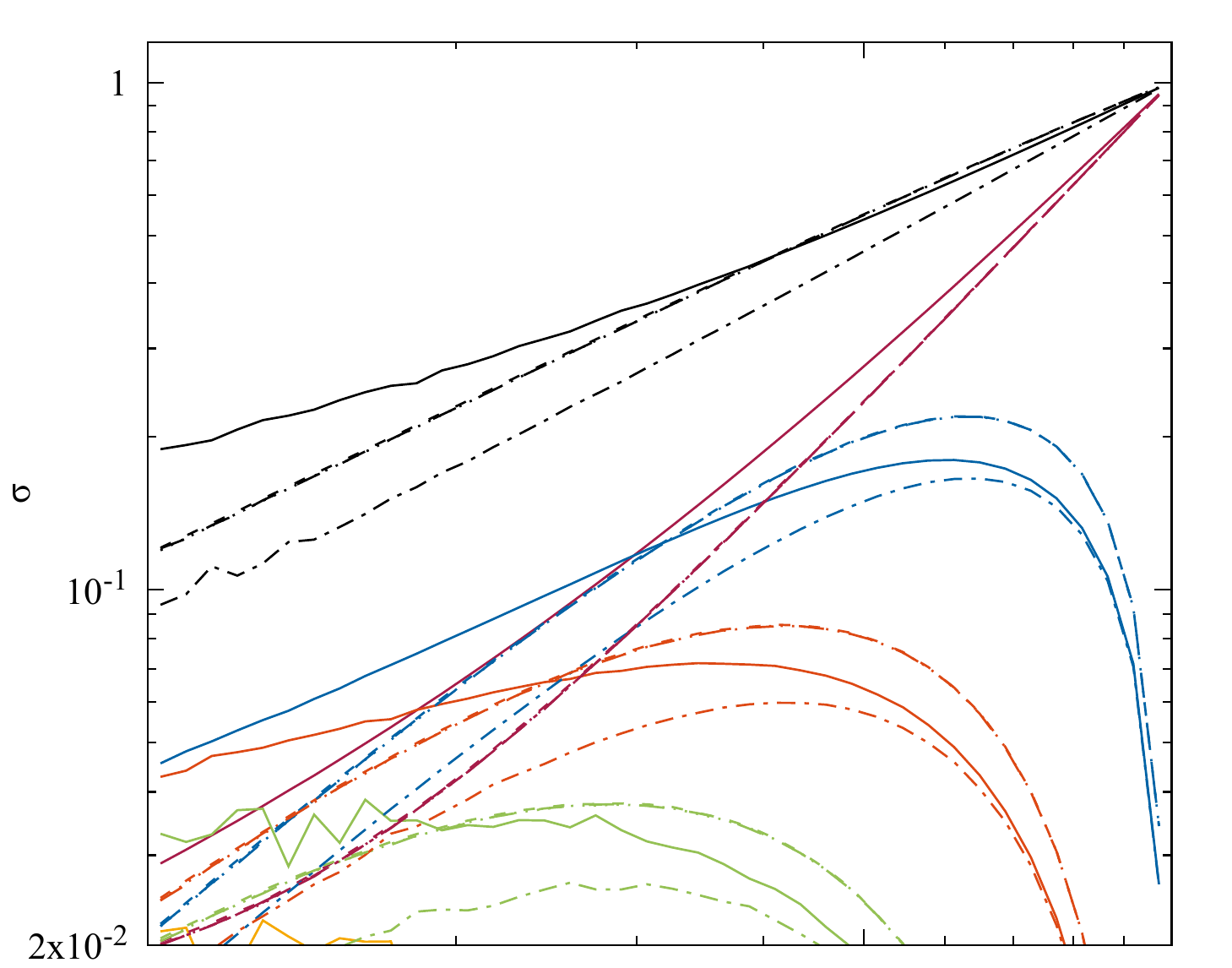}
\end{subfigure} \hfill
\begin{subfigure}[t]{0.9\textwidth}
\centering
\includegraphics[width=1.0\textwidth]{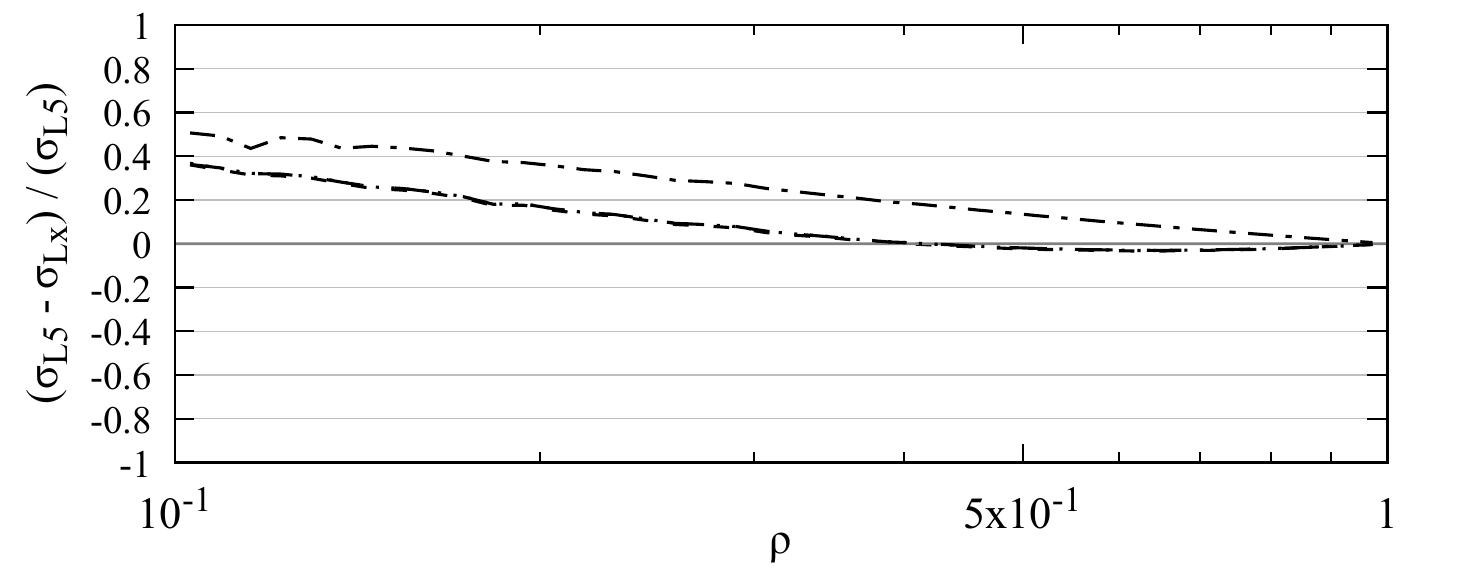}
\end{subfigure}
\caption{The veto cross-section for the $u$-channel quark exchange contribution to $qg \to qg$ in the asymmetric configuration. Solid: Full colour (L5), Dash-dotted: \LC\ + FCR (L4), Long-dashed: \LC\ + LCR + singlets (L3), Dotted: \LC\ + LCR (L2), Short-dashed: strict LC (L1). Evolution starts from the full-colour hard-scatter matrix.}
\label{fig:qg2qg-u-FCHa}
\end{figure}

\begin{figure}[t]
\centering
\begin{subfigure}[t]{0.9\textwidth}
\centering
\includegraphics[width=1.0\textwidth]{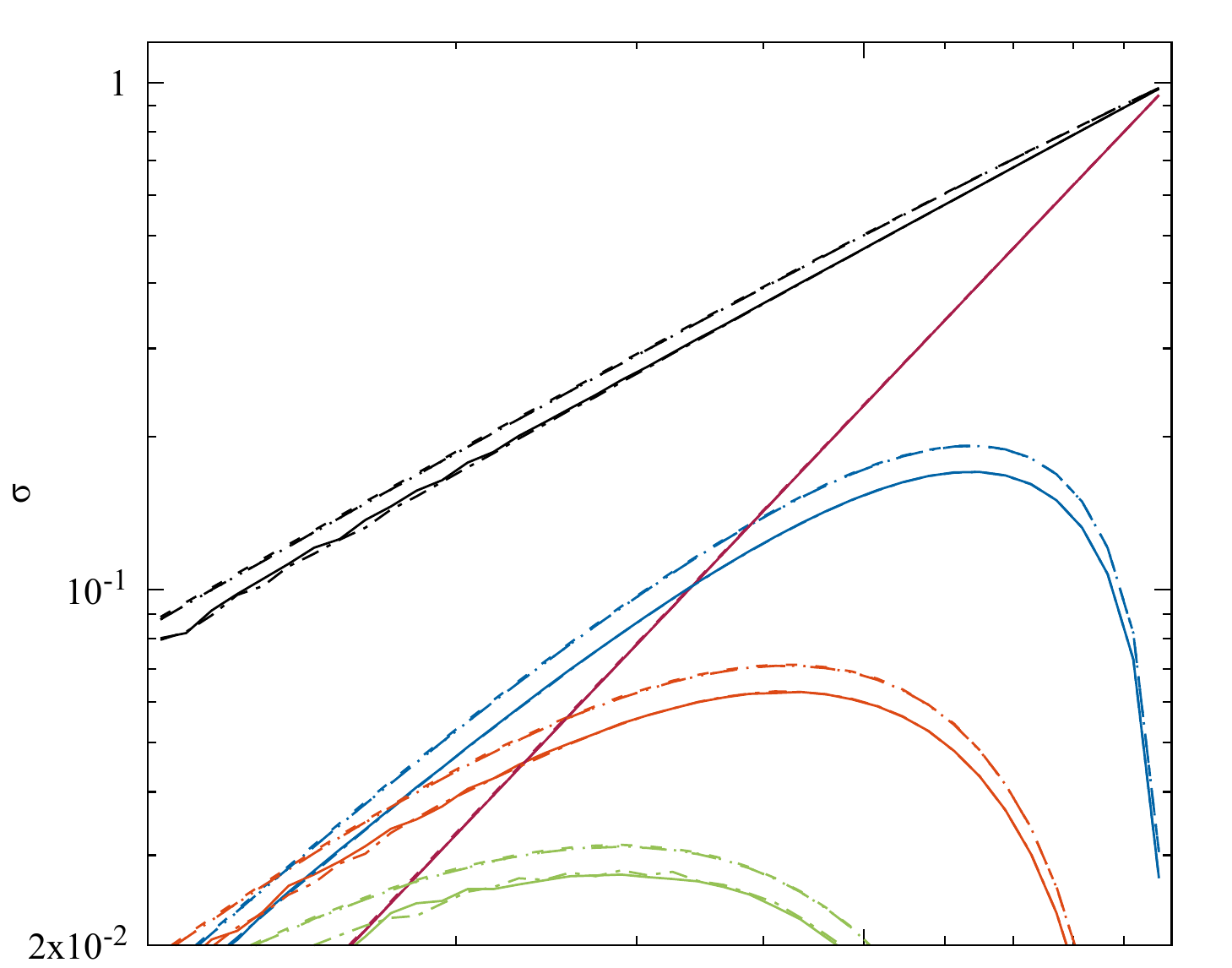}
\end{subfigure} \hfill
\begin{subfigure}[t]{0.9\textwidth}
\centering
\includegraphics[width=1.0\textwidth]{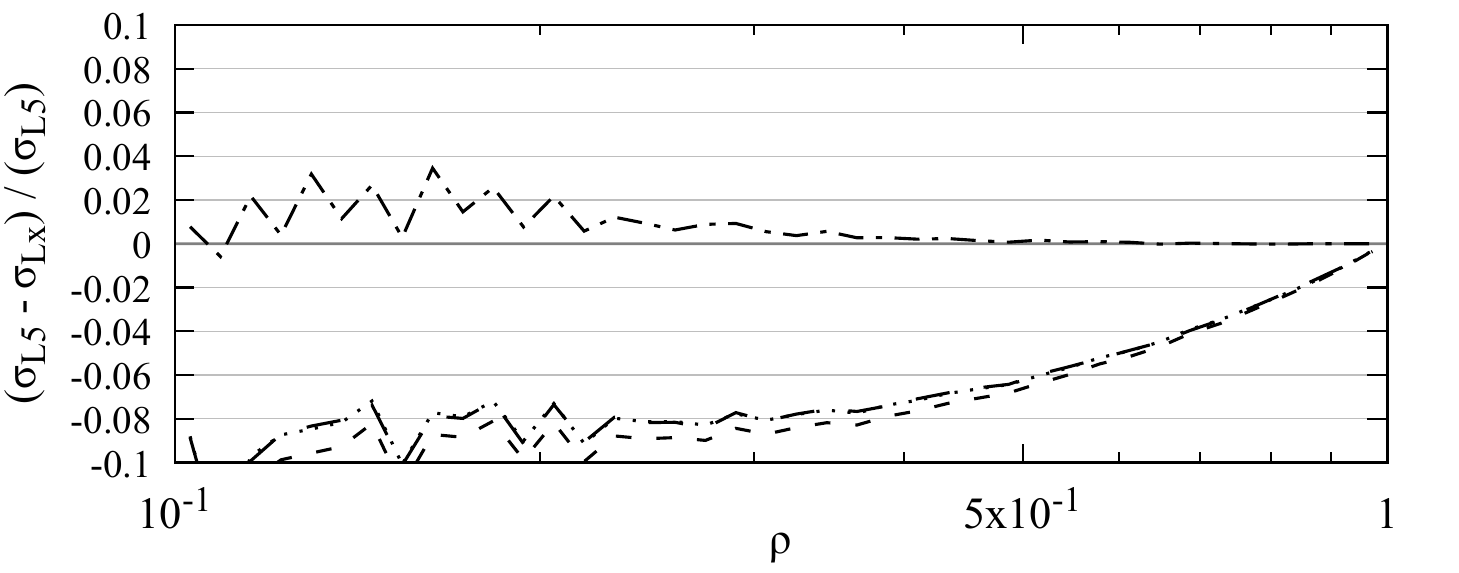}
\end{subfigure}
\caption{The veto cross-section for the $st$-channel interference contribution to $qg \to qg$ in the asymmetric configuration. Solid: Full colour (L5), Dash-dotted: \LC\ + FCR (L4), Long-dashed: \LC\ + LCR + singlets (L3), Dotted: \LC\ + LCR (L2), Short-dashed: strict LC (L1). Evolution starts from the full-colour hard-scatter matrix.}
\label{fig:qg2qg-st-FCHa}
\end{figure}

\begin{figure}[t]
\centering
\begin{subfigure}[t]{0.9\textwidth}
\centering
\includegraphics[width=1.0\textwidth]{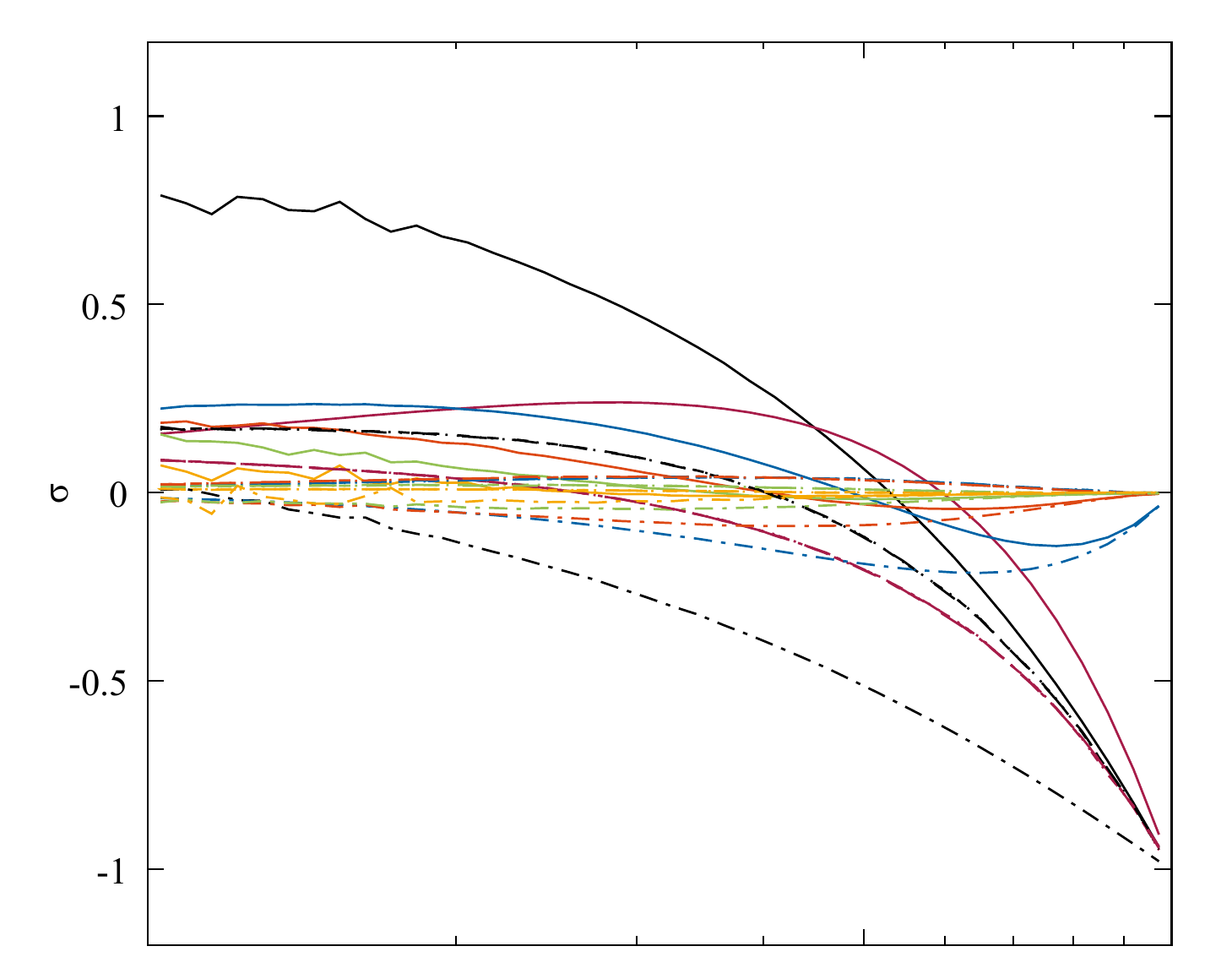}
\end{subfigure} \hfill
\begin{subfigure}[t]{0.9\textwidth}
\centering
\includegraphics[width=1.0\textwidth]{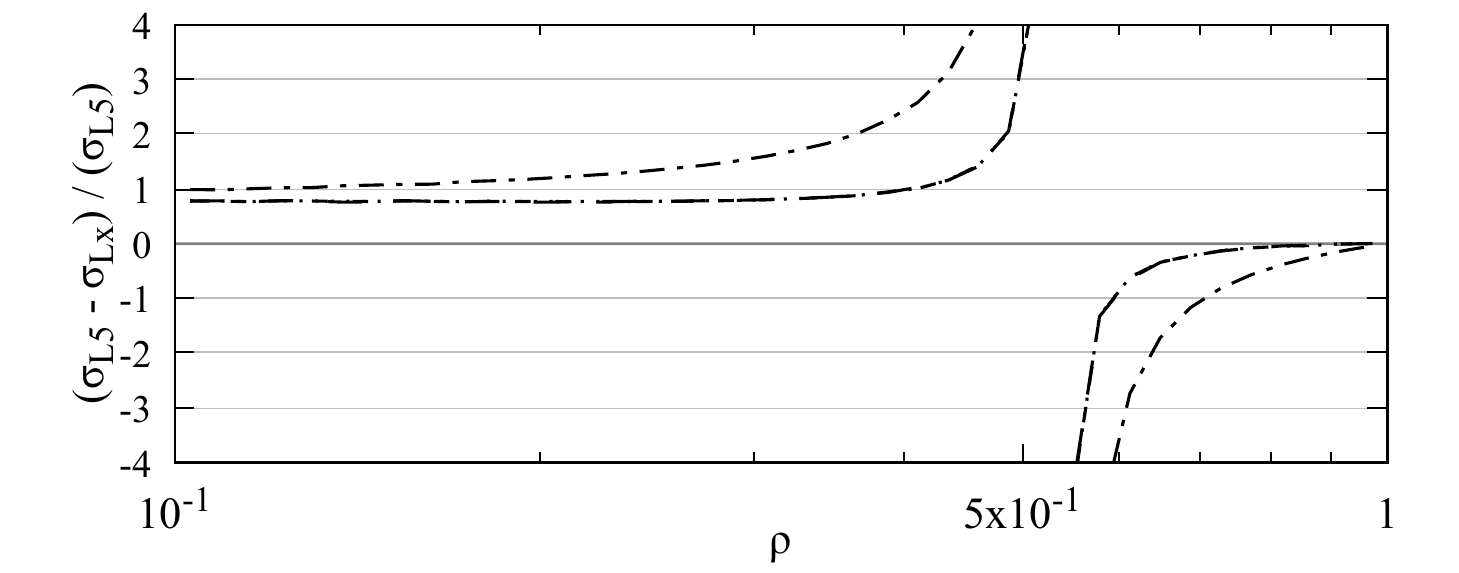}
\end{subfigure}
\caption{The veto cross-section for the $su$-channel interference contribution to $qg \to qg$ in the asymmetric configuration. Solid: Full colour (L5), Dash-dotted: \LC\ + FCR (L4), Long-dashed: \LC\ + LCR + singlets (L3), Dotted: \LC\ + LCR (L2), Short-dashed: strict LC (L1). Evolution starts from the full-colour hard-scatter matrix.}
\label{fig:qg2qg-su-FCHa}
\end{figure}

\begin{figure}[t]
\centering
\begin{subfigure}[t]{0.9\textwidth}
\centering
\includegraphics[width=1.0\textwidth]{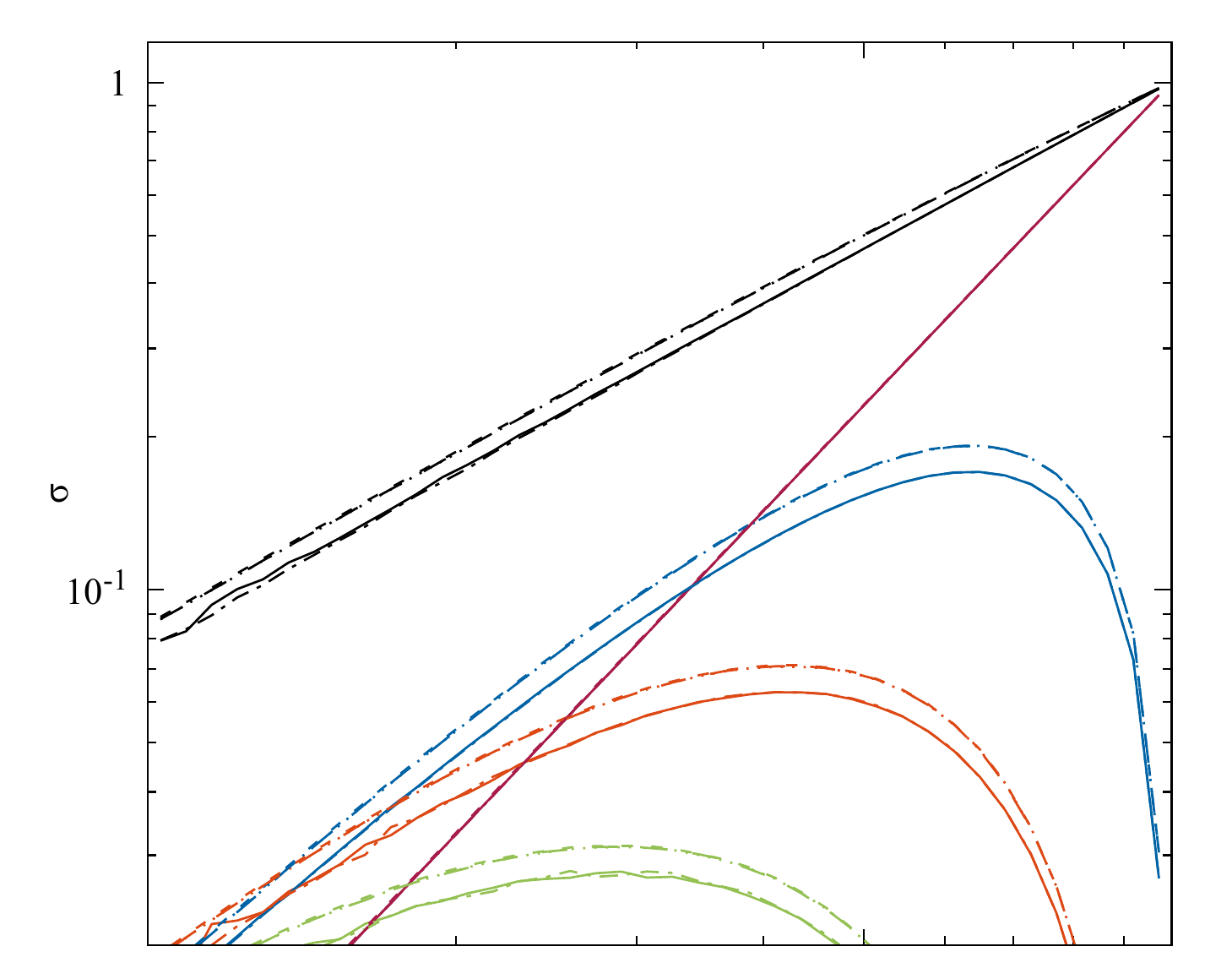}
\end{subfigure} \hfill
\begin{subfigure}[t]{0.9\textwidth}
\centering
\includegraphics[width=1.0\textwidth]{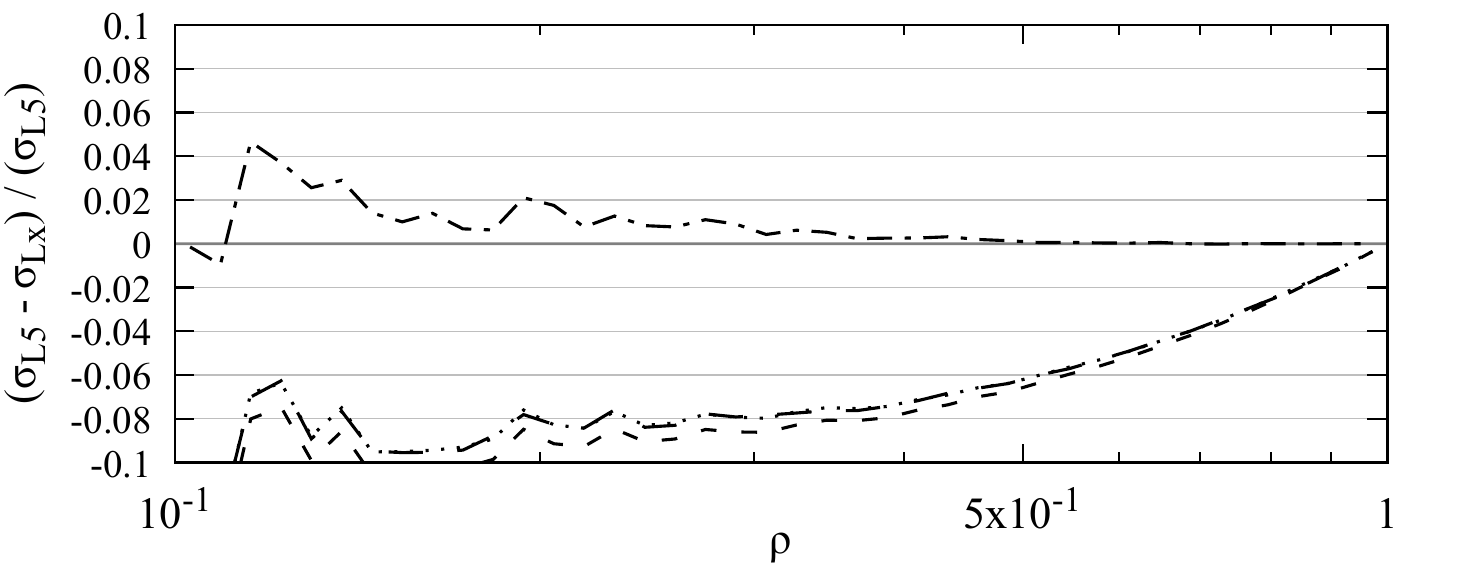}
\end{subfigure}
\caption{The veto cross-section for the $tu$-channel interference contribution to $qg \to qg$ in the asymmetric configuration. Solid: Full colour (L5), Dash-dotted: \LC\ + FCR (L4), Long-dashed: \LC\ + LCR + singlets (L3), Dotted: \LC\ + LCR (L2), Short-dashed: strict LC (L1). Evolution starts from the full-colour hard-scatter matrix.}
\label{fig:qg2qg-tu-FCHa}
\end{figure}

\begin{figure}[t]
\centering
\begin{subfigure}[t]{0.9\textwidth}
\centering
\includegraphics[width=1.0\textwidth]{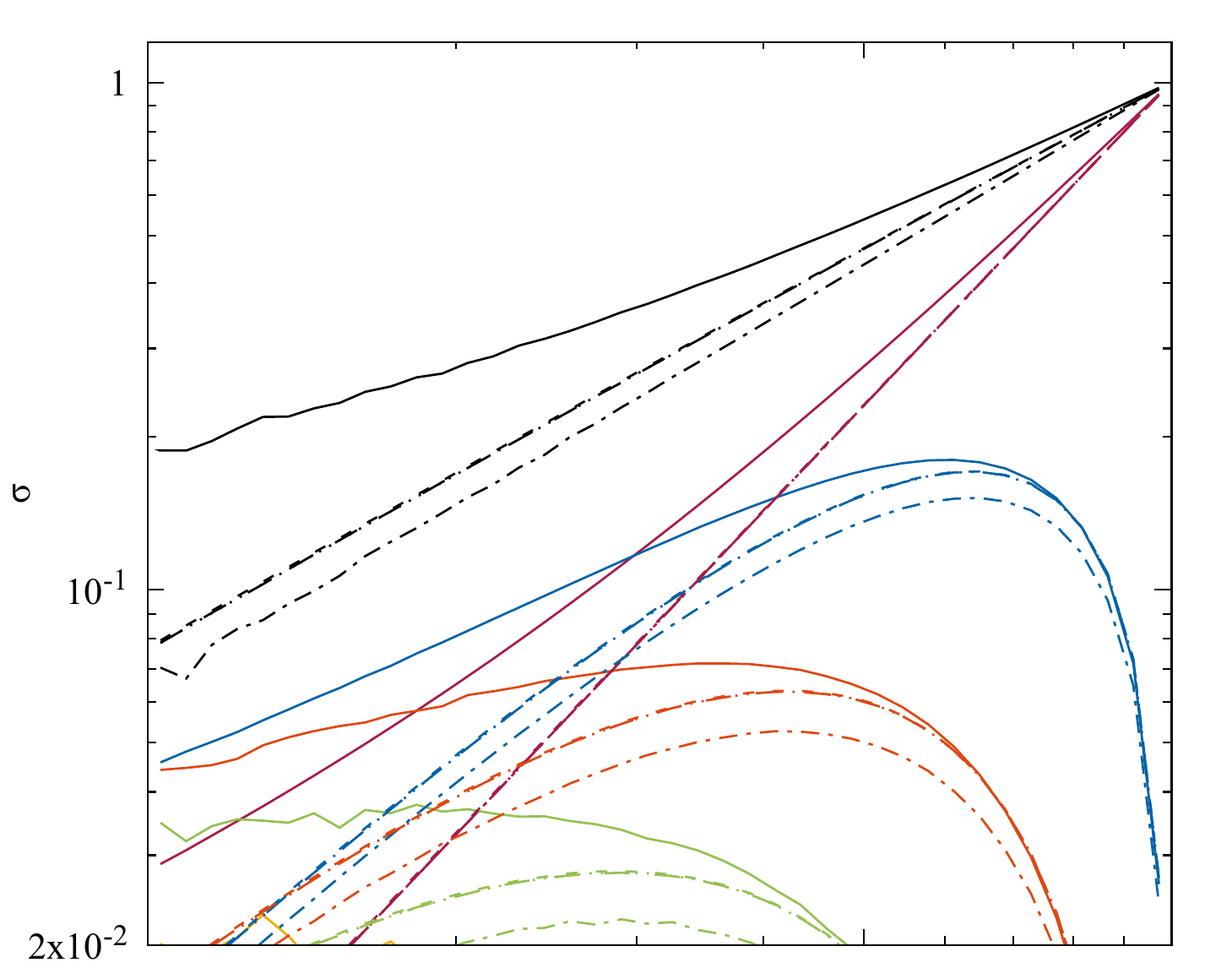}
\end{subfigure} \hfill
\begin{subfigure}[t]{0.9\textwidth}
\centering
\includegraphics[width=1.0\textwidth]{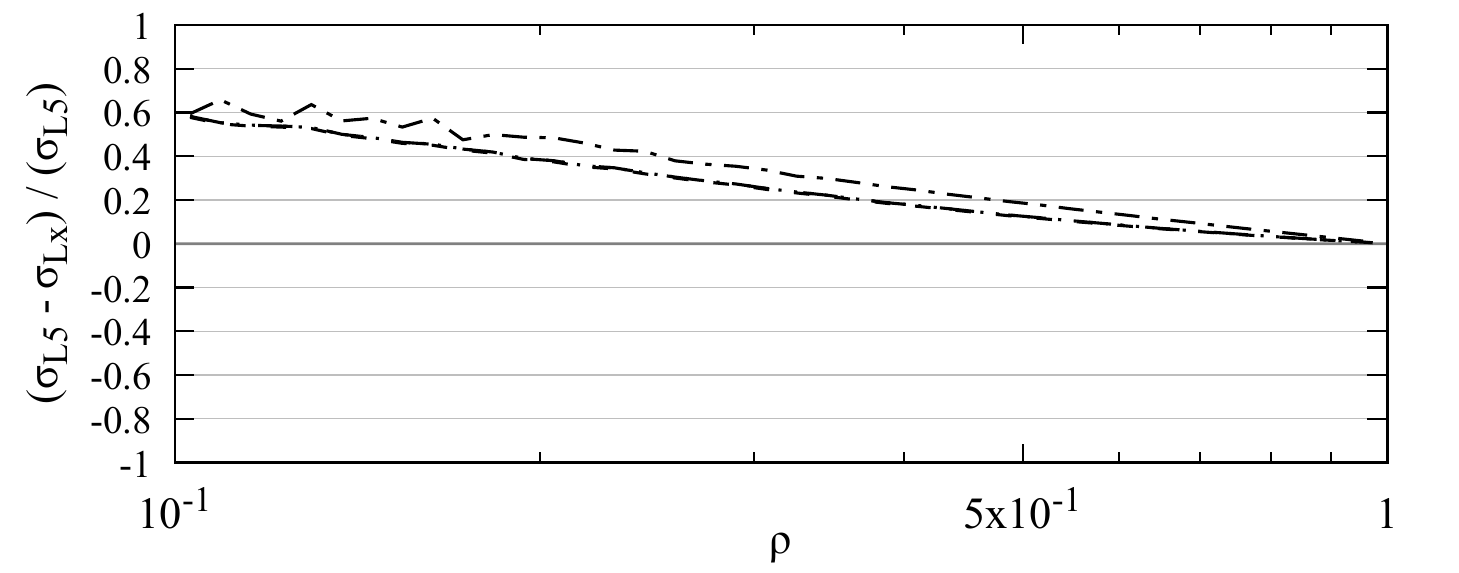}
\end{subfigure}
\caption{The veto cross-section for the $s$-channel quark exchange contribution to $qg \to qg$ in the asymmetric configuration. Solid: Full colour (L5), Dash-dotted: \LC\ + FCR (L4), Long-dashed: \LC\ + LCR + singlets (L3), Dotted: \LC\ + LCR (L2), Short-dashed: strict LC (L1). Evolution starts from the leading-colour hard-scatter matrix.}
\label{fig:qg2qg-s-LCHa}
\end{figure}

\begin{figure}[t]
\centering
\begin{subfigure}[t]{0.9\textwidth}
\centering
\includegraphics[width=1.0\textwidth]{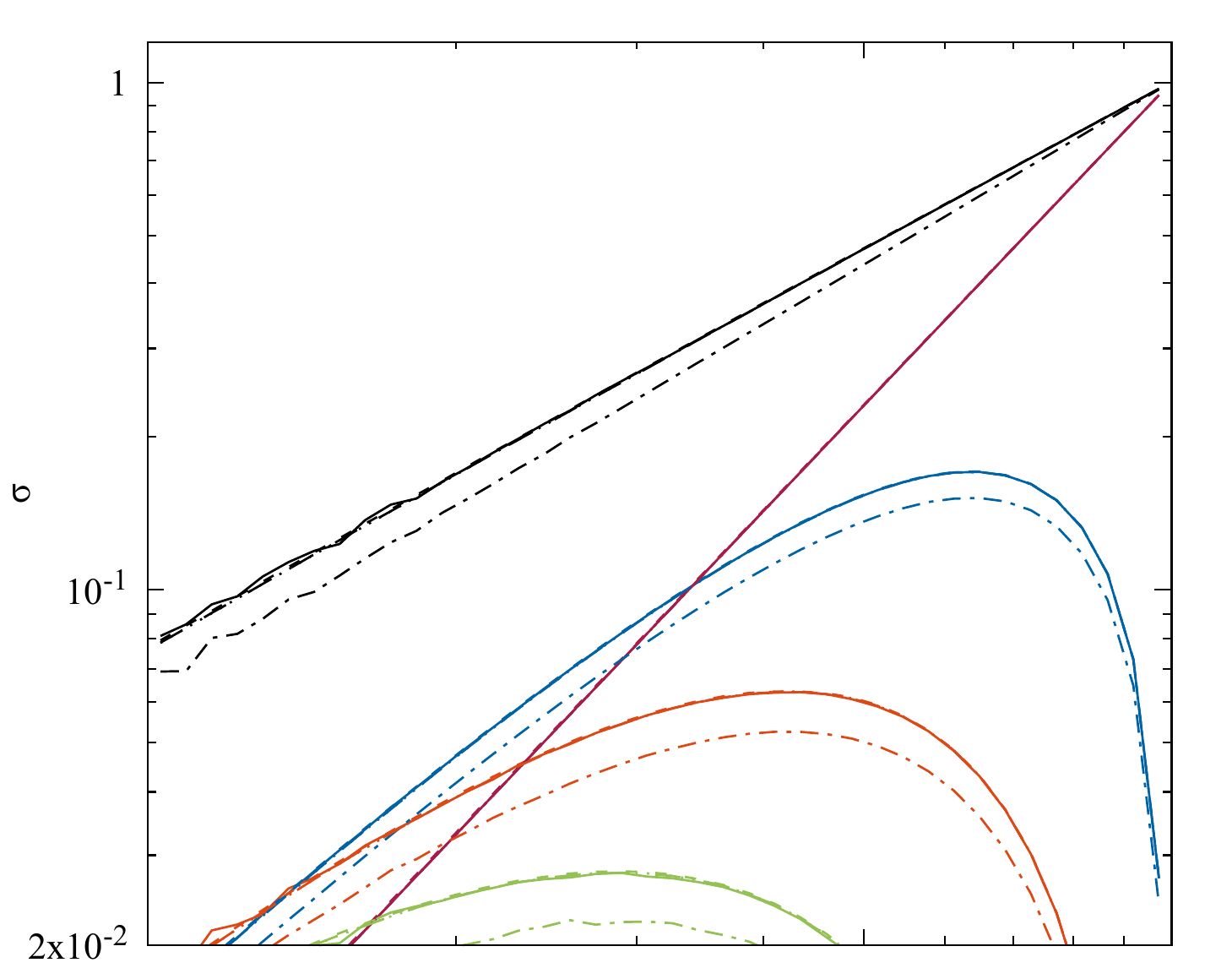}
\end{subfigure} \hfill
\begin{subfigure}[t]{0.9\textwidth}
\centering
\includegraphics[width=1.0\textwidth]{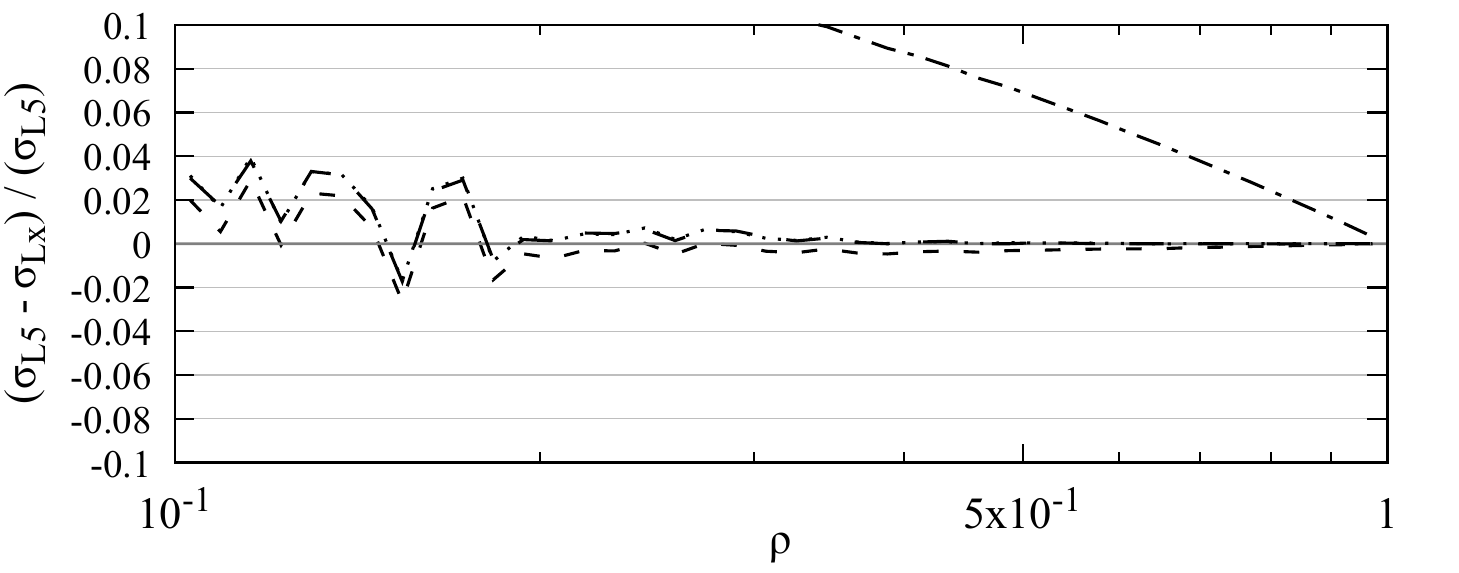}
\end{subfigure}
\caption{The veto cross-section for the $t$-channel gluon exchange contribution to $qg \to qg$ in the asymmetric configuration. Solid: Full colour (L5), Dash-dotted: \LC\ + FCR (L4), Long-dashed: \LC\ + LCR + singlets (L3), Dotted: \LC\ + LCR (L2), Short-dashed: strict LC (L1). Evolution starts from the leading-colour hard-scatter matrix.}
\label{fig:qg2qg-t-LCHa}
\end{figure}

\begin{figure}[t]
\centering
\begin{subfigure}[t]{0.9\textwidth}
\centering
\includegraphics[width=1.0\textwidth]{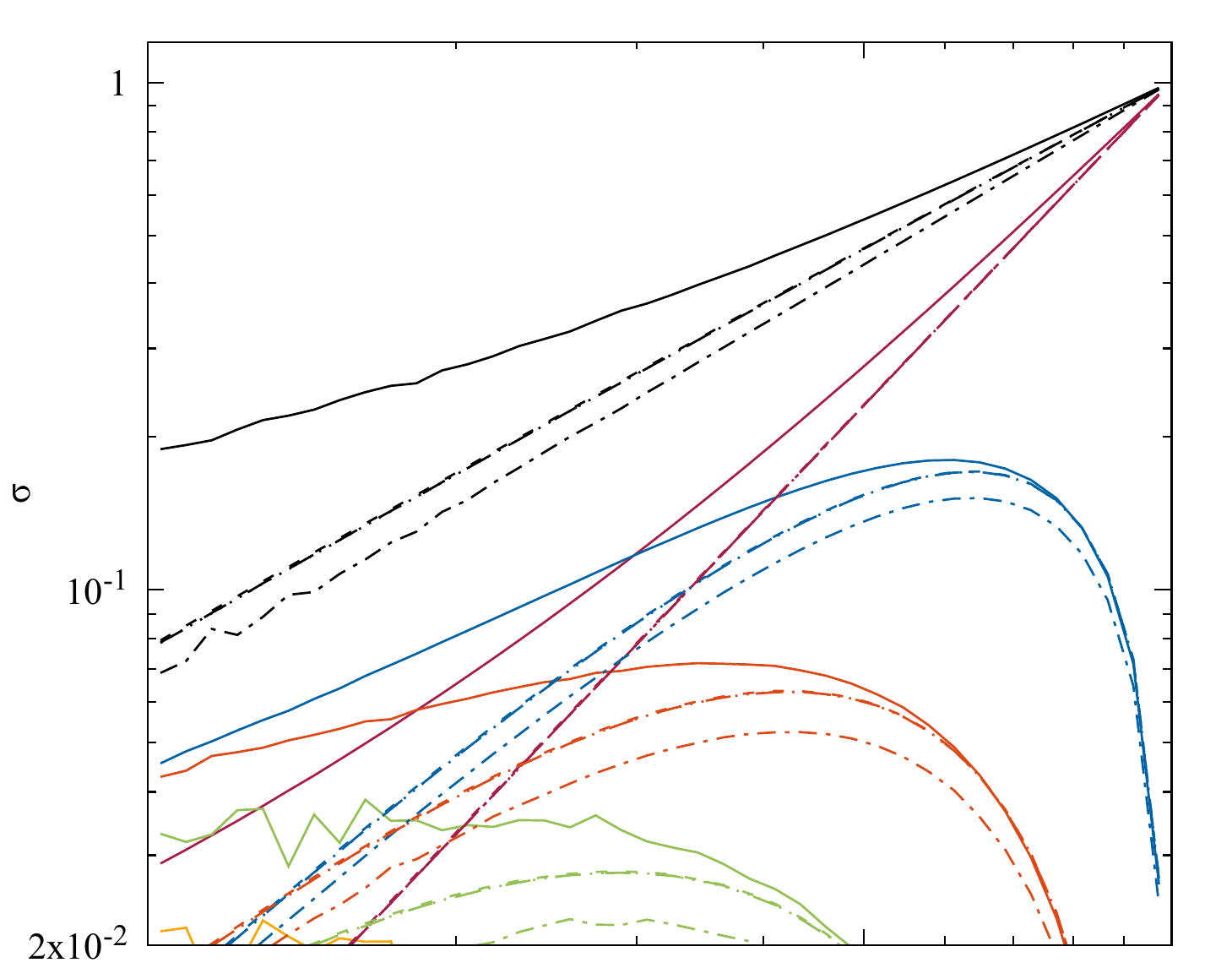}
\end{subfigure} \hfill
\begin{subfigure}[t]{0.9\textwidth}
\centering
\includegraphics[width=1.0\textwidth]{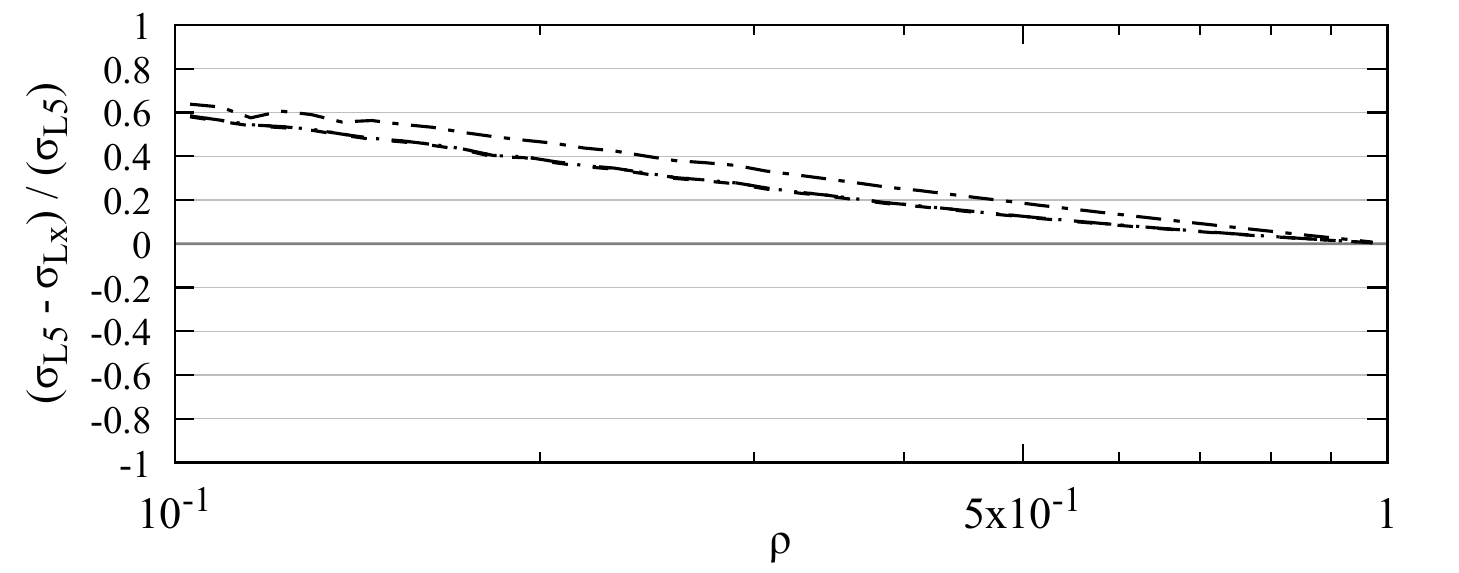}
\end{subfigure}
\caption{The veto cross-section for the $u$-channel quark exchange contribution to $qg \to qg$ in the asymmetric configuration. Solid: Full colour (L5), Dash-dotted: \LC\ + FCR (L4), Long-dashed: \LC\ + LCR + singlets (L3), Dotted: \LC\ + LCR (L2), Short-dashed: strict LC (L1). Evolution starts from the leading-colour hard-scatter matrix.}
\label{fig:qg2qg-u-LCHa}
\end{figure}

\begin{figure}[t]
\centering
\begin{subfigure}[t]{0.9\textwidth}
\centering
\includegraphics[width=1.0\textwidth]{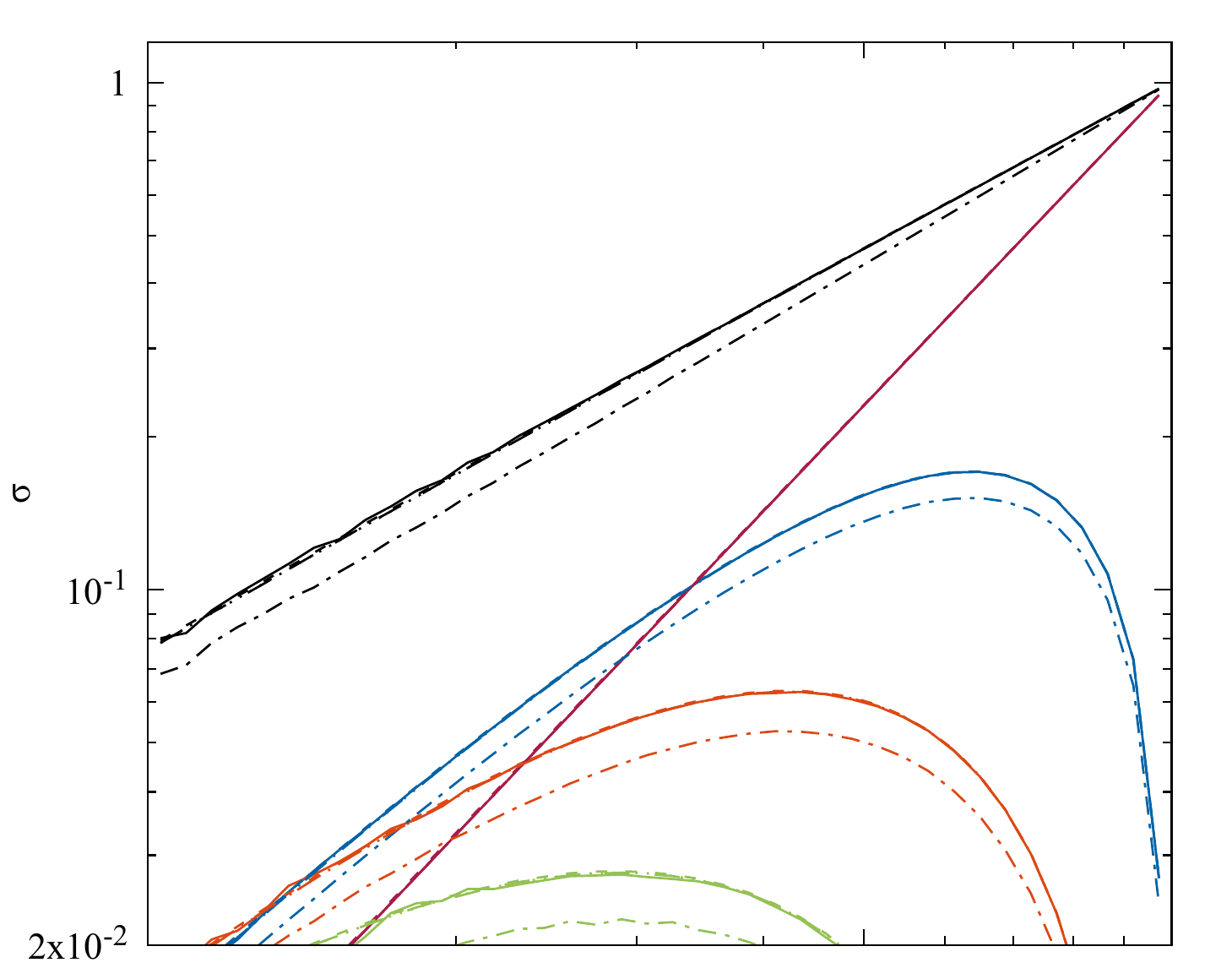}
\end{subfigure} \hfill
\begin{subfigure}[t]{0.9\textwidth}
\centering
\includegraphics[width=1.0\textwidth]{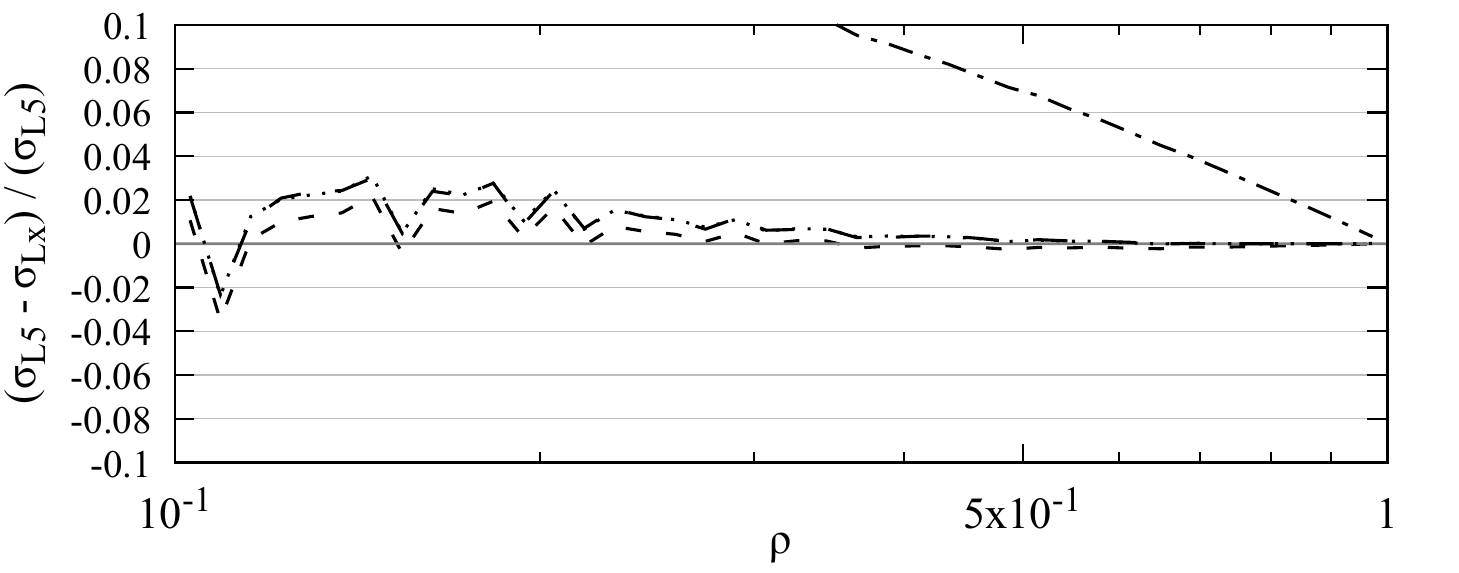}
\end{subfigure}
\caption{The veto cross-section for the $st$-channel interference contribution to $qg \to qg$ in the asymmetric configuration. Solid: Full colour (L5), Dash-dotted: \LC\ + FCR (L4), Long-dashed: \LC\ + LCR + singlets (L3), Dotted: \LC\ + LCR (L2), Short-dashed: strict LC (L1). Evolution starts from the leading-colour hard-scatter matrix.}
\label{fig:qg2qg-st-LCHa}
\end{figure}

\begin{figure}[t]
\centering
\begin{subfigure}[t]{0.9\textwidth}
\centering
\includegraphics[width=1.0\textwidth]{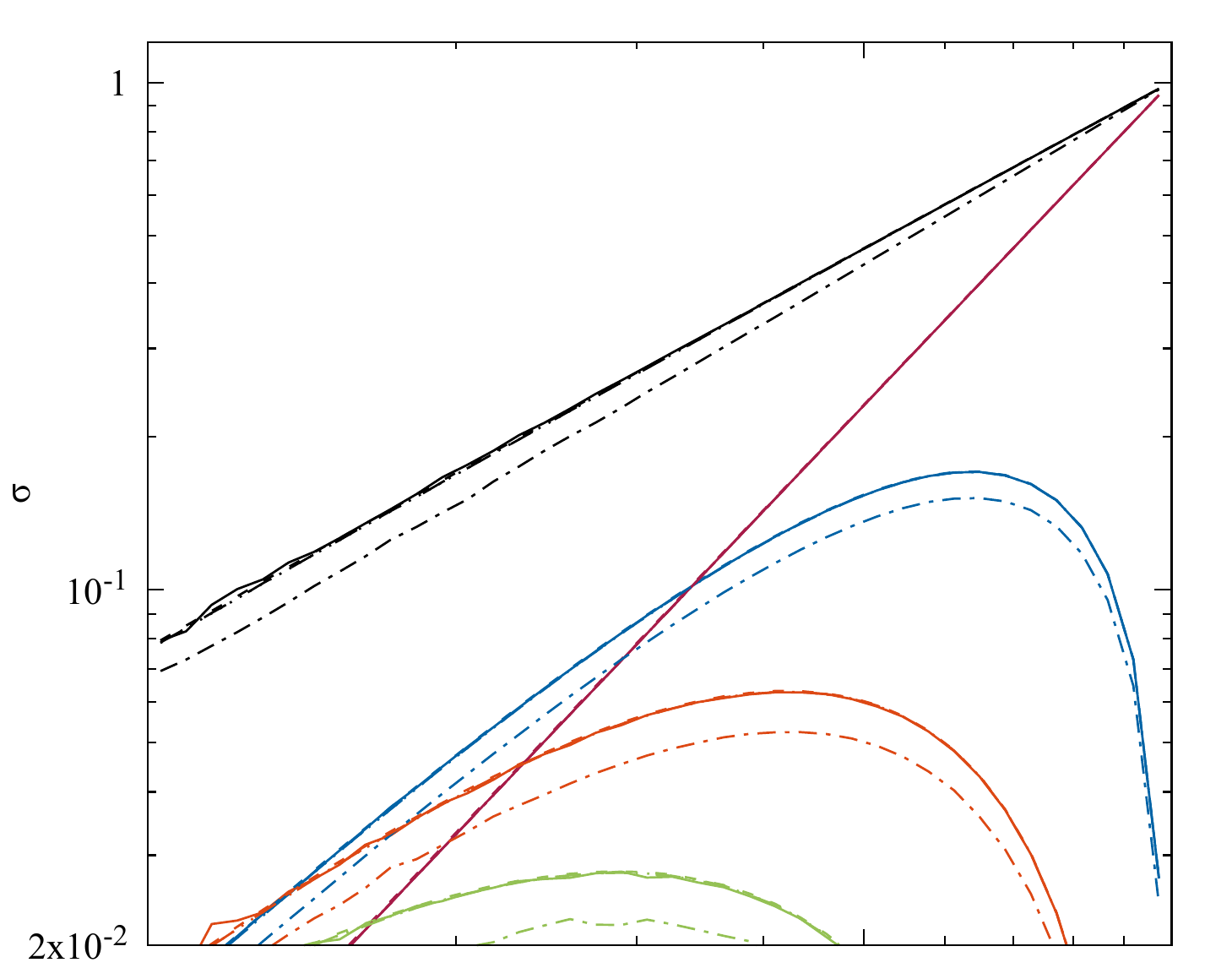}
\end{subfigure} \hfill
\begin{subfigure}[t]{0.9\textwidth}
\centering
\includegraphics[width=1.0\textwidth]{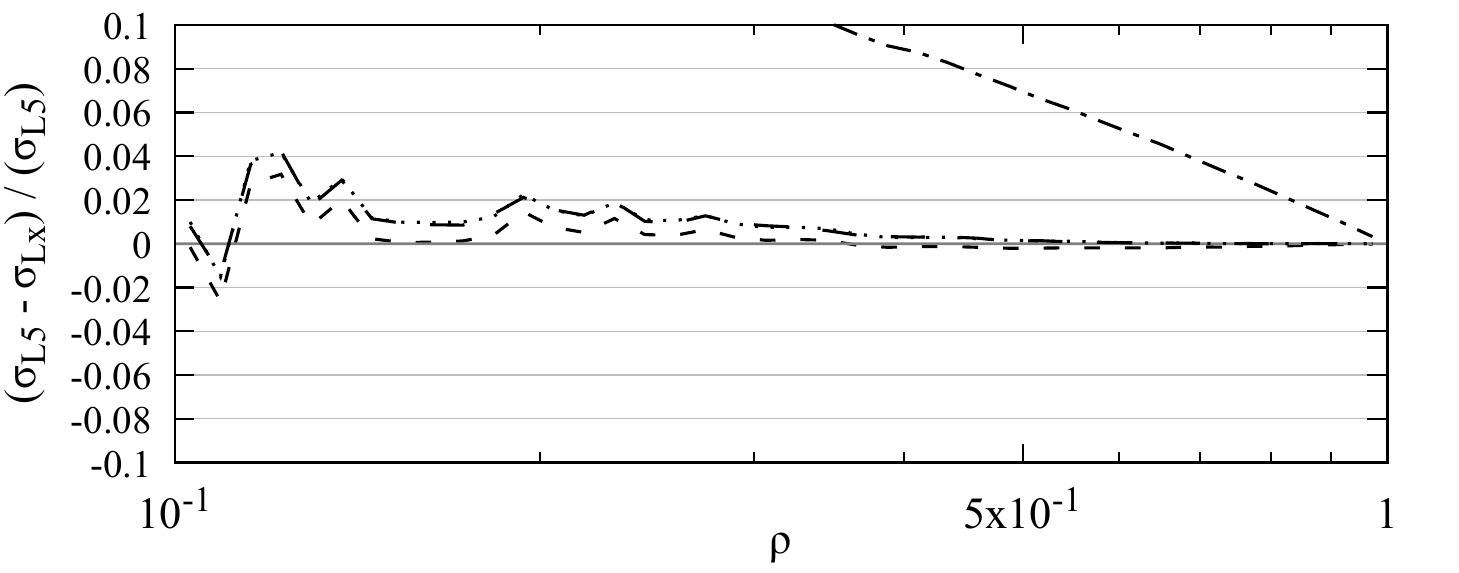}
\end{subfigure}
\caption{The veto cross-section for the $tu$-channel interference contribution to $qg \to qg$ in the asymmetric configuration. Solid: Full colour (L5), Dash-dotted: \LC\ + FCR (L4), Long-dashed: \LC\ + LCR + singlets (L3), Dotted: \LC\ + LCR (L2), Short-dashed: strict LC (L1). Evolution starts from the leading-colour hard-scatter matrix.}
\label{fig:qg2qg-tu-LCHa}
\end{figure}

Now we turn our attention to quark-gluon scattering. There are 3 diagrams to consider, as illustrated in Fig.~\ref{fig:qg2qgFeynman} and the corresponding hard scatter matrices are now rank 6. In total there are 6 distinct contributions corresponding to the $s$, $t$ and $u$-channel contributions plus the three ($st$, $su$ and $tu$) interference contributions. In what follows we will be systematic and explore the contributions to the veto cross-section for each of these. We will do this starting from full-colour in the hard process (FCH) and also leading-colour in the hard processes (LCH). And also for the symmetric and asymmetric kinematic configurations that we explored in the previous section. In total this means we will present 24 plots in Fig.~\ref{fig:qg2qg-s-FCHs} to Fig.~\ref{fig:qg2qg-tu-LCHa}. Table~\ref{tab:qg2qg} summarizes the key features of all 24 plots.

\begin{table}[t]
    \centering
    \begin{tabular}{|c|cc|cc|}
        \hline
        \multirow{2}{*}{Channel} & \multicolumn{2}{c|}{FCH} & \multicolumn{2}{c|}{LCH} \\
        \cline{2-5}
        & symmetric & asymmetric & symmetric & asymmetric \\
        \hline
        $s$ & none & none & none & none \\
        \arrayrulecolor{gray!70} \hline \arrayrulecolor{black}
        $t$ & $\approx$L4 & =L4 & $=$L1 \cite{Hatta:2020wre} & $=$L1, L2, L3 \\
        \arrayrulecolor{gray!70} \hline \arrayrulecolor{black}
        $u$ & none & none & none & none \\
        \arrayrulecolor{gray!70} \hline \arrayrulecolor{black}
        $st$ & $\approx$L4 & $\approx$L4 & $\approx$L1 & $\approx$L1, L2, L3 \\
        \arrayrulecolor{gray!70} \hline \arrayrulecolor{black}
        $su$ & none & none & N/A & N/A \\
        \arrayrulecolor{gray!70} \hline \arrayrulecolor{black}
        $tu$ & $\approx$L4 & $\approx$L4 & $\approx$L1, L3 & $\approx$L1, L2, L3 \\
        \hline
    \end{tabular}
    \caption{Summary of results for $qg \to qg$. In all cases, the comparison is with the full-colour evolution starting from the full-colour hard scatter matrix modulo an overall normalization.}
    \label{tab:qg2qg}
\end{table}

In the asymmetric configuration, we find that L1 $\approx$ L2 $\approx$ L3. This is because in this configuration the quark-quark pair is the closest in angle, suppressing that dipole relative to the others, and reducing the impact of singlet exchanges and emissions.

For the symmetric kinematic configuration, starting from full-colour in the hard scatter matrix none of the approximations L1--L3 is satisfactory whilst L4 is within 2\% for all contributions except for the $su$-channel interference where sub-leading colour effects are enormous.

Starting from the LCH, L1 does very well at approximating the L5 result for the $t$-, $st$-, and $tu$- channels. We also see L4 is within 2\% of the full colour result for the same channels starting from the FCH. For the other channels, none  of the approximations work. In the symmetric configuration, the approximation L2 is never satisfactory, while in the asymmetric configuration the approximations L1--L3 always perform similarly for the reasons explained before. We should also note that, for the channels where subleading colour effects cannot be described without L5 evolution, we see substantial subleading colour corrections: of order $10$--$20\%$ for the symmetric case and $30$--$50\%$ for the asymmetric case at $\rho = 0.1$, when compared to the best performing approximation, for the $s$- and $u$- channels. The $su$- interference has enormous effects and needs full colour evolution.

\subsection{\boldmath$gg \to gg$}
\label{sec:gg2gg}

\begin{figure}[t]
\centering
\begin{subfigure}[t]{0.9\textwidth}
\centering
\includegraphics[width=1.0\textwidth]{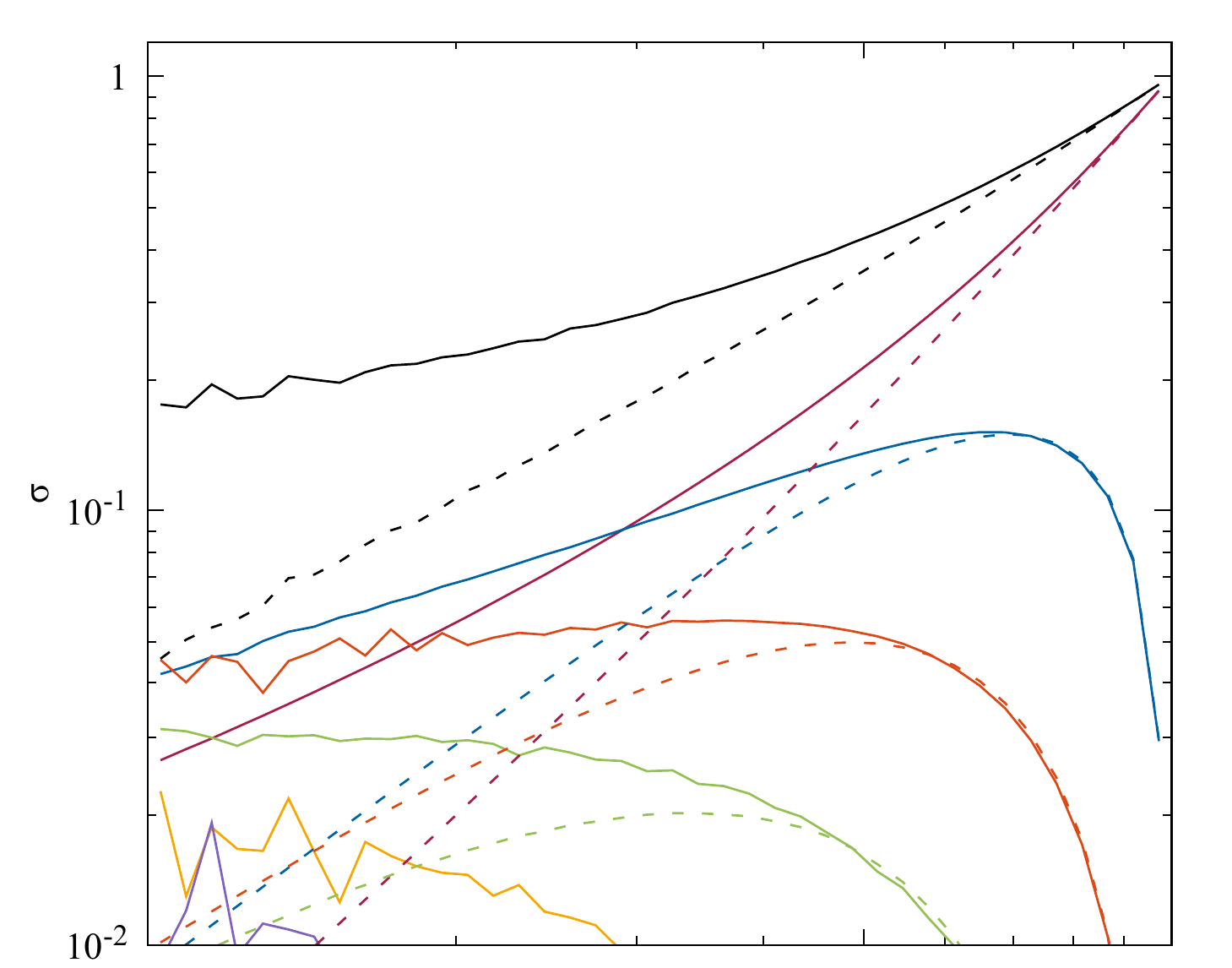}
\end{subfigure} \hfill
\begin{subfigure}[t]{0.9\textwidth}
\centering
\includegraphics[width=1.0\textwidth]{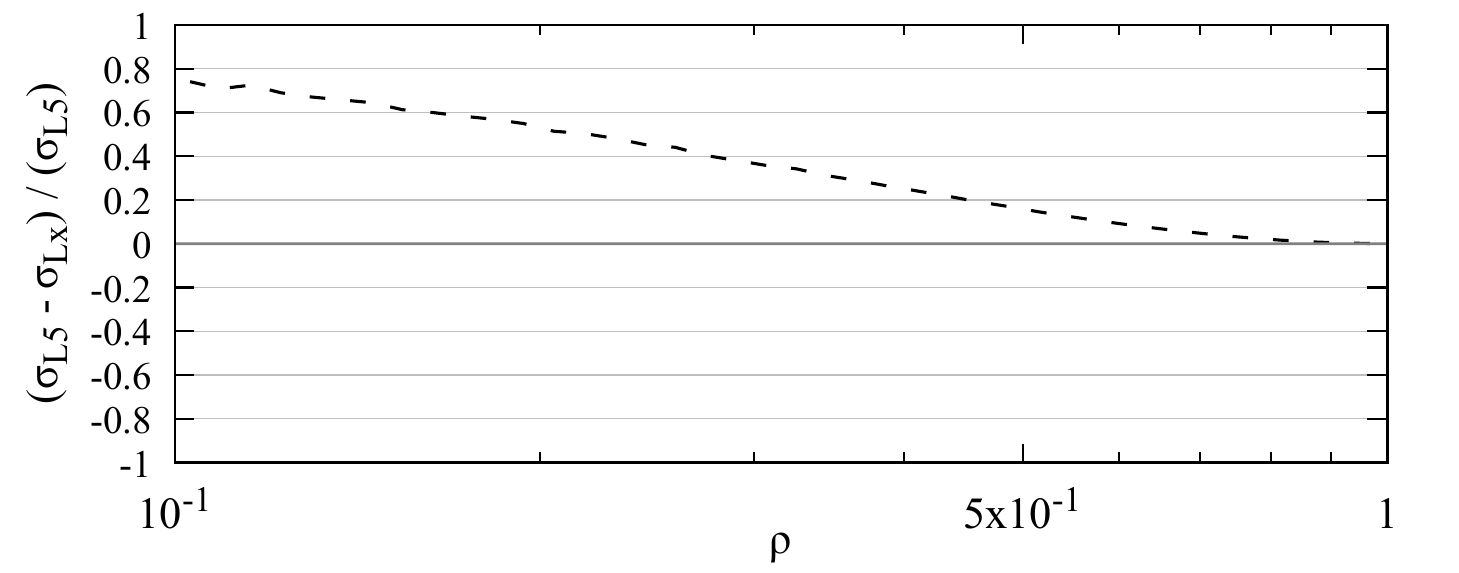}
\end{subfigure}
\caption{The veto cross-section for the $s$-channel contribution to $gg \to gg$. Solid: Full colour, Short-dashed: Leading colour.}
\label{fig:gg2gg-s}
\end{figure}

\begin{figure}[t]
\centering
\begin{subfigure}[t]{0.9\textwidth}
\centering
\includegraphics[width=1.0\textwidth]{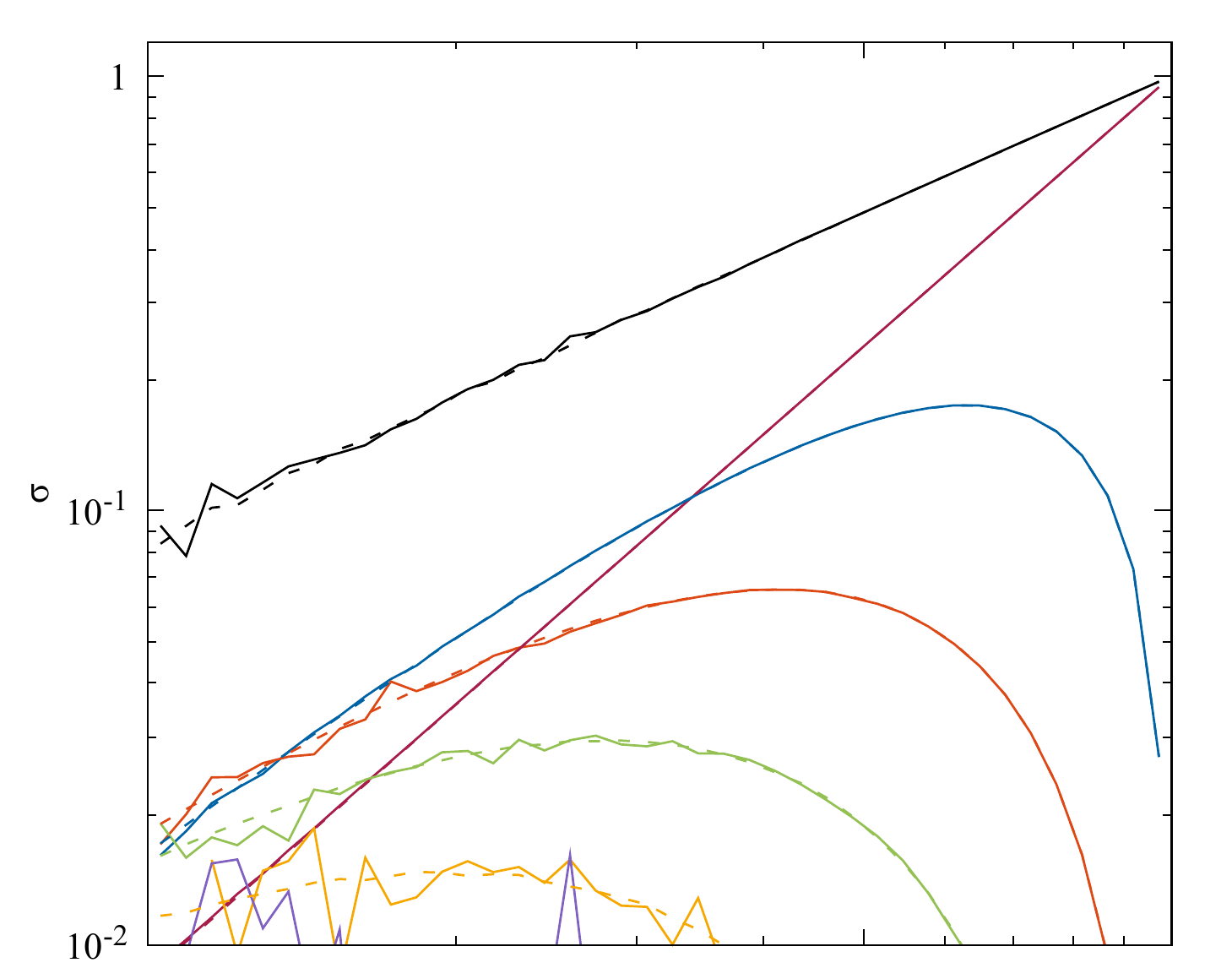}
\end{subfigure} \hfill
\begin{subfigure}[t]{0.9\textwidth}
\centering
\includegraphics[width=1.0\textwidth]{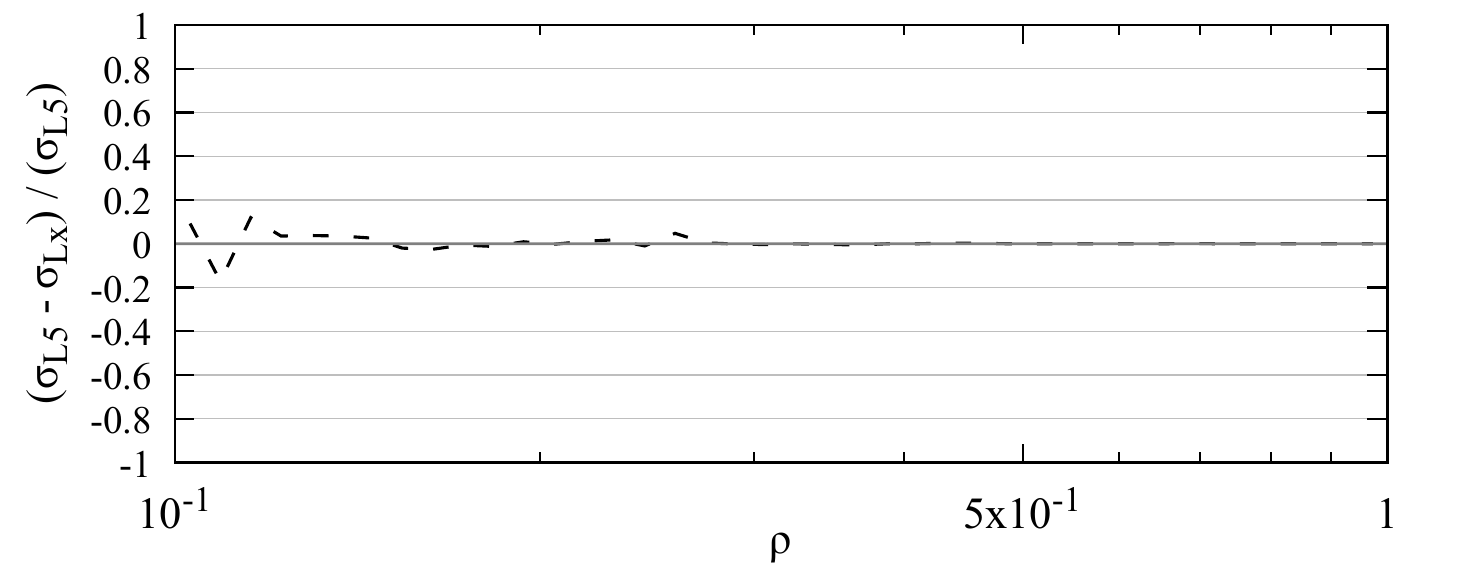}
\end{subfigure}
\caption{The veto cross-section for the $t$-channel contribution to $gg \to gg$. Solid: Full colour, Short-dashed: Leading colour.}
\label{fig:gg2gg-t}
\end{figure}

\begin{figure}[t]
\centering
\begin{subfigure}[t]{0.9\textwidth}
\centering
\includegraphics[width=1.0\textwidth]{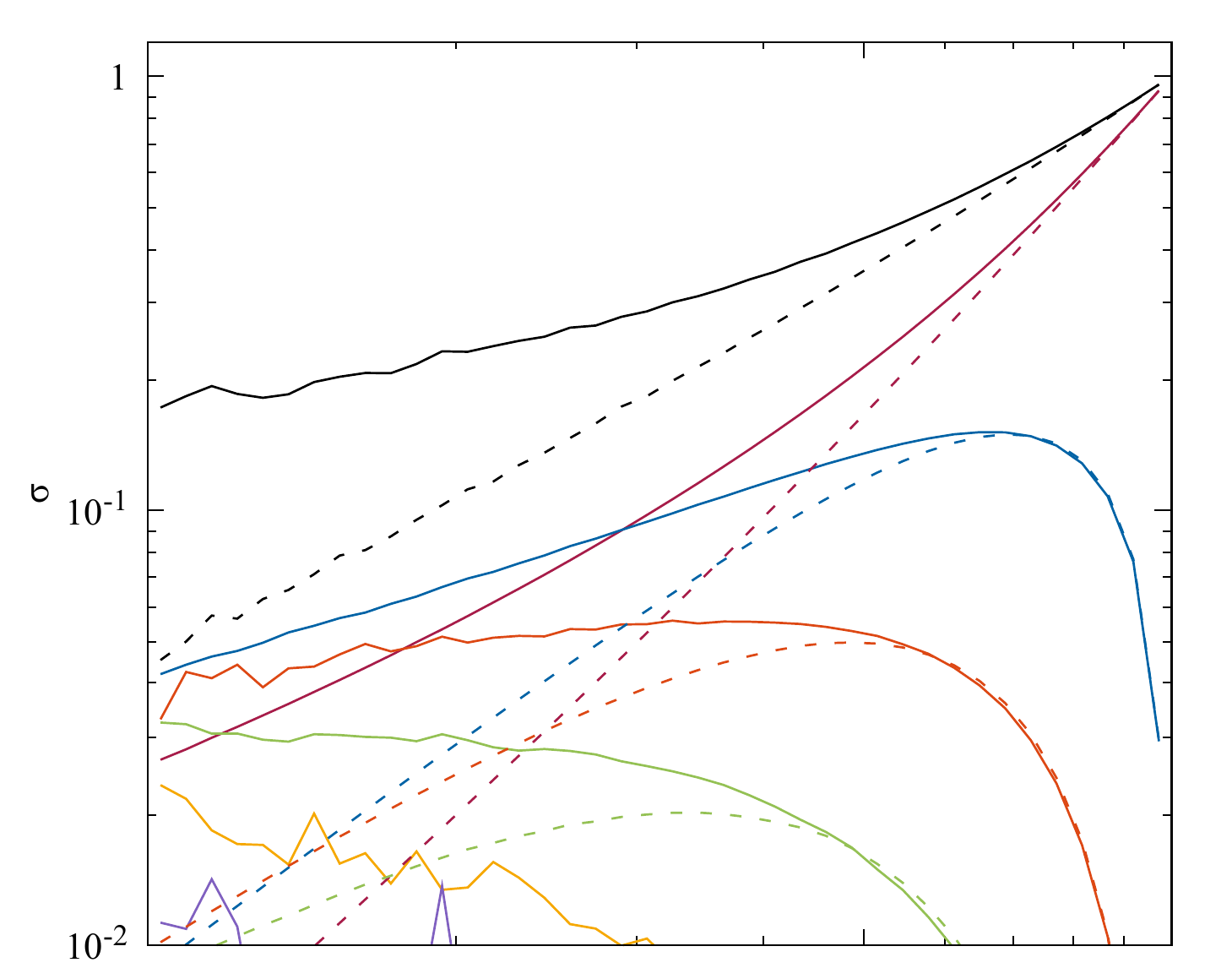}
\end{subfigure} \hfill
\begin{subfigure}[t]{0.9\textwidth}
\centering
\includegraphics[width=1.0\textwidth]{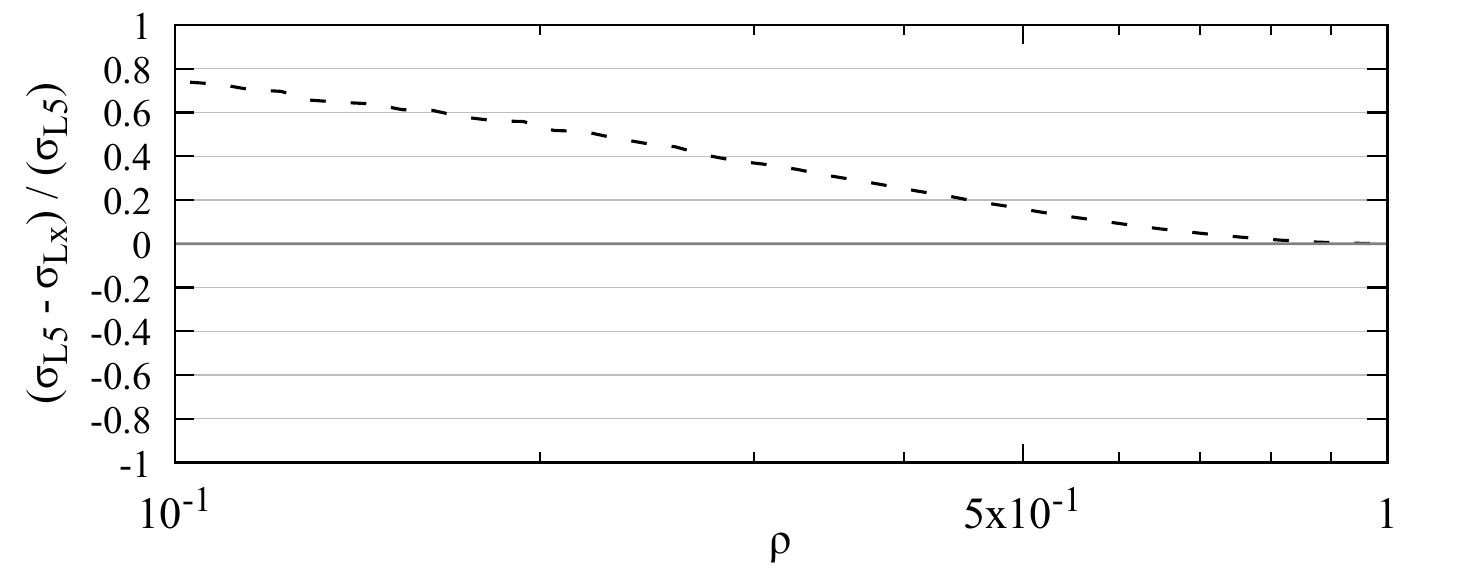}
\end{subfigure}
\caption{The veto cross-section for the $u$-channel contribution to $gg \to gg$. Solid: Full colour, Short-dashed: Leading colour.}
\label{fig:gg2gg-u}
\end{figure}

\begin{figure}[t]
\centering
\begin{subfigure}[t]{0.9\textwidth}
\centering
\includegraphics[width=1.0\textwidth]{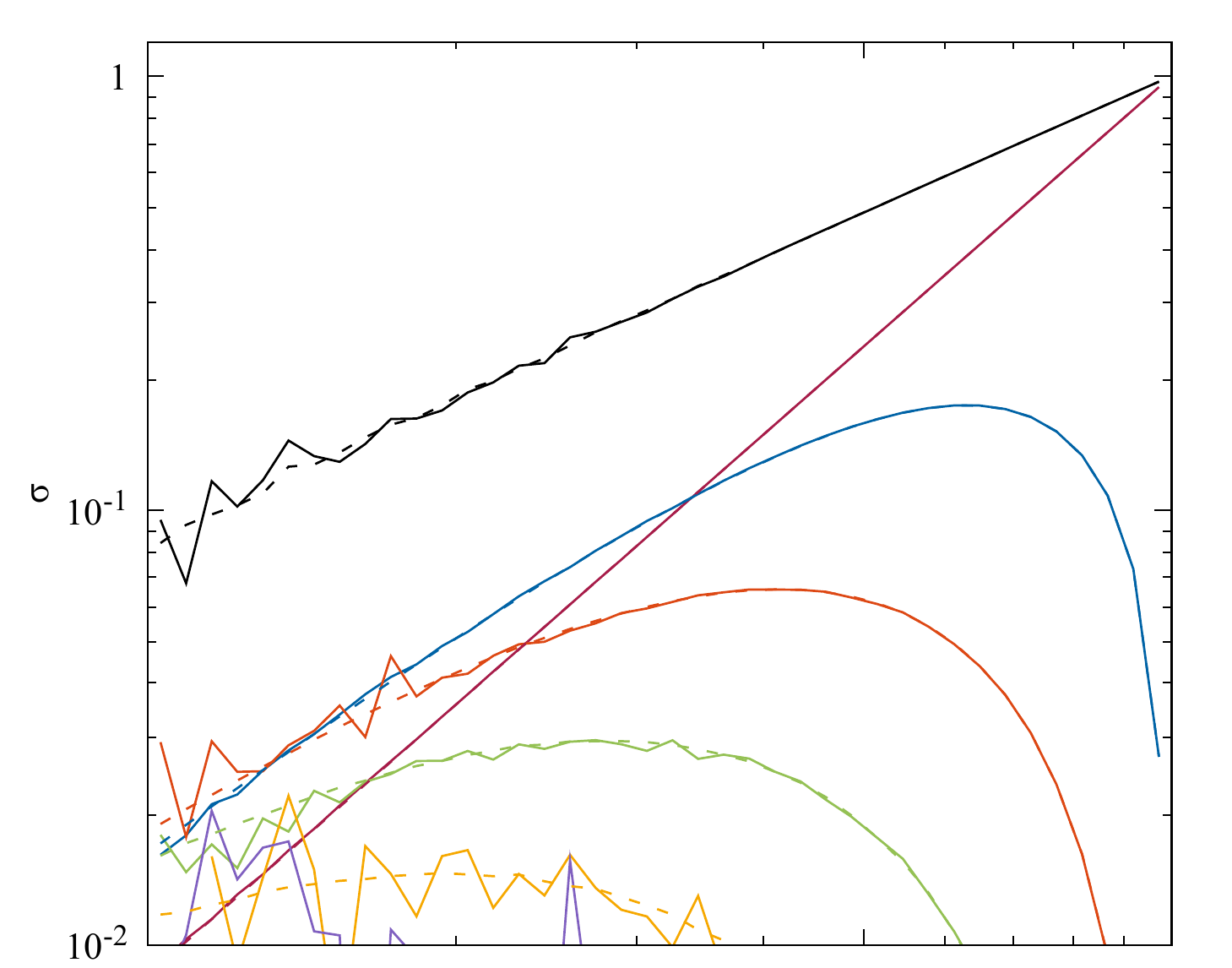}
\end{subfigure} \hfill
\begin{subfigure}[t]{0.9\textwidth}
\centering
\includegraphics[width=1.0\textwidth]{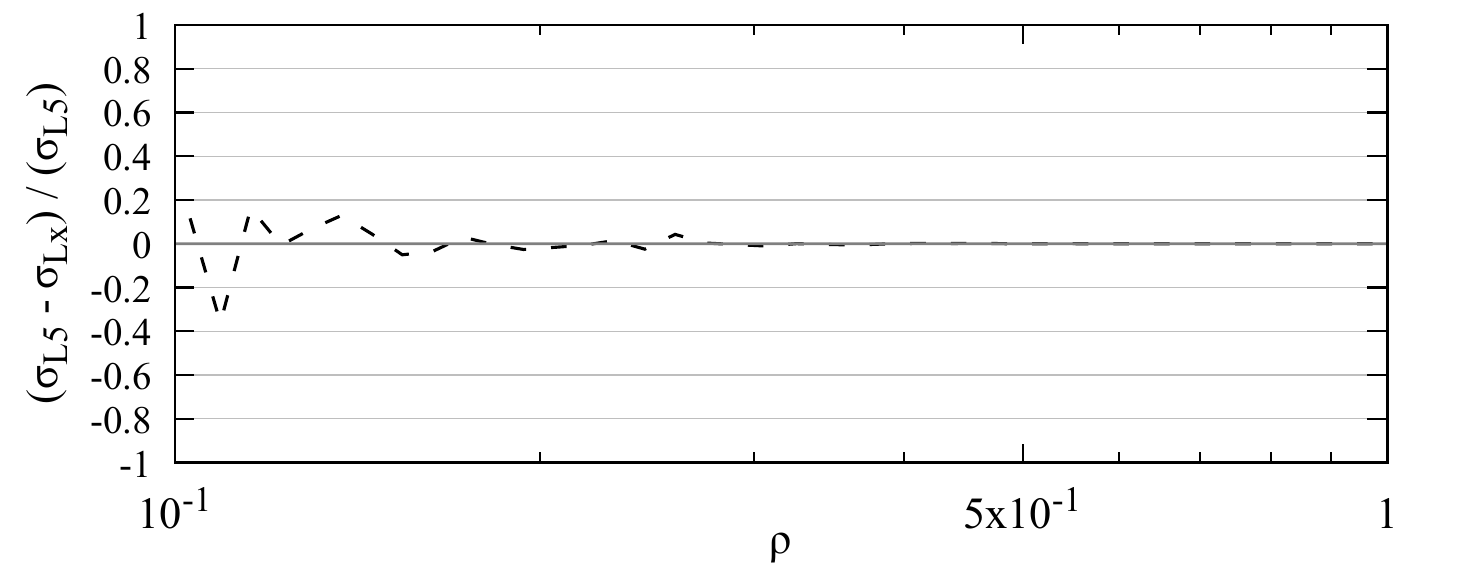}
\end{subfigure}
\caption{The veto cross-section for the $st$-interference contribution to $gg \to gg$. Solid: Full colour, Short-dashed: Leading colour.}
\label{fig:gg2gg-st}
\end{figure}

\begin{figure}[h]
\centering
\begin{subfigure}[t]{0.9\textwidth}
\centering
\includegraphics[width=1.0\textwidth]{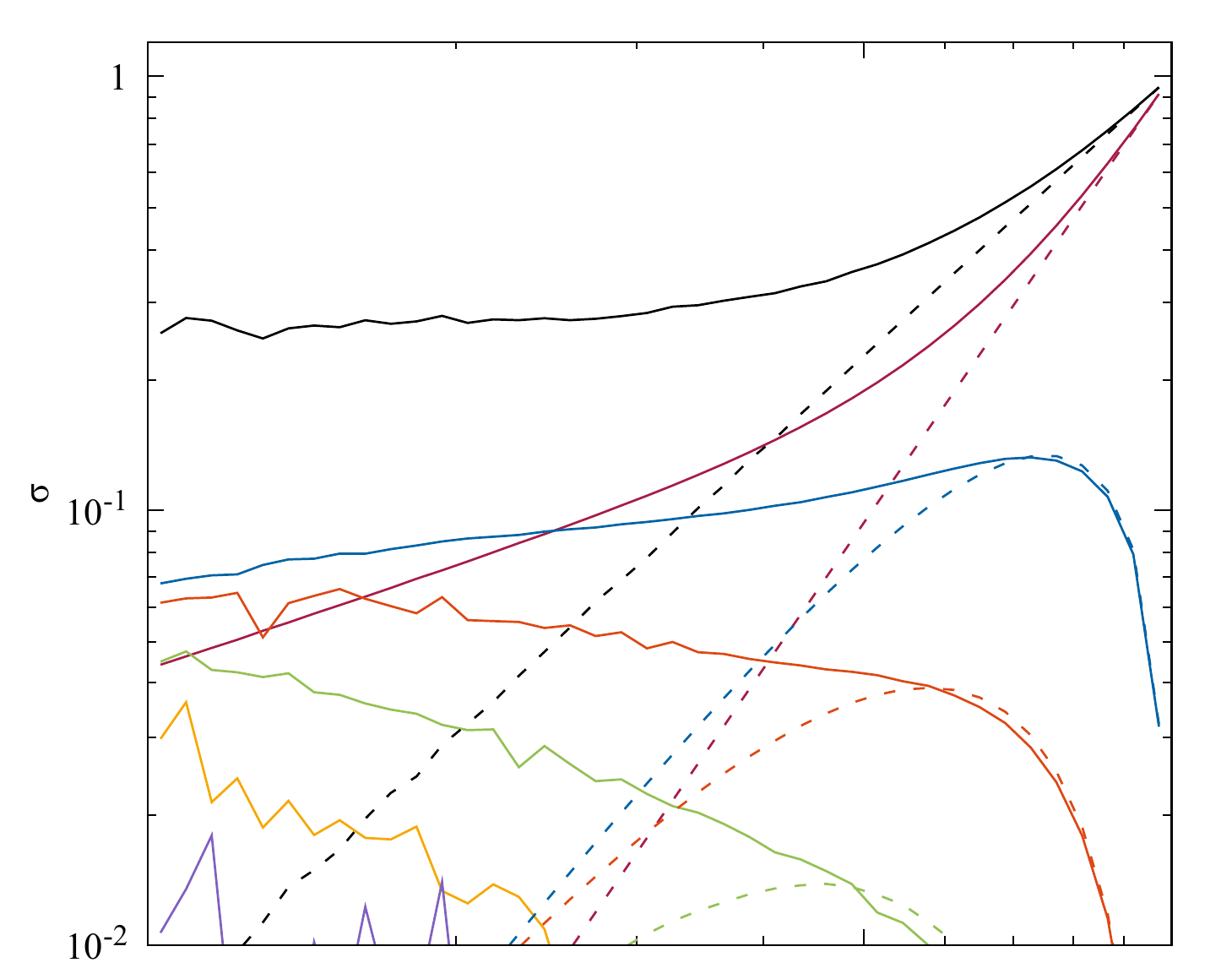}
\end{subfigure} \hfill
\begin{subfigure}[t]{0.9\textwidth}
\centering
\includegraphics[width=1.0\textwidth]{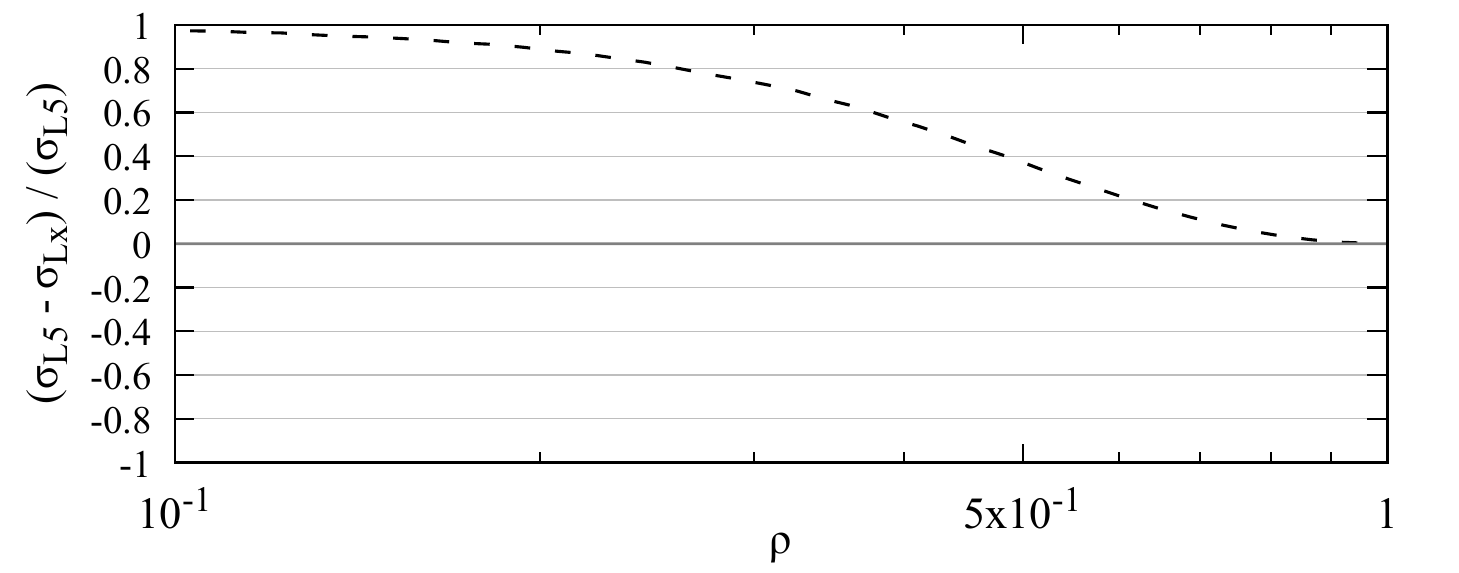}
\end{subfigure}
\caption{The veto cross-section for the $su$-interference contribution to $gg \to gg$. Solid: Full colour, Short-dashed: Leading colour.}
\label{fig:gg2gg-su}
\end{figure}

\begin{figure}[t]
\centering
\begin{subfigure}[t]{0.9\textwidth}
\centering
\includegraphics[width=1.0\textwidth]{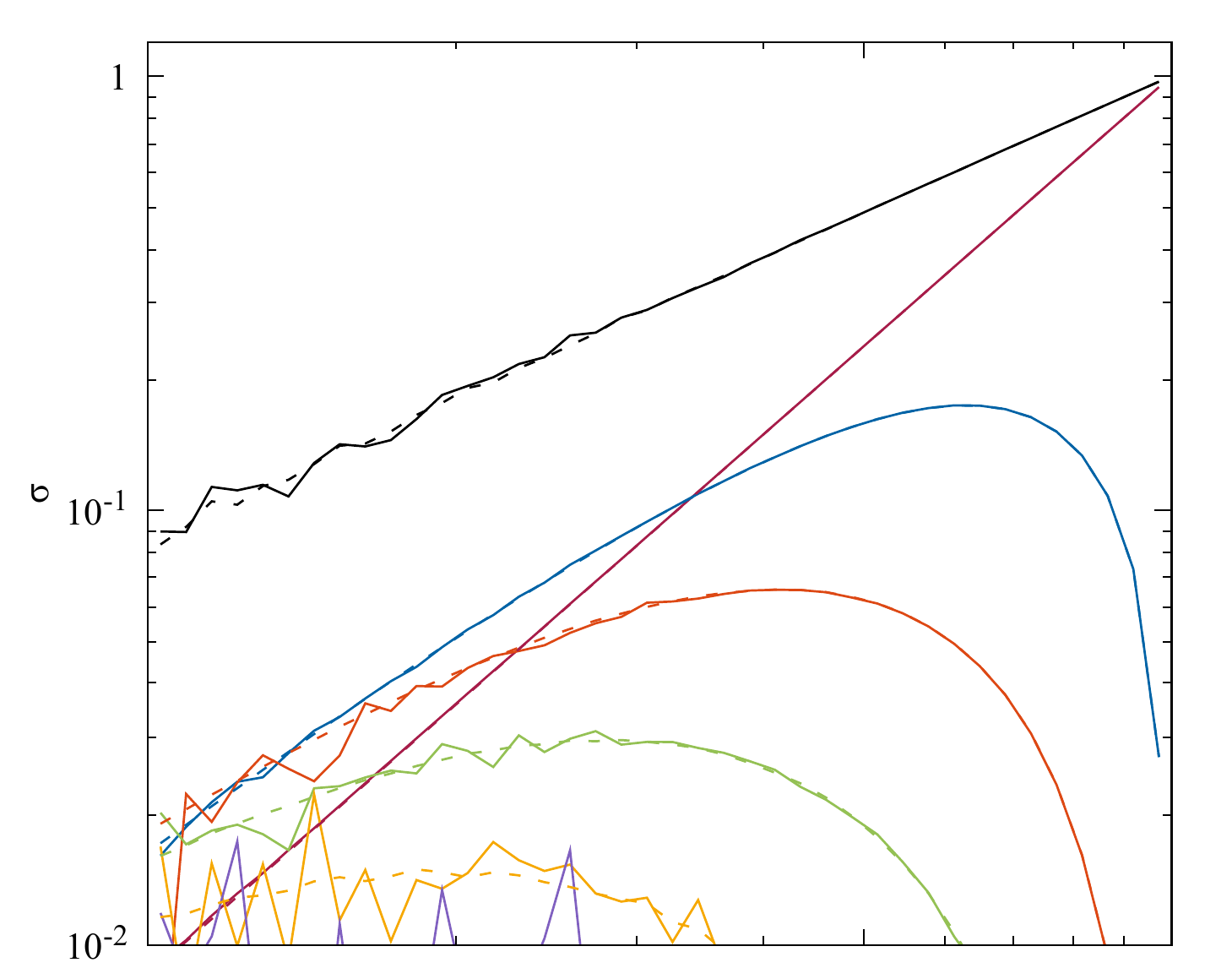}
\end{subfigure} \hfill
\begin{subfigure}[t]{0.9\textwidth}
\centering
\includegraphics[width=1.0\textwidth]{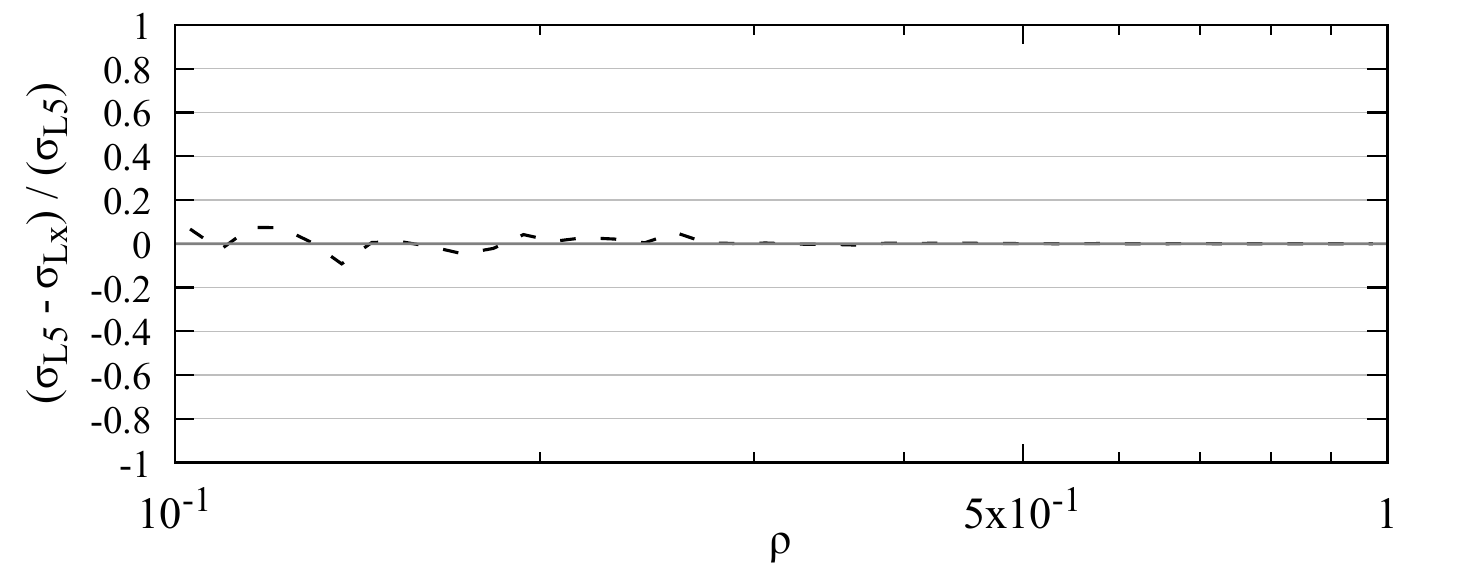}
\end{subfigure}
\caption{The veto cross-section for the $tu$-interference contribution to $gg \to gg$. Solid: Full colour, Short-dashed: Leading colour.}
\label{fig:gg2gg-tu}
\end{figure}

To complete our analysis of two-to-two scattering we study $gg \to gg$. In this case, we compare the full-colour result (FCR,FCV,FCH) with the strictly leading-colour approximation (LCR,LCV,LCH). Since the results do not depend much on whether we consider the symmetric or asymmetric veto cross-section we show results only for the asymmetric case in Fig.~\ref{fig:gg2gg-s} to Fig.~\ref{fig:gg2gg-tu}. The remarkable agreement between strict leading colour and full colour for $t$-channel processes persists and holds even for the $tu$ and $st$ interference contributions. Given the complexity of the subleading colour effects, the success of the vastly simpler leading colour approximation in the $t$-channel gluon exchange processes is quite remarkable, especially since we do observe very large differences in the evolution between $d=0$, $d=1$ and $d=2$ and emissions from rings and strings, which are absent in the leading colour approximation, greatly increase the possible trajectories in colour space. It seems there is a widespread and very non-trivial cancellation of sub-leading colour effects for these observables (subject to an overall normalization) the structure of which we will be investigating in more detail in an accompanying paper \cite{CVolver:eventgenerator}.

\section{Colour singlet production of four jets}
\label{sec:reconnect}

For our final study we will look at the production of four jets, as might be produced in $ ZZ \to q\bar{q} q\bar{q}$. In this case we fix the kinematics as in Table~\ref{tab:reconnect}.

\begin{table}[t]
    \centering
    \begin{tabular}{|c|c|c|}
        \hline
        & $\theta$ & $\phi$ \\
        \hline
   $q$ & 0 & 0 \\
   $\bar{q}$ & $2\pi/3$ & $5\pi/12$ \\
      $q$ & $11\pi/12$ & $\pi/2$ \\
         $\bar{q}$ & $3\pi/4$ & $\pi/4$ \\
        \hline
    \end{tabular}
    \caption{The directions of the quarks and anti-quarks produced in $ZZ \to q\bar{q}q\bar{q}$. The veto cross-section is defined by vetoing radiation with $E > Q_0$ outside of cones centred on each parton and with an opening angle $\pi/4$.}
    \label{tab:reconnect}
\end{table}

\begin{figure}[t]
\centering
\begin{subfigure}[t]{0.9\textwidth}
\centering
\includegraphics[width=1.0\textwidth]{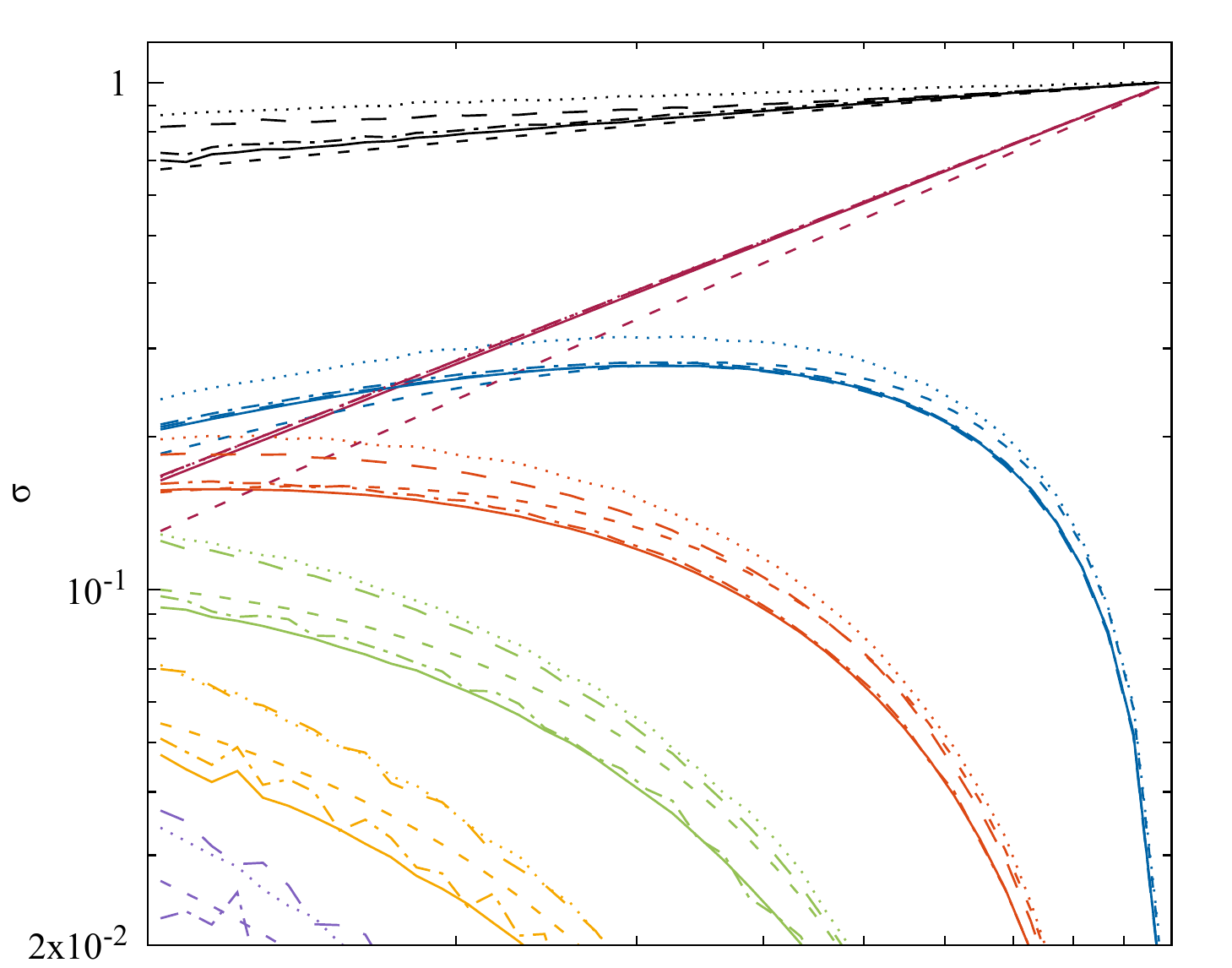}
\end{subfigure} \hfill
\begin{subfigure}[t]{0.9\textwidth}
\centering
\includegraphics[width=1.0\textwidth]{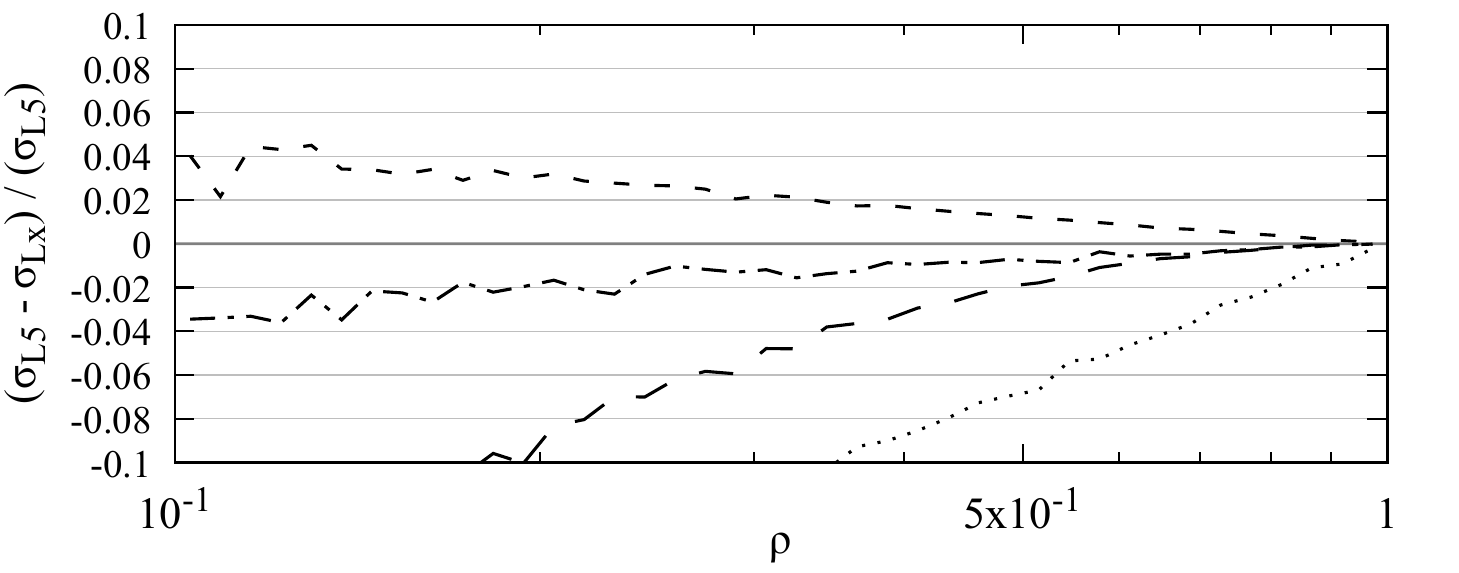}
\end{subfigure}
\caption{The veto cross-section for the $|01\>\<01|$ contribution to $ZZ \to q\bar{q}q\bar{q}$. Solid: Full colour (L5), Dash-dotted: \LC\ + FCR (L4), Long-dashed: \LC\ + LCR + singlets (L3), Dotted: \LC\ + LCR (L2), Short-dashed: strict LC (L1).}
\label{fig:qqqq0101}
\end{figure}

\begin{figure}[t]
\centering
\begin{subfigure}[t]{0.9\textwidth}
\centering
\includegraphics[width=1.0\textwidth]{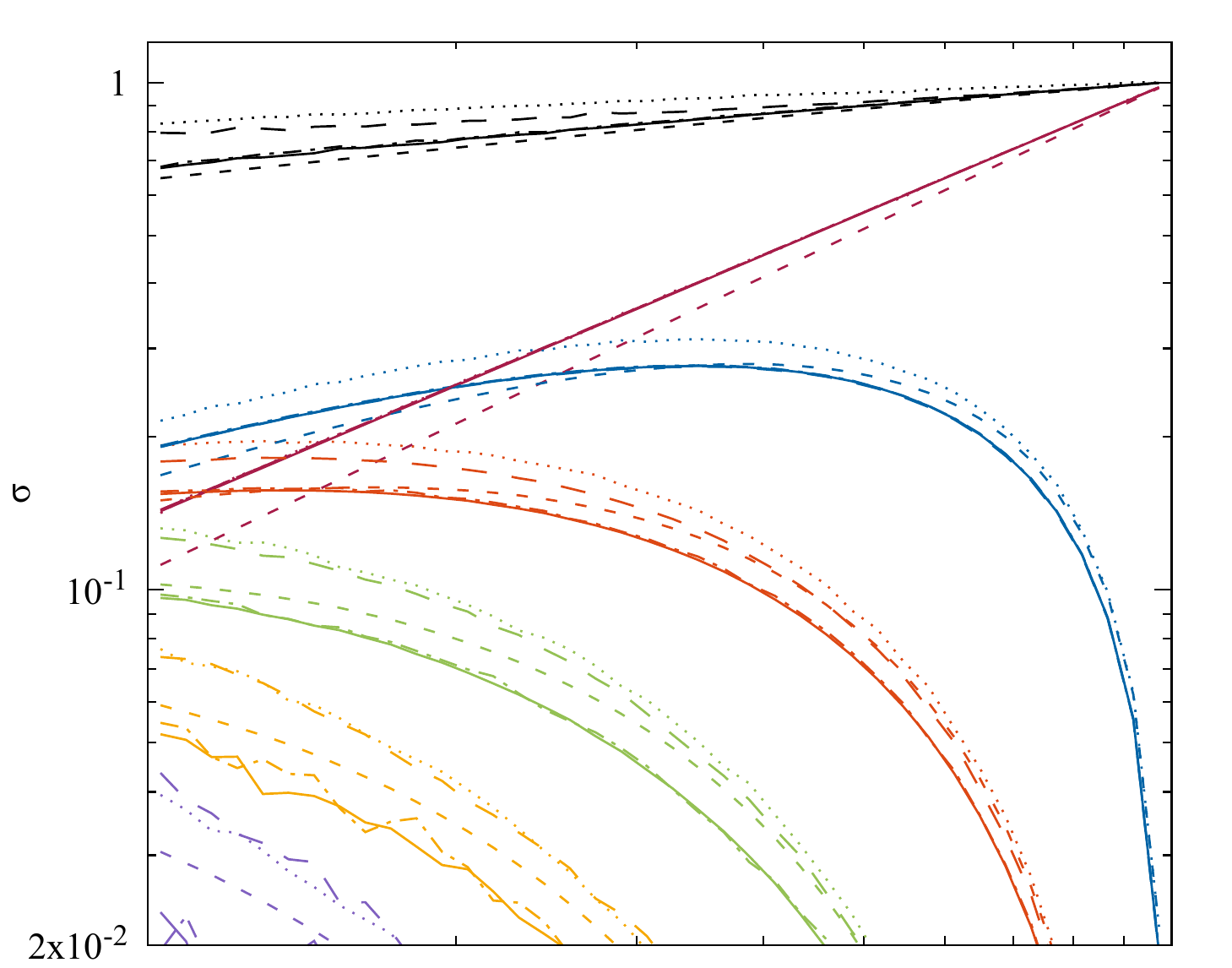}
\end{subfigure} \hfill
\begin{subfigure}[t]{0.9\textwidth}
\centering
\includegraphics[width=1.0\textwidth]{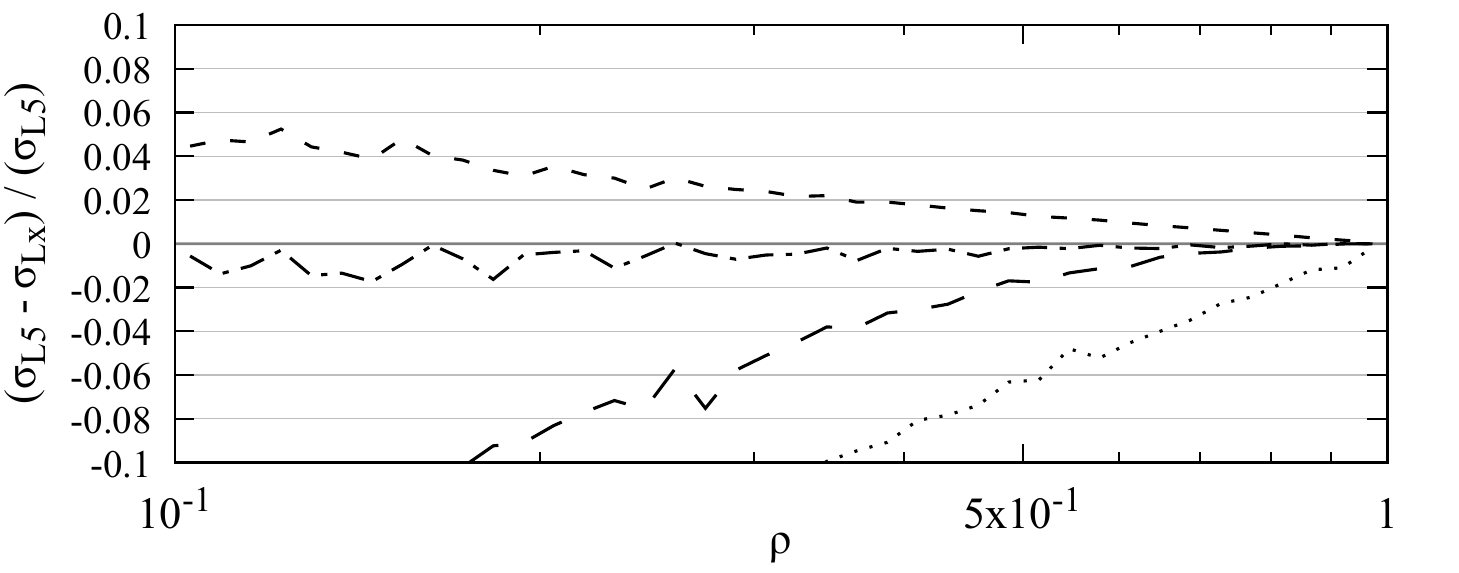}
\end{subfigure}
\caption{The veto cross-section for the $|10\>\<10|$ contribution to $ZZ \to q\bar{q}q\bar{q}$. Solid: Full colour (L5), Dash-dotted: \LC\ + FCR (L4), Long-dashed: \LC\ + LCR + singlets (L3), Dotted: \LC\ + LCR (L2), Short-dashed: strict LC (L1).}
\label{fig:qqqq1010}
\end{figure}

\begin{figure}[t]
\centering
\begin{subfigure}[t]{0.9\textwidth}
\centering
\includegraphics[width=1.0\textwidth]{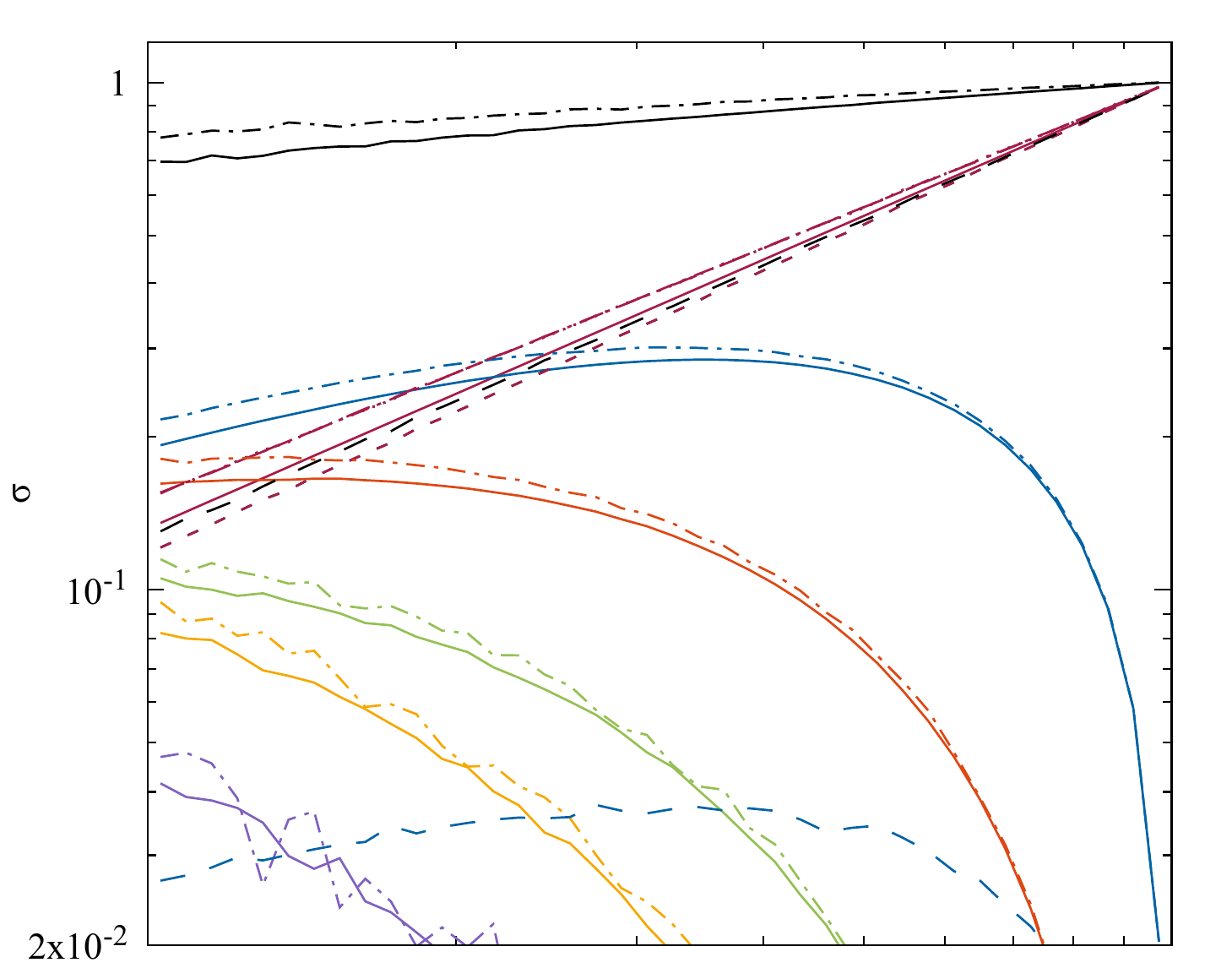}
\end{subfigure} \hfill
\begin{subfigure}[t]{0.9\textwidth}
\centering
\includegraphics[width=1.0\textwidth]{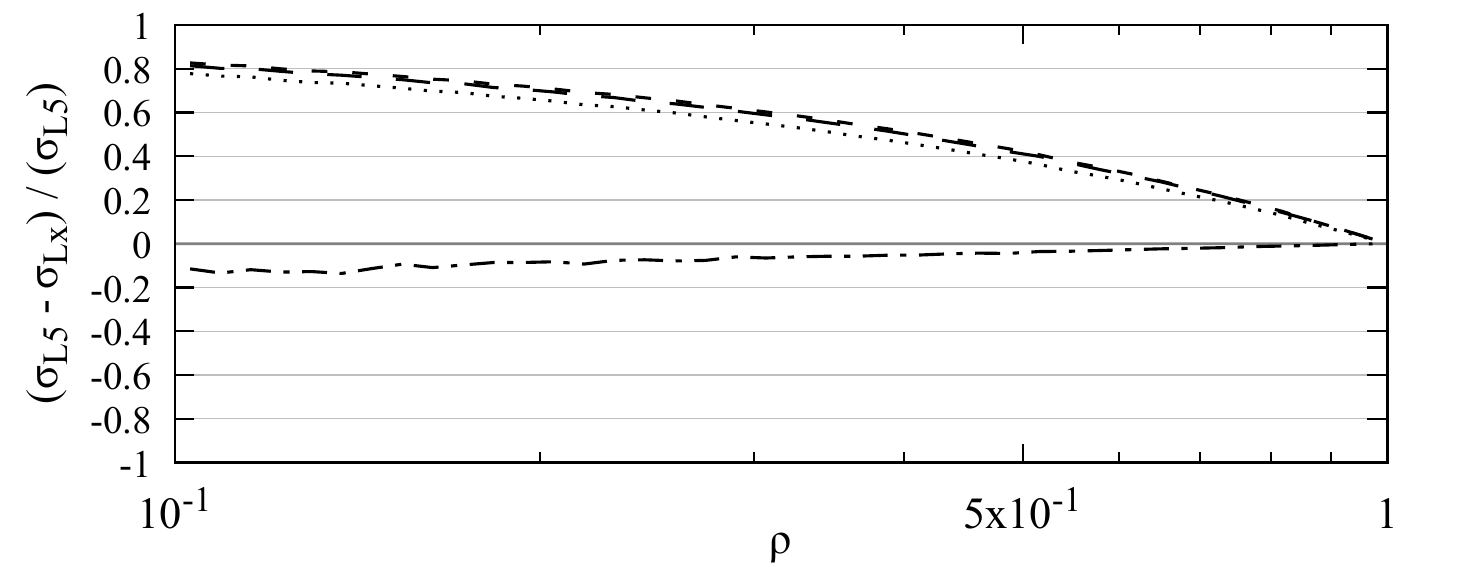}
\end{subfigure}
\caption{The veto cross-section for the $|01\>\<10|$ (interference) contribution to $ZZ \to q\bar{q}q\bar{q}$. Solid: Full colour (L5), Dash-dotted: \LC\ + FCR (L4), Long-dashed: \LC\ + LCR + singlets (L3), Dotted: \LC\ + LCR (L2), Short-dashed: strict LC (L1). The L3, 1 emission cuve (blue, long-dashed) is negative and the absolute value is plotted.}
\label{fig:qqqq1001}
\end{figure}

There are two possible colour flows, corresponding to the two possible production amplitudes: $|01\>$ and $|10\>$. In the former, the first two particles in Table~\ref{tab:reconnect} are colour connected and so are the final two. In the latter colour flow, it is the first and last particles (and the second and third particles) in the table that are colour connected. Fig.~\ref{fig:qqqq0101} shows that the L1 and L4 approximations are closest to the full colour result for the evolution of the $|01\rangle \langle 01|$ hard process but both differ by $4\%$ at $\rho=0.1$. For evolution starting from the crossed-channel $|10\rangle \langle 10|$ contribution, the L4 approximation is within 1\% of the full colour result (Fig.~\ref{fig:qqqq1010}). Finally, in Fig.~\ref{fig:qqqq1001} we once again see that none of the approximations is adequate and that approximations L1--L3 differ by 80\% from the full colour result.

\begin{figure}[t]
\centering
\begin{subfigure}[t]{0.9\textwidth}
\centering
\includegraphics[width=1.0\textwidth]{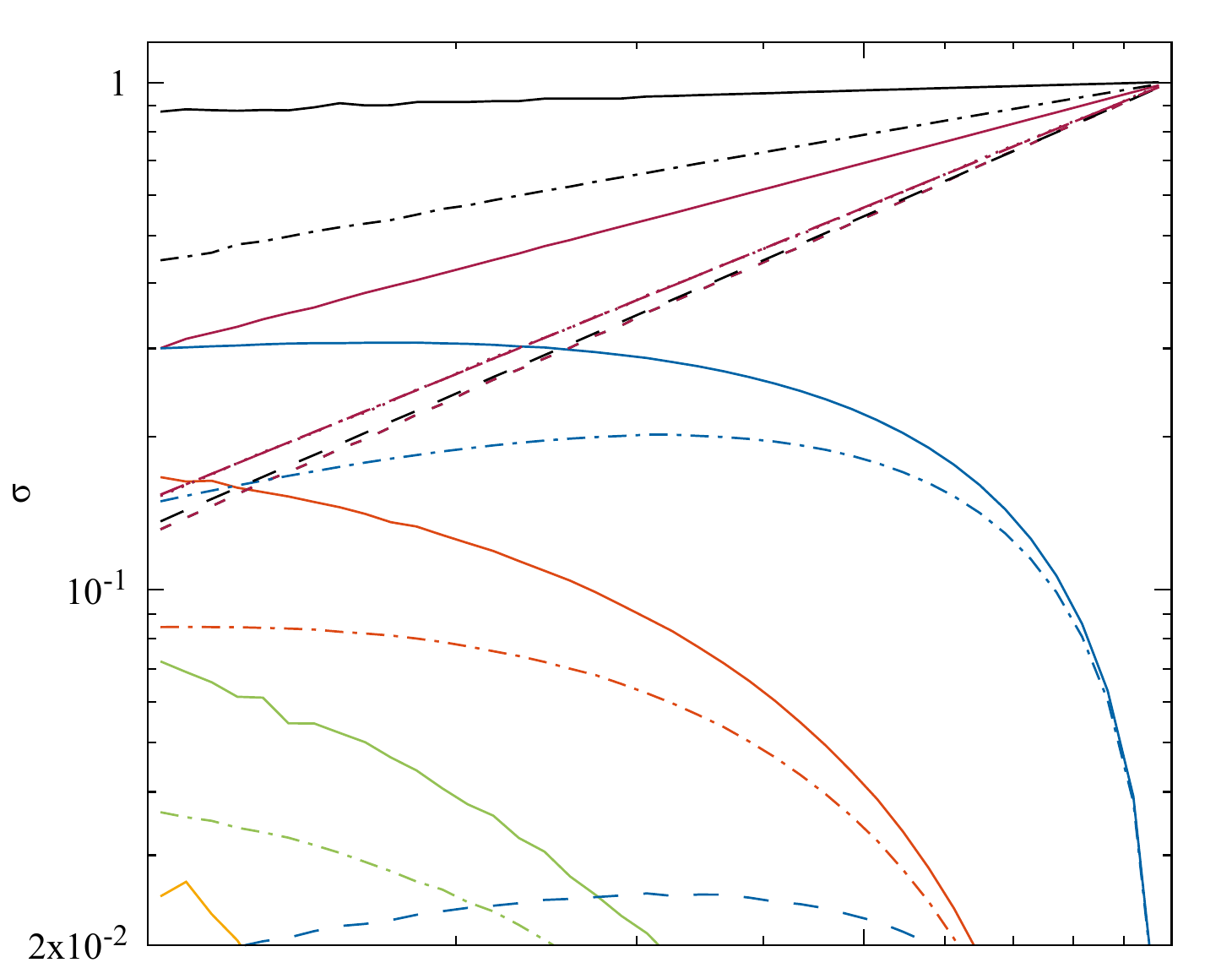}
\end{subfigure} \hfill
\begin{subfigure}[t]{0.9\textwidth}
\centering
\includegraphics[width=1.0\textwidth]{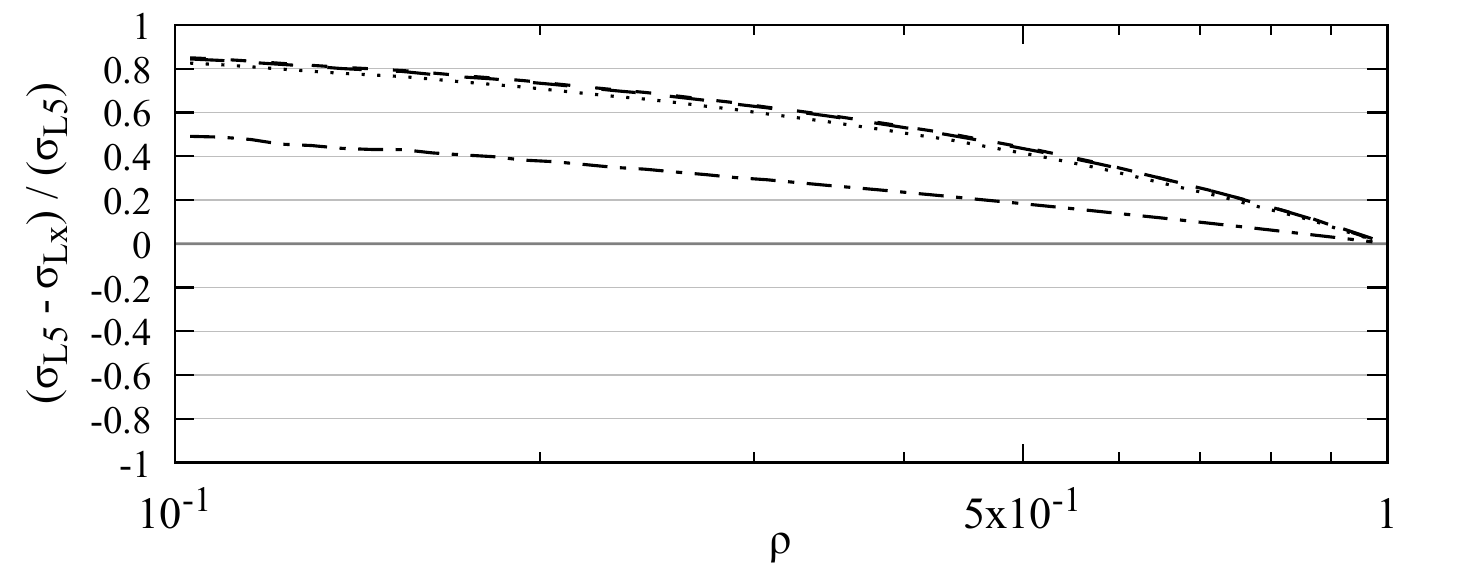}
\end{subfigure}
\caption{The veto cross-section for the $|10\>\<01|$ (interference) contribution to $q\bar{q} \to q\bar{q}$. Solid: Full colour (L5), Dash-dotted: \LC\ + FCR (L4), Long-dashed: \LC\ + LCR + singlets (L3), Dotted: \LC\ + LCR (L2), Short-dashed: strict LC (L1). The L3, 1 emission cuve (blue, long-dashed) is negative and the absolute value is plotted.}
\label{fig:qq2qq1001}
\end{figure}

\begin{figure}[t]
\centering
\begin{subfigure}[t]{0.9\textwidth}
\centering
\includegraphics[width=1.0\textwidth]{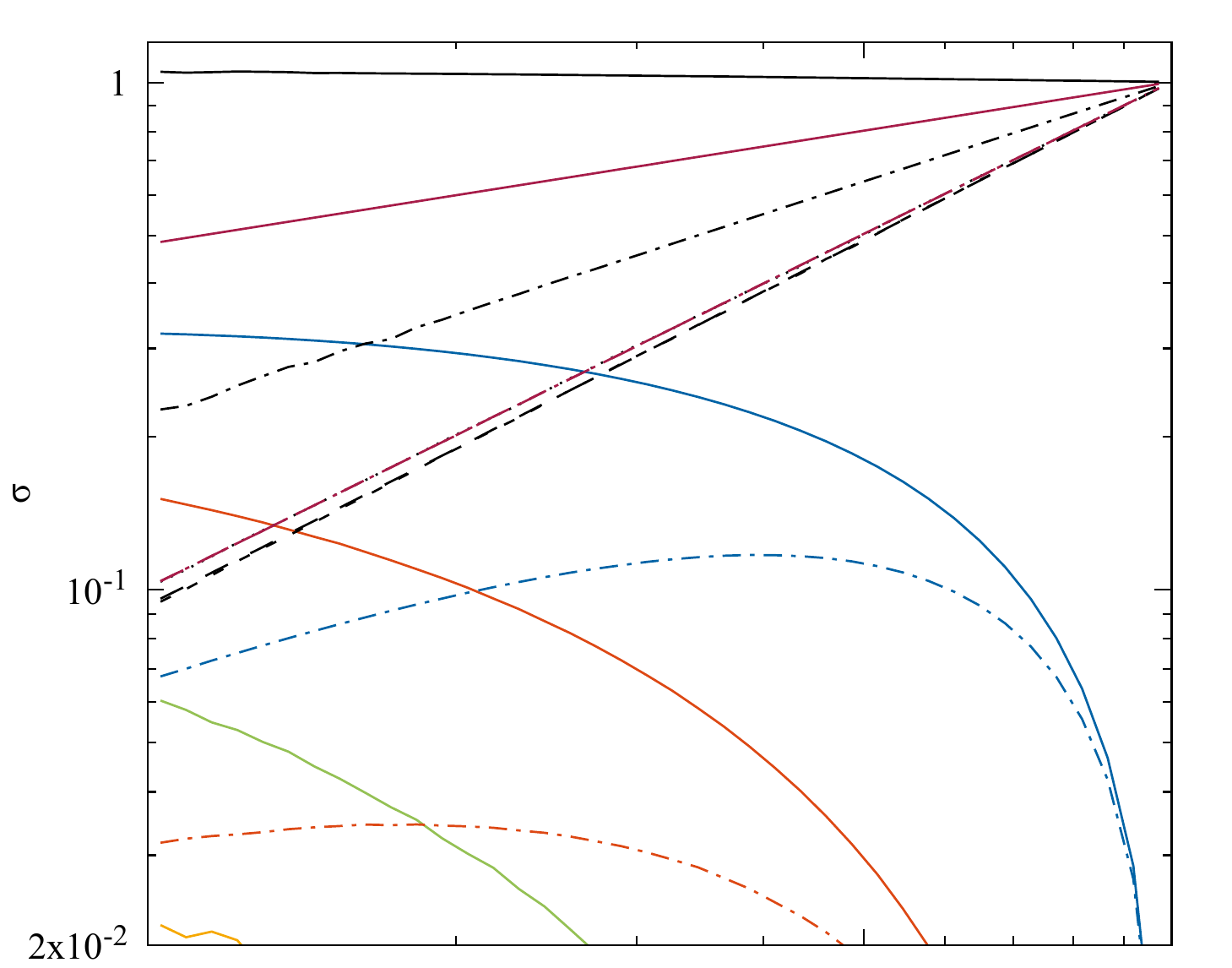}
\end{subfigure} \hfill
\begin{subfigure}[t]{0.9\textwidth}
\centering
\includegraphics[width=1.0\textwidth]{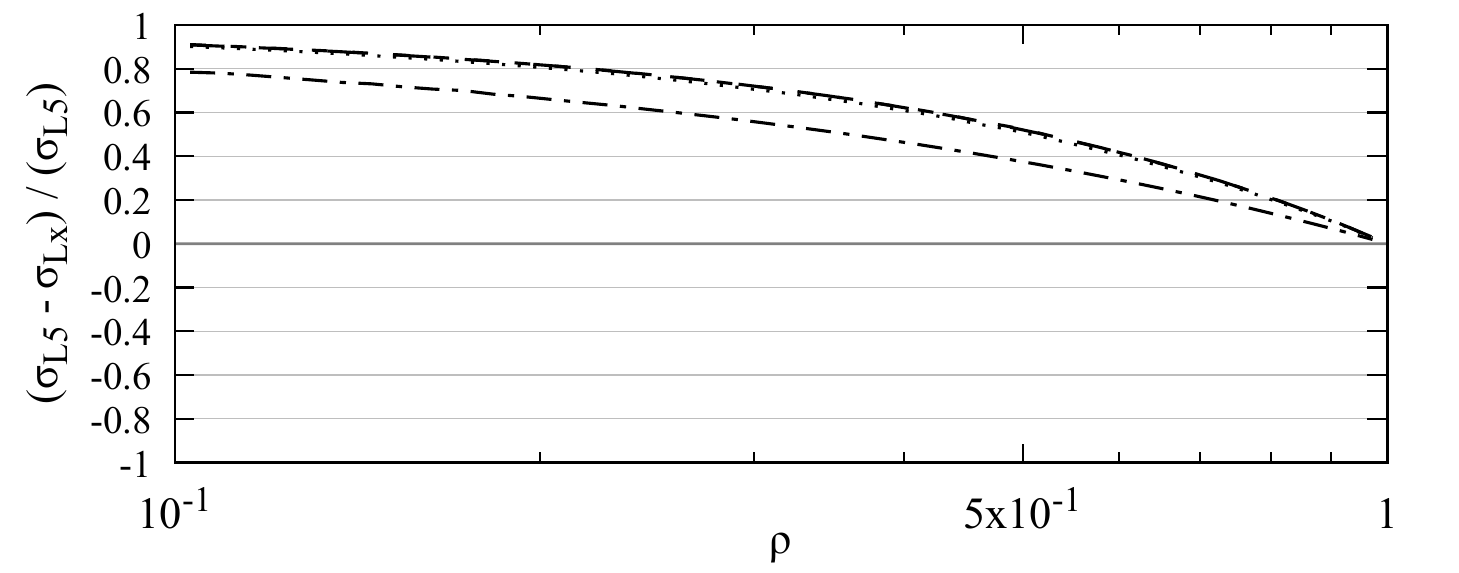}
\end{subfigure}
\caption{The veto cross-section for the $|10\>\<01|$ (interference) contribution to $q\bar{q} \to q\bar{q}$ in the asymmetric configuration. Solid: Full colour (L5), Dash-dotted: \LC\ + FCR (L4), Long-dashed: \LC\ + LCR + singlets (L3), Dotted: \LC\ + LCR (L2), Short-dashed: strict LC (L1).}
\label{fig:qq2qq1001a}
\end{figure}

\begin{figure}[t]
\centering
\begin{subfigure}[t]{0.9\textwidth}
\centering
\includegraphics[width=1.0\textwidth]{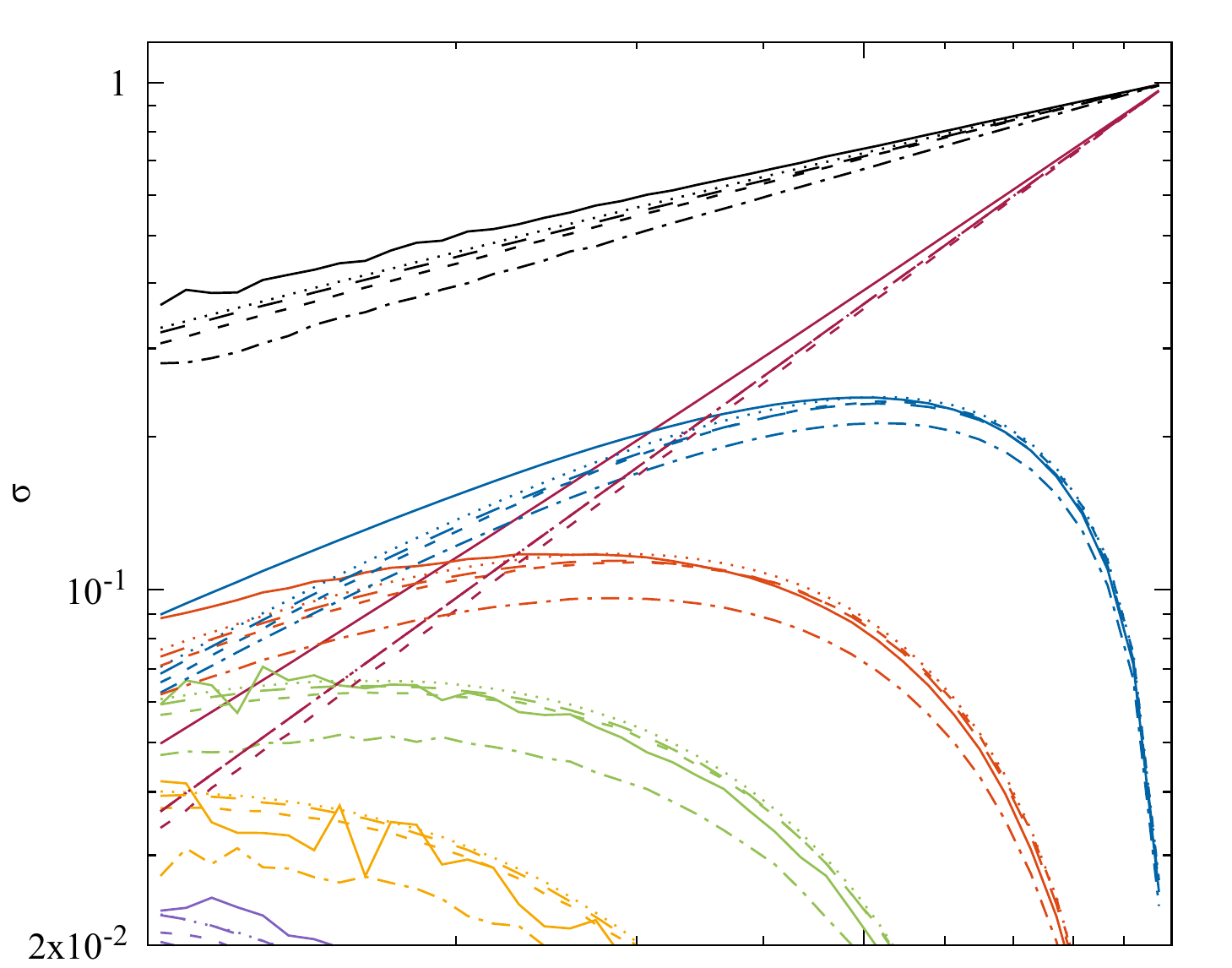}
\end{subfigure} \hfill
\begin{subfigure}[t]{0.9\textwidth}
\centering
\includegraphics[width=1.0\textwidth]{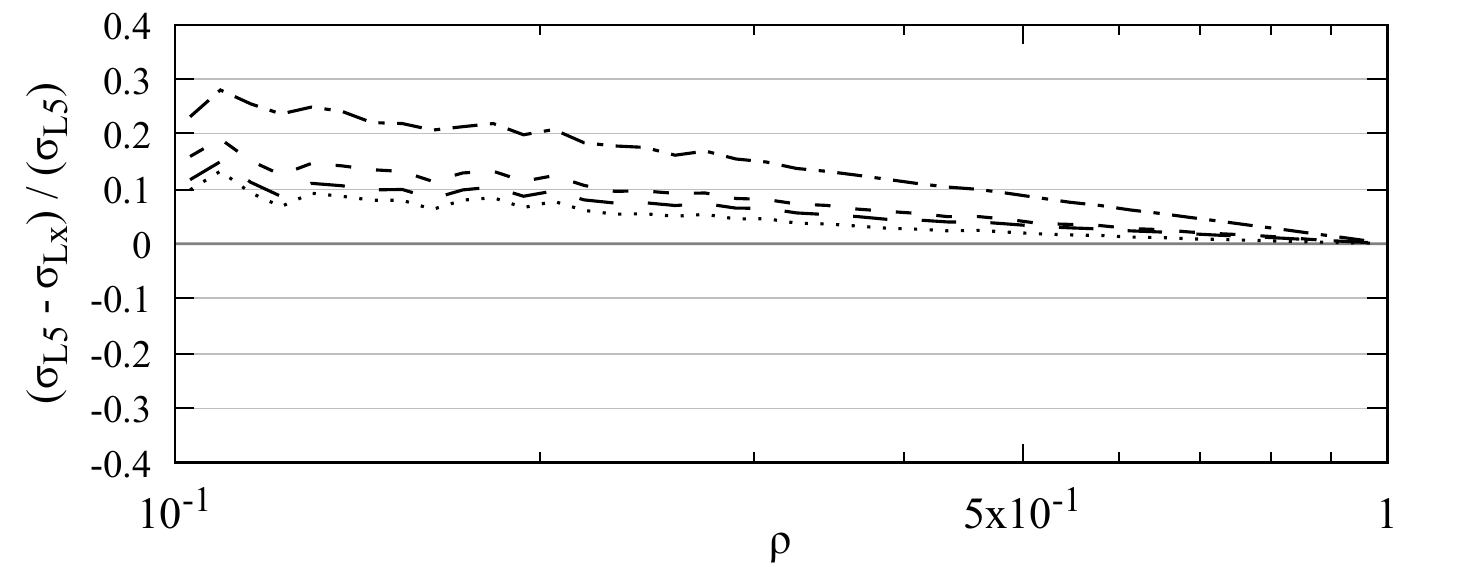}
\end{subfigure}
\caption{The veto cross-section for the $s$-channel quark exchange contribution to $qg \to qg$ in the symmetric configuration. Solid: Full colour (L5), Dash-dotted: \LC\ + FCR (L4), Long-dashed: \LC\ + LCR + singlets (L3), Dotted: \LC\ + LCR (L2), Short-dashed: strict LC (L1). Evolution starts from the leading-colour hard-scatter matrix.}
\label{fig:qg2qg-s-LCHs}
\end{figure}

\begin{figure}[t]
\centering
\begin{subfigure}[t]{0.9\textwidth}
\centering
\includegraphics[width=1.0\textwidth]{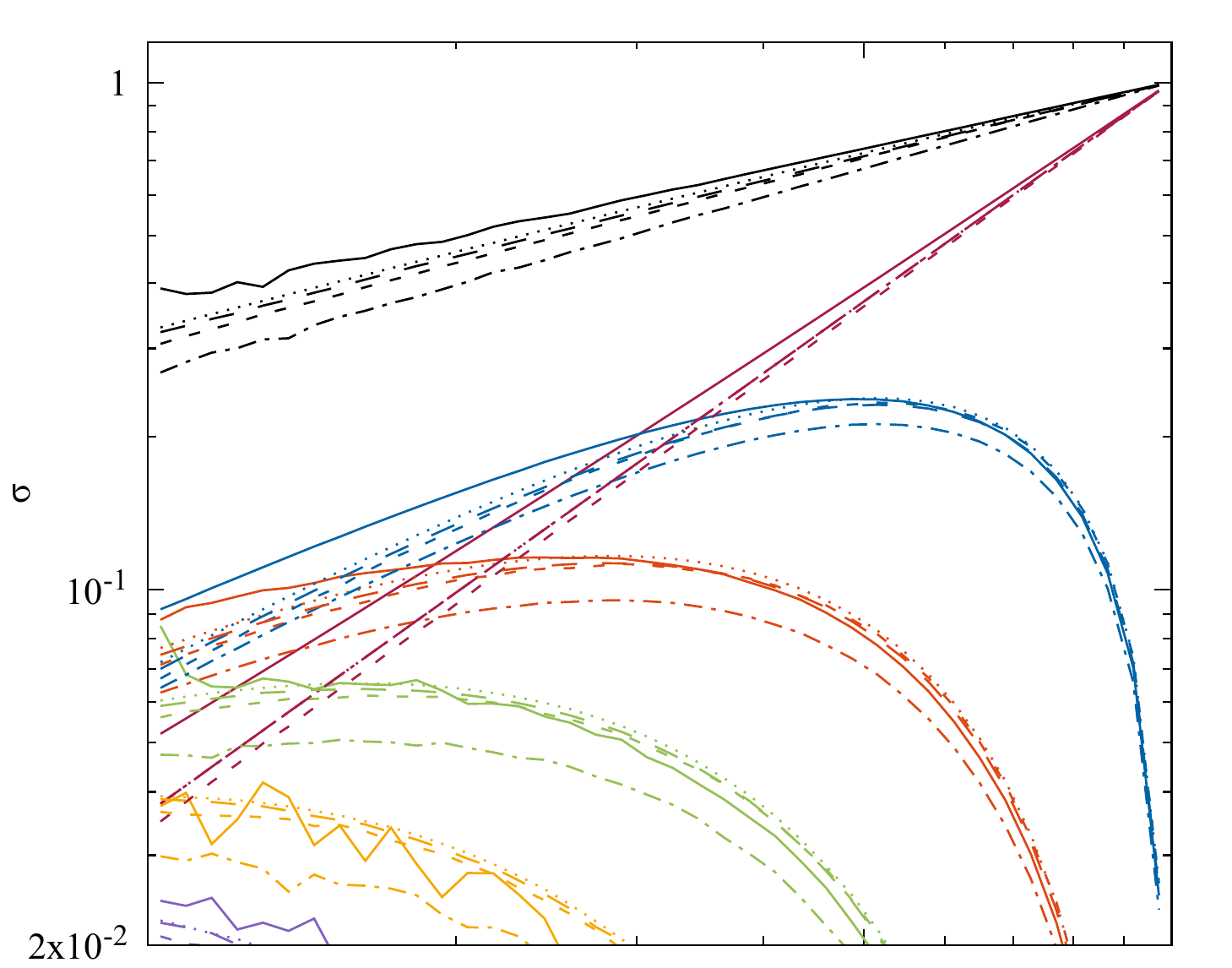}
\end{subfigure} \hfill
\begin{subfigure}[t]{0.9\textwidth}
\centering
\includegraphics[width=1.0\textwidth]{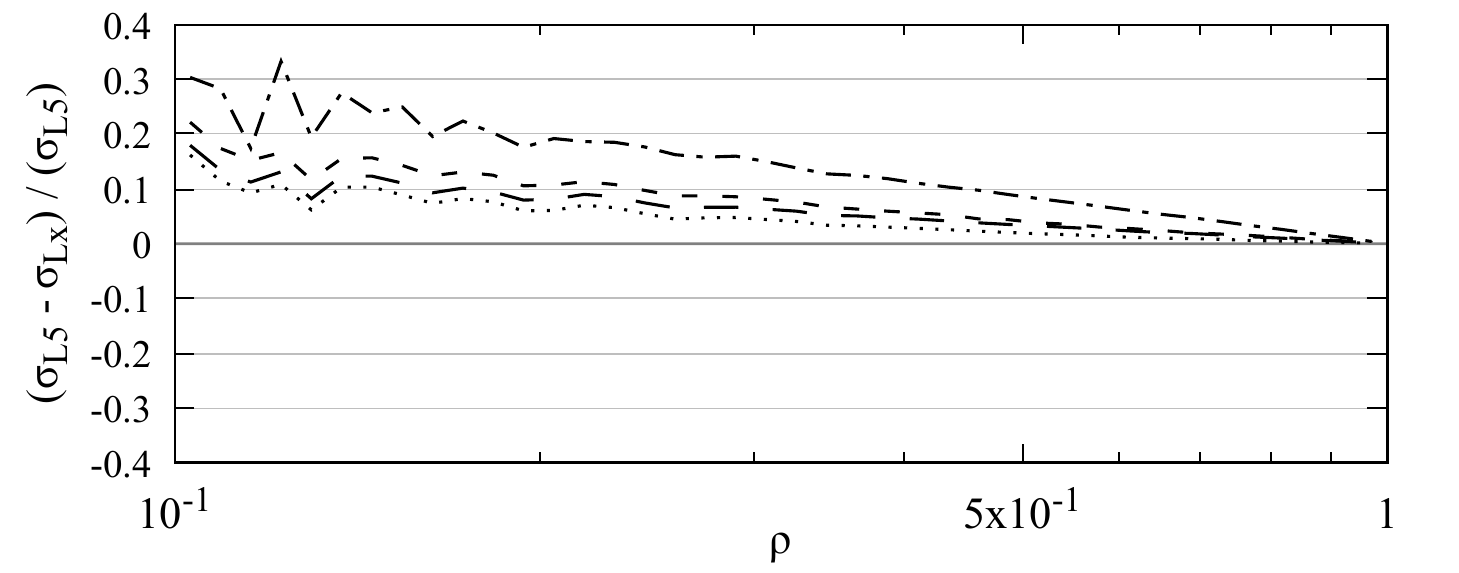}
\end{subfigure}
\caption{The veto cross-section for the $u$-channel quark exchange contribution to $qg \to qg$ in the symmetric configuration. Solid: Full colour (L5), Dash-dotted: \LC\ + FCR (L4), Long-dashed: \LC\ + LCR + singlets (L3), Dotted: \LC\ + LCR (L2), Short-dashed: strict LC (L1). Evolution starts from the leading-colour hard-scatter matrix.}
\label{fig:qg2qg-u-LCHs}
\end{figure}

There is a systematic hierarchy between L1, L2 and L3 present in all processes with quarks. L3 is always closer to L1 than L2 is, which implies that adding real singlet emissions counterbalances part of the effect of \LC . Therefore, in every diagram where L1 performs well then L3 will perform better than L2. The only cases where L2 performs better than L3 are those where L1 is not good and the overcorrection of \LC \ is in the right direction: Fig. \ref{fig:qq2qq0101}, where L2 performs best, and some cases where none of the approximations work well, specifically Figs. \ref{fig:qg2qg-s-LCHs} and \ref{fig:qg2qg-u-LCHs} and the evolutions of individiual interference terms, Figs. \ref{fig:qq2qq1001}, \ref{fig:qq2qq1001a} and \ref{fig:qqqq1001}.

\section{Conclusions}
\label{sec:conclusions}

The systematic calculation of sub-leading colour effects in general scattering processes is now within reach. In this paper we used \CVolver's soft gluon plugin to assess the size of sub-leading colour effects in the jet veto cross section for fixed parton kinematics and found substantial effects in many instances. We have also seen that in some specific processes there is a intriguing and highly-nontrivial cancellation of subleading colour corrections, as far as the veto scale dependence is concerned. In an accompanying paper \cite{CVolver:eventgenerator} we will show the results for other observables, obtained using a \CVolver\ plugin which operates in differential mode closer to a full-fledged event generator. For the near future, we aim to go beyond the soft gluon approximation and include hard-collinear emissions and incoming hadrons \cite{Forshaw:2019ver}. \CVolver\ will also able to explore the impact of collinear and higher order anomalous dimensions, and sophisticated models of hadronization following the factorised approach outlined in \cite{Platzer:2022jny}. Apart from allowing studies in collider phenomenology, these improvements will also allow us to explore the fascinating role of Coulomb gluons.

\acknowledgments This work has received funding from the U.K. Science and Technology Facilities Council grant no. ST/X00077X/1. FTG is supported by the Royal Society through Grant URF/R1/201500. We thank Jack Holguin for fruitful discussions, and Matthew De Angelis for earlier contributions to the program. The numerical results presented in this paper have been obtained on the computing clusters of the Particle Physics Groups of Universit\"at Wien and The University of Manchester. We are grateful for having been able to use these facilities. FTG acknowledges the kind hospitality of the Theoretical Physics Group of the Institute of Physics of the Universit\"at Graz.

\bibliography{refs}

\providecommand{\href}[2]{#2}\begingroup\raggedright\begin{thebibliography}{10}

\bibitem{Bewick:2023tfi}
G.~Bewick et~al., \emph{{Herwig 7.3 release note}},
  \href{https://doi.org/10.1140/epjc/s10052-024-13211-9}{\emph{Eur. Phys. J. C}
  {\bfseries 84} (2024) 1053}
  [\href{https://arxiv.org/abs/2312.05175}{{\ttfamily 2312.05175}}].

\bibitem{Sherpa:2024mfk}
{\scshape Sherpa} collaboration, \emph{{Event generation with Sherpa 3}},
  \href{https://doi.org/10.1007/JHEP12(2024)156}{\emph{JHEP} {\bfseries 12}
  (2024) 156} [\href{https://arxiv.org/abs/2410.22148}{{\ttfamily
  2410.22148}}].

\bibitem{Bierlich:2022pfr}
C.~Bierlich et~al., \emph{{A comprehensive guide to the physics and usage of
  PYTHIA 8.3}},
  \href{https://doi.org/10.21468/SciPostPhysCodeb.8}{\emph{SciPost Phys.
  Codeb.} {\bfseries 2022} (2022) 8}
  [\href{https://arxiv.org/abs/2203.11601}{{\ttfamily 2203.11601}}].

\bibitem{vanBeekveld:2024wws}
M.~van Beekveld et~al., \emph{{New Standard for the Logarithmic Accuracy of
  Parton Showers}},
  \href{https://doi.org/10.1103/PhysRevLett.134.011901}{\emph{Phys. Rev. Lett.}
  {\bfseries 134} (2025) 011901}
  [\href{https://arxiv.org/abs/2406.02661}{{\ttfamily 2406.02661}}].

\bibitem{Nagy:2012bt}
Z.~Nagy and D.E.~Soper, \emph{{Parton shower evolution with subleading color}},
  \href{https://doi.org/10.1007/JHEP06(2012)044}{\emph{JHEP} {\bfseries 06}
  (2012) 044} [\href{https://arxiv.org/abs/1202.4496}{{\ttfamily 1202.4496}}].

\bibitem{Nagy:2015hwa}
Z.~Nagy and D.E.~Soper, \emph{{Effects of subleading color in a parton
  shower}}, \href{https://doi.org/10.1007/JHEP07(2015)119}{\emph{JHEP}
  {\bfseries 07} (2015) 119}
  [\href{https://arxiv.org/abs/1501.00778}{{\ttfamily 1501.00778}}].

\bibitem{Nagy:2019pjp}
Z.~Nagy and D.E.~Soper, \emph{{Parton showers with more exact color
  evolution}}, \href{https://doi.org/10.1103/PhysRevD.99.054009}{\emph{Phys.
  Rev. D} {\bfseries 99} (2019) 054009}
  [\href{https://arxiv.org/abs/1902.02105}{{\ttfamily 1902.02105}}].

\bibitem{Platzer:2013fha}
S.~Pl\"atzer, \emph{{Summing Large-$N$ Towers in Colour Flow Evolution}},
  \href{https://doi.org/10.1140/epjc/s10052-014-2907-2}{\emph{Eur. Phys. J. C}
  {\bfseries 74} (2014) 2907}
  [\href{https://arxiv.org/abs/1312.2448}{{\ttfamily 1312.2448}}].

\bibitem{DeAngelis:2020rvq}
M.~De~Angelis, J.R.~Forshaw and S.~Pl\"atzer, \emph{{Resummation and Simulation
  of Soft Gluon Effects beyond Leading Color}},
  \href{https://doi.org/10.1103/PhysRevLett.126.112001}{\emph{Phys. Rev. Lett.}
  {\bfseries 126} (2021) 112001}
  [\href{https://arxiv.org/abs/2007.09648}{{\ttfamily 2007.09648}}].

\bibitem{CVolver:eventgenerator}
{Forshaw, Jeffrey R.}, {Pl\"atzer, Simon} and {{Torre Gonz\'alez}, Fernando},
  ``{Differential Resummation of Soft Gluon Effects at Amplitude Level}.''
  2025.

\bibitem{AngelesMartinez:2018cfz}
R.~\'Angeles~Mart\'\i{}nez, M.~De~Angelis, J.R.~Forshaw, S.~Pl\"atzer and
  M.H.~Seymour, \emph{{Soft gluon evolution and non-global logarithms}},
  \href{https://doi.org/10.1007/JHEP05(2018)044}{\emph{JHEP} {\bfseries 05}
  (2018) 044} [\href{https://arxiv.org/abs/1802.08531}{{\ttfamily
  1802.08531}}].

\bibitem{Platzer:2022jny}
S.~Pl\"atzer, \emph{{Colour evolution and infrared physics}},
  \href{https://doi.org/10.1007/JHEP07(2023)126}{\emph{JHEP} {\bfseries 07}
  (2023) 126} [\href{https://arxiv.org/abs/2204.06956}{{\ttfamily
  2204.06956}}].

\bibitem{Forshaw:2021mtj}
J.R.~Forshaw, J.~Holguin and S.~Pl\"atzer, \emph{{Rings and strings: a basis
  for understanding subleading colour and QCD coherence beyond the two-jet
  limit}}, \href{https://doi.org/10.1007/JHEP05(2022)190}{\emph{JHEP}
  {\bfseries 05} (2022) 190}
  [\href{https://arxiv.org/abs/2112.13124}{{\ttfamily 2112.13124}}].

\bibitem{Hatta:2013iba}
Y.~Hatta and T.~Ueda, \emph{{Resummation of non-global logarithms at finite
  $N_c$}}, \href{https://doi.org/10.1016/j.nuclphysb.2013.06.021}{\emph{Nucl.
  Phys. B} {\bfseries 874} (2013) 808}
  [\href{https://arxiv.org/abs/1304.6930}{{\ttfamily 1304.6930}}].

\bibitem{Hatta:2020wre}
Y.~Hatta and T.~Ueda, \emph{{Non-global logarithms in hadron collisions at
  $N_c$ = 3}},
  \href{https://doi.org/10.1016/j.nuclphysb.2020.115273}{\emph{Nucl. Phys. B}
  {\bfseries 962} (2021) 115273}
  [\href{https://arxiv.org/abs/2011.04154}{{\ttfamily 2011.04154}}].

\bibitem{Forshaw:2006fk}
J.R.~Forshaw, A.~Kyrieleis and M.H.~Seymour, \emph{{Super-leading logarithms in
  non-global observables in QCD}},
  \href{https://doi.org/10.1088/1126-6708/2006/08/059}{\emph{JHEP} {\bfseries
  08} (2006) 059} [\href{https://arxiv.org/abs/hep-ph/0604094}{{\ttfamily
  hep-ph/0604094}}].

\bibitem{Forshaw:2008cq}
J.R.~Forshaw, A.~Kyrieleis and M.H.~Seymour, \emph{{Super-leading logarithms in
  non-global observables in QCD: Colour basis independent calculation}},
  \href{https://doi.org/10.1088/1126-6708/2008/09/128}{\emph{JHEP} {\bfseries
  09} (2008) 128} [\href{https://arxiv.org/abs/0808.1269}{{\ttfamily
  0808.1269}}].

\bibitem{Becher:2021zkk}
T.~Becher, M.~Neubert and D.Y.~Shao, \emph{{Resummation of Super-Leading
  Logarithms}},
  \href{https://doi.org/10.1103/PhysRevLett.127.212002}{\emph{Phys. Rev. Lett.}
  {\bfseries 127} (2021) 212002}
  [\href{https://arxiv.org/abs/2107.01212}{{\ttfamily 2107.01212}}].

\bibitem{Becher:2023mtx}
T.~Becher, M.~Neubert, D.Y.~Shao and M.~Stillger, \emph{{Factorization of
  non-global LHC observables and resummation of super-leading logarithms}},
  \href{https://doi.org/10.1007/JHEP12(2023)116}{\emph{JHEP} {\bfseries 12}
  (2023) 116} [\href{https://arxiv.org/abs/2307.06359}{{\ttfamily
  2307.06359}}].

\bibitem{Boer:2024hzh}
P.~B\"oer, P.~Hager, M.~Neubert, M.~Stillger and X.~Xu,
  \emph{{Renormalization-group improved resummation of super-leading
  logarithms}}, \href{https://doi.org/10.1007/JHEP08(2024)035}{\emph{JHEP}
  {\bfseries 08} (2024) 035}
  [\href{https://arxiv.org/abs/2405.05305}{{\ttfamily 2405.05305}}].

\bibitem{Becher:2024nqc}
T.~Becher, P.~Hager, G.~Martinelli, M.~Neubert, D.~Schwienbacher and
  M.~Stillger, \emph{{Super-leading logarithms in pp $\to$ 2 jets}},
  \href{https://doi.org/10.1007/JHEP01(2025)171}{\emph{JHEP} {\bfseries 01}
  (2025) 171} [\href{https://arxiv.org/abs/2411.12742}{{\ttfamily
  2411.12742}}].

\bibitem{Forshaw:2019ver}
J.R.~Forshaw, J.~Holguin and S.~Pl\"atzer, \emph{{Parton branching at amplitude
  level}}, \href{https://doi.org/10.1007/JHEP08(2019)145}{\emph{JHEP}
  {\bfseries 08} (2019) 145}
  [\href{https://arxiv.org/abs/1905.08686}{{\ttfamily 1905.08686}}].

\end{thebibliography}\endgroup

\clearpage

\clearpage

\end{document}